\documentclass[aps,prd,twocolumn,floatfix,nofootinbib,superscriptaddress,tightenlines]{revtex4}
\usepackage[dvipsnames]{xcolor}
\usepackage{amsmath}
\usepackage{dcolumn}
\usepackage{lipsum}
\usepackage{amssymb}
\usepackage{soul}
\usepackage{url}
\usepackage{epsfig}
\usepackage{graphicx}
\usepackage{amsmath}
\usepackage{bm}
\usepackage{setspace}
\usepackage{appendix}
\usepackage{lscape}
\usepackage{amsthm}
\usepackage{bbold}
\usepackage{dcolumn}
\usepackage{epsfig}
\usepackage{graphics}
\usepackage{graphicx}
\usepackage[utf8]{inputenc}

\usepackage{natbib}
\usepackage{graphicx}
\usepackage{dcolumn}
\usepackage{bm}
\usepackage{amsmath}
\usepackage{float}
\usepackage{multirow}
\usepackage{slashed}
\usepackage{xcolor}
\usepackage{physics}
\usepackage{multirow}
\usepackage{gensymb}
\usepackage{mathtools,braket}
\usepackage{subcaption}
\usepackage{lipsum}  
\usepackage{color}
\usepackage{soul}

\usepackage{bm}
\usepackage{xspace}
\usepackage{cancel}
\usepackage{float}
\usepackage{multirow}
\definecolor{darkgreen}{rgb}{0,0.5,0}
\definecolor{purple}{rgb}{0.5,0,0.5}
\definecolor{nblue}{rgb}{0.0,0.0,0.50}
\definecolor{scarlet}{rgb}{1.0,0.2,0}
\definecolor{darkmagenta}{rgb}{0.55, 0.0, 0.55}
\definecolor{darkolivegreen}{rgb}{0.33, 0.42, 0.18}
\definecolor{darkcandyapplered}{rgb}{0.64, 0.0, 0.0}

\usepackage[colorlinks=true, 
linkcolor=purple, citecolor= purple, urlcolor=blue]{hyperref}

\newcommand{\be}{\begin{equation}}

\newcommand{\bfk}{\textcolor{black}{\textbf{k}^2_\perp}}

\newcommand{\ee}{\end{equation}}
\newcommand{\bea}{\begin{eqnarray}}
\newcommand{\eea}{\end{eqnarray}}
\newcommand{\beas}{\begin{eqnarray*}}
\newcommand{\eeas}{\end{eqnarray*}}

\begin{document}
\title{Valence quark distribution of rho meson using light-front quark model}

\author{Satyajit Puhan}
\email{puhansatyajit@gmail.com}
\affiliation{ Computational High Energy Physics Lab, Department of Physics, Dr. B.R. Ambedkar National
	Institute of Technology, Jalandhar 144008, India}

\author{Shubham Sharma}
\email{s.sharma.hep@gmail.com}
\affiliation{Laboratory for Advanced Scientific Technologies of Mega-Science Facilities and Experiments, Moscow Institute of Physics and Technology (MIPT), Dolgoprudny 141700, Russia}  

\author{Narinder Kumar}
\email{narinderhep@gmail.com}
\affiliation{Computational Theoretical High Energy Physics Lab, Department of Physics, Doaba College, Jalandhar 144004, India}
  
\author{Harleen Dahiya}
\email{dahiyah@nitj.ac.in}
\affiliation{ Computational High Energy Physics Lab, Department of Physics, Dr. B.R. Ambedkar National
	Institute of Technology, Jalandhar, 144008, India}

\begin{abstract}
We investigate the partonic structure of the $\rho$ meson, the lightest spin-$1$ vector meson, within the light-front quark model (LFQM). To explore the sensitivity to model assumptions, we employ two distinct types of spin wave functions in the LFQM. Using light-front helicity wave functions, we derive explicit expressions for the leading-twist and subleading-twist quark parton distribution functions (PDFs), and evolve the leading-twist PDFs to higher scales with next-to-leading order (NLO) Dokshitzer--Gribov--Lipatov--Altarelli--Parisi (DGLAP) evolution. We have also calculated the Mellin moment from the evolved PDFs using a simple neural network frame and compared with available theoretical predictions. Furthermore, we compute the full set of nine leading-twist transverse-momentum-dependent distributions (TMDs) for the valence quark in the $\rho$ meson, including three tensor TMDs that arise from spin-$1$ tensor polarization of the hadron. Positivity constraints for the PDFs and TMDs are examined within this framework. Our findings highlight the crucial role of tensor polarization in shaping the three-dimensional partonic structure of vector mesons.
\end{abstract}

\maketitle
\vspace{0.5em}

 \section{Introduction}
\label{intro}
For decades, scientists have sought to unravel the subatomic world, probing the constituents of matter through a variety of experimental and theoretical approaches \cite{Arbuzov:2020cqg,Accardi:2012qut,Anderle:2021wcy,ATLAS:2008xda,Lai:2010vv,Evans:2008zzb,CMS:2008xjf}. Among hadrons, nucleons have naturally received the greatest attention owing to their stability and direct relevance to nuclear matter \cite{Aidala:2012mv,Accardi:2012qut,Diehl:2015uka,Anderle:2021wcy,Bacchetta:2016ccz} 
\par
In contrast, other hadrons, particularly those with spin greater than $1/2$ are comparatively less explored. However, these systems possess a much richer 
internal structure and provide unique opportunities to deepen our understanding of Quantum Chromodynamics (QCD) \cite{Hagler:2009ni,Brodsky:1997de,Gross:2022hyw,Achenbach:2023pba}. Quarks and gluons, collectively named as partons, are the primary degrees of freedom of QCD and make up the hadron through quark confinement. The distributions of these partons through low-energy non-perturbative models from first principles is still a challenging task. Compared to spin-$1/2$ nucleons, spin-1 hadrons and nuclei provide unique opportunities to explore novel effects due to high spin content. Deuteron, the lightest nuclei in nature, has gained a lot of attention in spin-1 hadrons due to its complex six-quark system. Many well-established theories have been reported for the deuteron in Refs. \cite{Berger:2001zb,Hoodbhoy:1988am,Brodsky:1983vf,Garcon:2001sz,Carlson:1997bs,Arnold:1980zj,Close:1990zw,Mamedov:2024tth}. However, in this work, we are focusing on the lightest spin-$1$ vector meson, the $\rho$ meson, which has a simpler internal structure compared to the deuteron \cite{Kumano:2010vz}.

The distribution of the partons inside a hadron is very complicated to calculate due to the many degrees of freedom. Theoretically, these distributions are calculated through one-dimensional parton distribution functions (PDFs), three-dimensional transverse momentum parton distribution functions (TMDs) \cite{Diehl:2015uka,Angeles-Martinez:2015sea,Pasquini:2008ax,Boussarie:2023izj}, three-dimensional generalized parton distribution functions (GPDs), and five-dimensional generalized parton distribution functions (GTMDs). These distribution functions provide a quantitative description of the momentum and spin correlations of quarks and gluons inside hadrons \cite{Diehl:2015uka,Boussarie:2023izj,Gross:2022hyw,AbdulKhalek:2021gbh,Diehl:2003ny,Belitsky:2005qn}. While significant progress has been achieved to study these distributions for the nucleons \cite{Sharma:2022ylk,Sharma:2023wha,Sharma:2023tre,Sharma:2023ibp,Sharma:2024lal,Sharma:2024arf,Jain:2024lsj,Boffi:2007yc,Diehl:2001pm}, the extension to spin-$1$ systems is still in its early stages \cite{Berger:2001zb,Abidin:2008ku,Cano:2003ju,Cosyn:2019aio,Polyakov:2019lbq,Winter:2017bfs}. In this work, we only deal with PDFs and TMDs for the case of $\rho$ meson. Compared to three PDFs of spin-$1/2$ nucleons and one PDF of spin-$0$ mesons, the spin-1 mesons have a total of four PDFs at the leading twist. The most notable among these PDFs is the tensor $f_{1LL}(x)$ PDF, which is directly connected to the tensor structure functions $b_1$ and $b_2$. Similarly, at the sub-leading twist, there are a total of five PDFs, out of which two are tensor PDFs. Looking into the leading twist TMDs, the tensor $f_{1LL}$, $f_{1TT}$, and $f_{1LT}$ quark TMDs play an important role to describe the tensor polarizations of the spin-1 system, which are absent for the spin-$1/2$ nucleon case.

Various approaches have been proposed to investigate the leading twist quark PDFs and TMDs of spin-$1$ mesons, including light-front holographic model (LFHM) \cite{Kaur:2020emh,Puhan:2023hio}, Nambu–Jona-Lasinio (NJL) model \cite{Ninomiya:2017ggn,Zhang:2024plq,Zhang:2024nxl}, light-cone quark model (LCQM) \cite{Tanisha:2025qda,Kaur:2020emh,Puhan:2023hio,Acharyya:2024tql}, Bethe-Salpeter wave functions method \cite{Shi:2022erw}, Light-front designed wave functions \cite{Li:2021cwv} and QCD instanton vacuum \cite{Liu:2025fuf}. In Refs. \cite{Kumano:2010vz,Kumano:2021fem,Kumano:2021xau,Kumano:2025rai}, the authors have derived all the possible leading and higher twist PDFs and TMDs possibles for spin-1 system along with PDF sum rules. However, there is no lattice simulation results available for the case of PDFs and TMDs. Nevertheless, a comprehensive analysis of both leading and subleading twist distributions through theoretical models for mesonic states remains scarce.  This gap motivates a dedicated study of the $\rho$ meson, where unique spin-1 effects can be systematically explored.
\par In this work, we have explored both leading and subleading twist PDFs for the case of light $\rho$ vector meson. As we are dealing with only quark antiquark system, so the gluon and seaquark contributions are neglected. So, the tilde terms arises at the subleading twist PDFs are coming to be zero. There are total five PDFs available at the leading twist for spin -1 mesons, out of which four are T-even ($f$, $g$, $h$ and $f_{1LL}$) and one is T-odd ($h_{1LT}$) in nature \cite{Kumano:2020ijt,Bacchetta:2001rb}. Similarly, there are total seven PDFs at the subleading twist, out of which five are T-even and two are T-odd in nature ($h_{LL}$ and $g_{LT}$) \cite{Kumano:2020ijt}. In this work, we have focused on the T-even quark PDFs and TMDs only. Similarly, for the case of TMDs, we have only calculated the leading twist quark TMDs. There are total eighteen quark TMDs present at the leading twist out of which nine are T-even and nine are T-odd in nature \cite{Kumano:2020ijt}. Among the eighteen TMDs, ten are tensor TMDs, which are absent for spin-$1/2$ and 0 systems. In this work, we have calculated all the quark PDFs and TMDs by taking care of all polarizations of meson along with the helicities of the quark antiquarks by solving the quark-quark correlation functions. We have presented all the quark PDFs and TMDs using the overlap form of light-front wave functions (LFWFs) amplitude in a $6 \times 6$ matrix, which explicitly described in the Appendix. The positivity constraints of the PDFs and TMDs have also been studied in this work. For all the calculations, we have used light-front quark model (LFQM) with two two different spin methods.
\par LFQM is a non-perturbative approach framework describing the structure and properties of hadrons as well as the internal structure in terms of the constituents \cite{Arifi:2024mff,Ji:1992yf,Choi:1999nu,Choi:2007yu,Chen:2021ywv,Choi:2015ywa,Chung:1988mu,Yabu:1993xq,Ke:2019lcf,Hwang:2010iq}. It is gauge invariant and relativistic by nature. This model is among the most successful and efficient non-perturbative quark models utilized to investigate hadronic features, including form factors and decay constants. This framework of light-front quantization seeks to establish a clear Fock state expansion of hadronic wave functions by eliminating the intricate zero-modes. The light-front wave functions characterize hadrons through their fundamental quarks and gluons degrees of freedom, thus ensuring explicit Lorentz invariance. The meson mass are calculated by solving the bound state Hamiltonial through variational principle. 

The remainder of this paper is organized as follows. In Sec. \ref{lfqm}, we have discussed the LFQM along with two different spin wave functions.
In Sec.~\ref{dis}, we have presented the results of leading twist and subleading twist quark PDFs. We have derived the LFWFs form and explicit form of all the quark PDFs. 
Sec.~ \ref{tmdss} is devoted to calculations of leading twist TMDs along with the positivity constraints obtained in this model. Finally, in Sec. \ref{Conclusion}, we have discussed the conclusion.

\section{Light-Front Quark Model (LFQM)}\label{lfqm}
In the LFQM, the meson Fock state ($|M\rangle$) is described as a bound state of quarks, gluons, and sea quarks as \cite{Pasquini:2023aaf,Ji:2003yj,Kaur:2018ewq,Puhan:2025kzz}
\begin{widetext}
    \begin{eqnarray}
|M\rangle &=& \sum |q\bar{q}\rangle \Psi_{q\bar{q}}
        + \sum
        |q\bar{q}g\rangle \Psi_{q\bar{q}g} + \sum
        |q\bar{q}gg\rangle \Psi_{q\bar{q}g g} + |q\bar{q}(q \bar q)_{sea}\rangle \Psi_{q\bar{q}(q \bar q)_{sea}} + \cdots  \, .
        \label{a1}
\end{eqnarray}
\end{widetext}
Representing the meson as a sum over multi-particle Fock states within light-front (LF) quantization, expressed in terms of the momentum and spin of its constituents, we can write
 \begin{widetext}
     \begin{eqnarray*}\label{fockstate}
|M(P, \lambda) \rangle
   &=&\sum_{n,i,j}\int\prod_{m=1}^n \frac{\mathrm{d} x_m \mathrm{d}^2
        \mathbf{k}_{\perp m}}{\sqrt{x_m}~16\pi^3}
 16\pi^3 ~ \delta\Big(1-\sum_{m=1}^n x_m\Big)\delta^{(2)}\Big(\sum_{m=1}^n \mathbf{k}_{\perp m}\Big) ~\Psi_{n/N}(x_m,\mathbf{k}_{\perp m})   | n ; \mathbf{k}^+_m, \mathbf{k}_{\perp m},
        i,j \rangle.
\end{eqnarray*}
 \end{widetext}
Here, $|M(P,\lambda)\rangle$ denotes the meson state with momentum $P=(P^+,P^-,P_\perp)$, and $\lambda$ is the spin projection of the hadron. The values $\lambda=\pm 1$ and $\lambda=0$ correspond to the transverse and longitudinal spin projections, respectively. The indices $i$, $j$, and $n$ denote the quark helicity, antiquark helicity, and number of flavors, respectively. The four-momentum of the $m^{th}$ constituent is $k_m=(k_m^+,k_m^-,\mathbf{k}_{\perp m})$, and $x_m=k_m^+/P^+$ is its longitudinal momentum fraction. Both the longitudinal momentum fractions and the transverse momenta of the constituents satisfy the momentum sum rules
\begin{eqnarray}
   \sum_{m=1}^n \mathbf{k}_{\perp m}=0, \qquad \sum_{m=1}^n x_m=1.
\end{eqnarray}
\par As we restrict our analysis to mesons without explicit gluonic components, the hadron Fock state in Eq.~(\ref{a1}) reduces to $|M\rangle=\sum |q\bar{q}\rangle \, \Psi_{q\bar{q}}$. Neglecting higher Fock-state contributions, the meson state expressed in terms of quark–antiquark helicities becomes
\begin{eqnarray}
|M(P, \lambda)\rangle &=& \sum_{i,j}\int
\frac{\mathrm{d}x \, \mathrm{d}^2
        \mathbf{k}_{\perp}}{\sqrt{x(1-x)}\,2(2 \pi)^3}
           \nonumber\\
           &&\times \Psi^{\lambda}_{i,j}(x,\mathbf{k}_{\perp}^2)\,
           |x,\mathbf{k}_{\perp}, i,j \rangle .
\label{meson}
\end{eqnarray}
Here, $x$ and $1-x$ are the longitudinal momentum fractions carried by the quark and antiquark, respectively. The total meson wave function $\Psi^{\lambda}_{i,j}(x,\mathbf{k}_{\perp}^2)$ combines spin and momentum-space components, and can be written as
\begin{eqnarray}
\Psi^{\lambda}_{i,j}(x,\mathbf{k}_{\perp}^2)
  = \mathcal{S}_\lambda(x,\mathbf{k}_\perp,i,j)\,\phi(x,\mathbf{k}_\perp).
\label{eq4e}
\end{eqnarray}
Here, $\mathcal{S}_\lambda(x,\mathbf{k}_\perp,i,j)$ represents the spin wave function, while $\phi(x,\mathbf{k}_\perp)$ denotes the radial wave function. For the radial wave function, we adopt the rotationally invariant Gaussian form \cite{Arifi:2024mff,Ji:1992yf,Choi:1999nu,Choi:2007yu}
\begin{equation}\label{GWk2}
	\phi (\mathbf{k}) = \frac{4\pi^{3/4}}{\beta^{3/2}} \,
        e^{-\mathbf{k}^2/ 2\beta^2},
\end{equation} 
where $\mathbf{k} = (k_z, \mathbf{k}_\perp)$ and $\beta$ is a variational parameter obtained from mass-spectroscopic analysis. This Gaussian wave function satisfies the normalization condition
\begin{equation}
\int \frac{dk_z \, d^2\mathbf{k}_\perp}{2(2\pi)^3}\; \big|\phi(\mathbf{k})\big|^2 =1.
\end{equation}
To ensure rotational invariance, the normalization of $\phi(\mathbf{k})$ can also be expressed in terms of $\phi(x,\mathbf{k}_\perp)$ by performing the variable transformation $(k_z, \mathbf{k}_\perp) \to (x,\mathbf{k}_\perp)$ as
\begin{equation}\label{eq:norm2}
 \int_0^1  \! dx \int \frac{d^2 \mathbf{k}_\perp}{2(2\pi)^3}\; \big| \phi(x, \mathbf{k}_\perp) \big|^2 =1.
\end{equation} 
The function $\phi(x, \mathbf{k}_\perp)$ includes the Jacobian factor,
\begin{equation}\label{HO1S2SJac}
    \phi(x,\mathbf{k}_\perp) = \sqrt{\frac{\partial k_z}{\partial x}}\,\phi(\mathbf{k}),
\end{equation}
with $k_z = \tfrac{1}{2}(2x-1) M_{q \bar q}$ and $\frac{\partial k_z}{\partial x} = \tfrac{M_{q \bar q}}{4x(1-x)}$ for the case of equal quark and antiquark masses. The invariant mass of the bound state is given by
\begin{equation}
M_{q \bar q} = \sqrt{\frac{\mathbf{k}_\perp^2 + m^2}{x(1-x)}}.
\end{equation}
In this work, we focus on the $\rho$ meson. Within the LFQM, the $\rho$ meson is treated under SU(2) isospin symmetry with equal constituent masses $m_q=m_{\bar q}=m$. We habe obtained $m=0.22~\text{GeV}$ and $\beta=0.3659~\text{GeV}$ using the variational principle through the solution of the bound state Hamiltonian. The $\rho$ meson mass is obtained to be $0.770$ GeV through these parameters.

The spin wave function $\mathcal{S}_\lambda(x,\mathbf{k}_\perp,i,j)$ in Eq.~(\ref{eq4e}) can be constructed through different approaches. In this work, we employ two distinct methods, corresponding to different meson vertex structures, which yield two independent forms of the spin wave function used in our analysis as \cite{Arifi:2025olq,Ji:1992yf,Ahmady:2020mht,Ahmady:2016ujw}
\begin{widetext}
  \begin{eqnarray}
    \mathcal{S}_{\lambda}(x,\textbf{k}_\perp, i, j) &=& \bar u (k_1,i) \Bigg[-\frac{1}{\sqrt{2}
\sqrt{M_{q \bar q}^2-(m_q-m_{\bar q})^2}}(\gamma^\mu-\frac{k_1^\mu-k_2^\mu}{M_{q\bar q}+m_q+m_{\bar q}})\epsilon^{\nu}_{\lambda}(P) \Bigg]v(k_2,j) \, , \ \ \ \ (S-1) \label{spin1}\\
\mathcal{S}_{\lambda}(x,\textbf{k}_\perp, i, j) &=& \bar u (k_1,i) \Bigg[\epsilon^{\nu}_{\lambda}(P). \gamma \Bigg]v(k_2,j) \, . \ \ \ \ (S-2)\label{spin2}
\end{eqnarray}
\end{widetext}
Here, $u(k_1,i)$ and $v(k_2,j)$ denote the Dirac spinors for the quark and antiquark with momenta $k_1$ and $k_2$, respectively, while $i$ and $j$ represent their helicities. The quantity $\gamma^\mu$ is the gamma matrix, and $\epsilon_{\lambda}(P)$ is the spin-$1$ polarization vector of the meson with momentum $P$. The polarization vector satisfies the transversality condition $\epsilon\!\cdot\!P=0$, which ensures current conservation at the meson--quark vertex. The transverse $(\lambda=\pm)$ and longitudinal $(\lambda=0)$ polarization vectors used in this work are given by
\begin{equation}
\begin{aligned}
{\epsilon}^\nu_{\lambda=\pm}(P) &= \left( 0, \; \frac{2}{P^+}\, \boldsymbol{\epsilon}_\perp(\pm)\cdot \bm{P}_\perp, \; \boldsymbol{\epsilon}_\perp(\pm)\right), \\
{\epsilon}^\nu_{\lambda=0}(P) &= \frac{1}{M_{q \bar q}}\left(P^+, \; \frac{\bm{P}_\perp^{2}-M^2_{q \bar q}}{P^+}, \; \bm{P}_\perp\right),
\end{aligned}
\end{equation}
with 
\[
\boldsymbol{\epsilon}_\perp(\pm) = \mp \frac{1}{\sqrt{2}} (1, \pm i).
\]
In Eqs.~(\ref{spin1}) and (\ref{spin2}), we label the first spin wave function as $S$-1 and the second as $S$-2, and these notations are used consistently throughout this work. Within the LFQM framework, the $S$-1 type spin wave functions are commonly employed, as they satisfy the standard normalization conditions. However, in this study we also consider the $S$-2 type spin wave functions, which allow us to probe tensor-like structures in quark PDFs and TMDs that vanish in the $S$-1 case. The $S$-2 spin wave functions have also been widely used in other theoretical approaches, such as the LFHM \cite{Ahmady:2020mht,Ahmady:2016ujw} and NJL model \cite{Zhang:2024nxl,Zhang:2024plq,Zhang:2022zim}.
\par Solving Eq.~(\ref{spin1}) with the explicit Dirac spinors and polarization vectors, the spin wave function $\mathcal{S}_\lambda(x,\mathbf{k}_\perp,i,j)$ for different quark--antiquark helicities is obtained as
\begin{widetext}
\begin{align}\label{s1}
\mathcal{S}_{\lambda=+1} &= \omega_{q \bar q}
\begin{pmatrix}
\mathbf{k}_\perp^2+BA & \textbf{k}_\perp^R(x M_{q\bar q}+m_q)\\
- \textbf{k}_\perp^R((1-x) M_{q\bar q}+m_{\bar q}) & - (\textbf{k}_\perp^R )^2
\end{pmatrix}, \nonumber \\[6pt]
\mathcal{S}_{\lambda=0}  &= \frac{\omega_{q \bar q}}{\sqrt{2}}
\begin{pmatrix}
\textbf{k}_\perp^L C & 2\mathbf{k}_\perp^2+BA \\
2\mathbf{k}_\perp^2+BA & -\textbf{k}_\perp^R C
\end{pmatrix}, \nonumber \\[6pt]
\mathcal{S}_{\lambda=-1} &= \omega_{q \bar q}
\begin{pmatrix}
-(\textbf{k}_\perp^L)^2 & \textbf{k}_\perp^L((1-x)M_{q \bar q}+m_{\bar q}) \\
- \textbf{k}_\perp^L(x M_{q \bar q}+m_q) & \mathbf{k}_\perp^2+BA
\end{pmatrix}.
\end{align}
\end{widetext}
%
%
%
%
with 
\begin{eqnarray}
\textbf{k}_\perp^{R(L)}&=& \textbf{k}_\perp^x \pm i \textbf{k}_\perp^y, \nonumber\\
A&=&(1-x)m_q+xm_{\bar q},\nonumber\\
B&=&M_{q \bar q}+m_q+m_{\bar q},\nonumber\\
C&=&(1-2x)M_{q\bar q}+(m_{\bar q}-m_q),\nonumber\\
\omega_{q \bar q}&=&\frac{1}{(M_{q \bar q}+m_q+m_{\bar q})\sqrt{x(1-x)[M_{q \bar q}^2-(m_q-m_{\bar q})^2]}}.\nonumber
\end{eqnarray}
The above $S$-1 spin wave function satisfies the orthogonality condition 
\begin{eqnarray}
   \sum_{i,j} \mathcal{S}^*_{\lambda^{\prime}}(x,\textbf{k}_\perp, i, j) \mathcal{S}_\lambda(x,\textbf{k}_\perp, i, j) = \delta_{\lambda^{\prime} \lambda}.
\end{eqnarray}
It is important to note that the $S$-1 type spin wave function has the same form as that obtained from the Melosh--Wigner rotation, which transforms the instant-form wave function into the front-form representation. 
\par In a similar way, the $S$-2 type spin wave function is found to be  
\begin{eqnarray}\label{s1}
\mathcal{S}_{\lambda=+1} &=& \frac{\sqrt{2}A_T}{\sqrt{x(1-x)}}
\begin{pmatrix}
m_q-x\delta m & x \textbf{k}_\perp^R\\
 -(1-x) \textbf{k}_\perp^R & 0\\
\end{pmatrix},
\nonumber\\
\mathcal{S}_{\lambda=0}  &=& \frac{A_L}{\sqrt{x(1-x)}}
\begin{pmatrix}
\textbf{k}_\perp^L \delta m& A_1\\
A_1  & -\textbf{k}_\perp^R \delta m	\\
\end{pmatrix},\\
\mathcal{S}_{\lambda=-1} &=& \frac{\sqrt{2}A_T}{\sqrt{x(1-x)}}
\begin{pmatrix}
0 & (1-x)\textbf{k}_\perp^L\\
-x\textbf{k}_\perp^L& m_q-x\delta m\\
\end{pmatrix}, \nonumber\qquad
\end{eqnarray}
with $A_1 = \mathbf{k}_\perp^2 + m_q m_{\bar q} + M^2_{q \bar q}\,x(1-x)$ and $\delta m = m_q - m_{\bar q}$. 

Here, $A_L$ and $A_T$ denote the normalization constants for the longitudinally and transversely polarized meson, respectively. These constants are calculated as \begin{eqnarray}
   \sum_{ij} \int \frac{dxd^2\textbf{k}_\perp}{2 (2\pi)^3}|\mathcal{S}_\lambda(x,\textbf{k}_\perp, i, j)|^2\phi(x,\bfk)^2=1,
\end{eqnarray}
yielding $A_L = 0.702$ and $A_T = 0.828$. 
\par The four-momenta of the constituent quark ($k_1$) and antiquark ($k_2$) used in this work are expressed as
\begin{eqnarray}
k_1&\equiv&\bigg(x P^+, \frac{\textbf{k}_\perp^2+m_q^2}{x P^+},\textbf{k}_\perp \bigg),\label{n1}\\
k_2&\equiv&\bigg((1-x) P^+, \frac{\textbf{k}_\perp^2+m_{\bar q}^2}{(1-x) P^+},-\textbf{k}_\perp \bigg).
\label{n3}
\end{eqnarray}
\begin{figure*}
\centering
\begin{minipage}[c]{0.98\textwidth}
(a)\includegraphics[width=7.5cm]{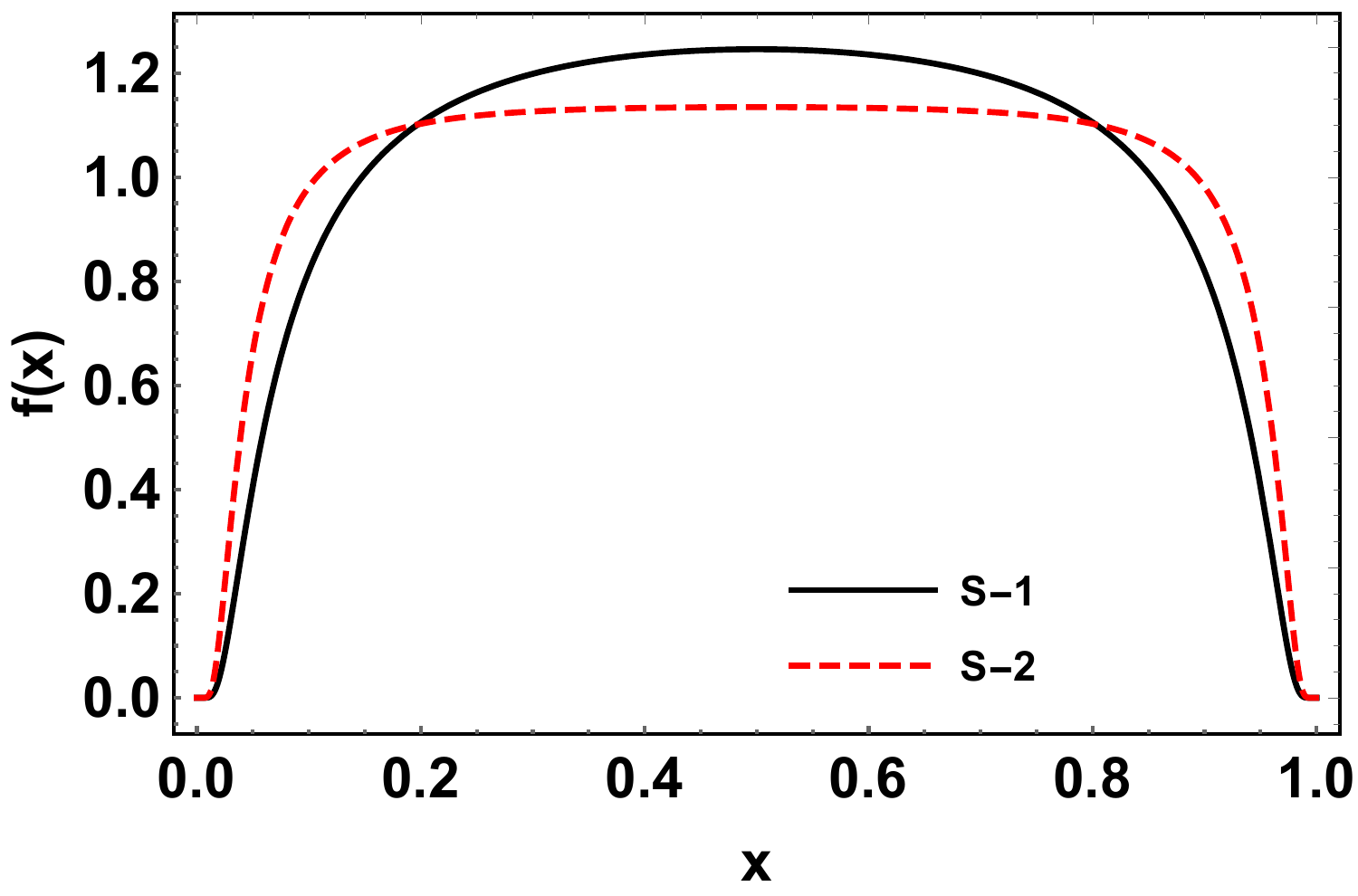}
\hspace{0.03cm}	
(b)\includegraphics[width=7.5cm]{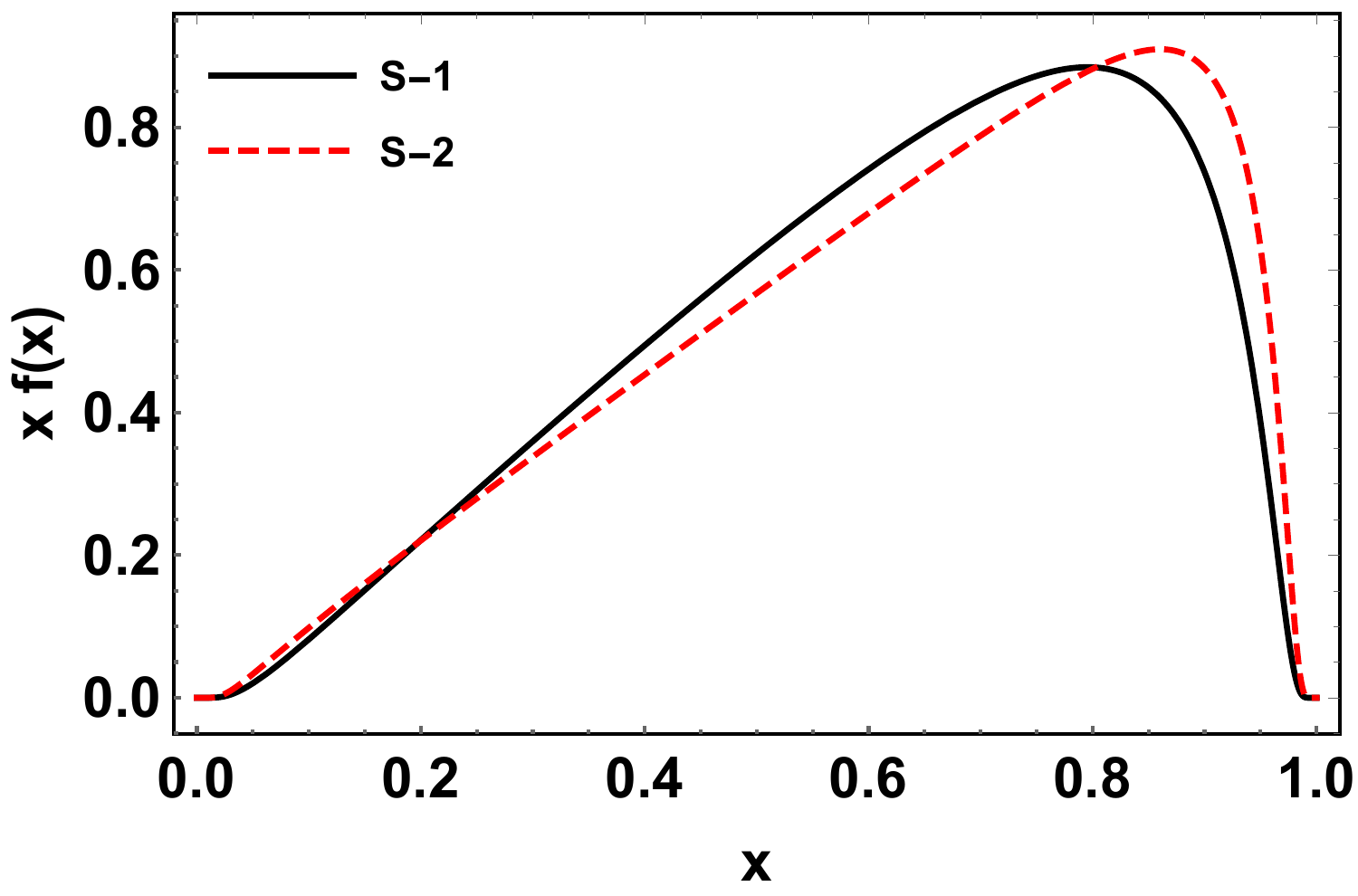} 
\hspace{0.03cm}
\end{minipage}
\centering
\begin{minipage}[c]{0.98\textwidth}
(c)\includegraphics[width=7.5cm]{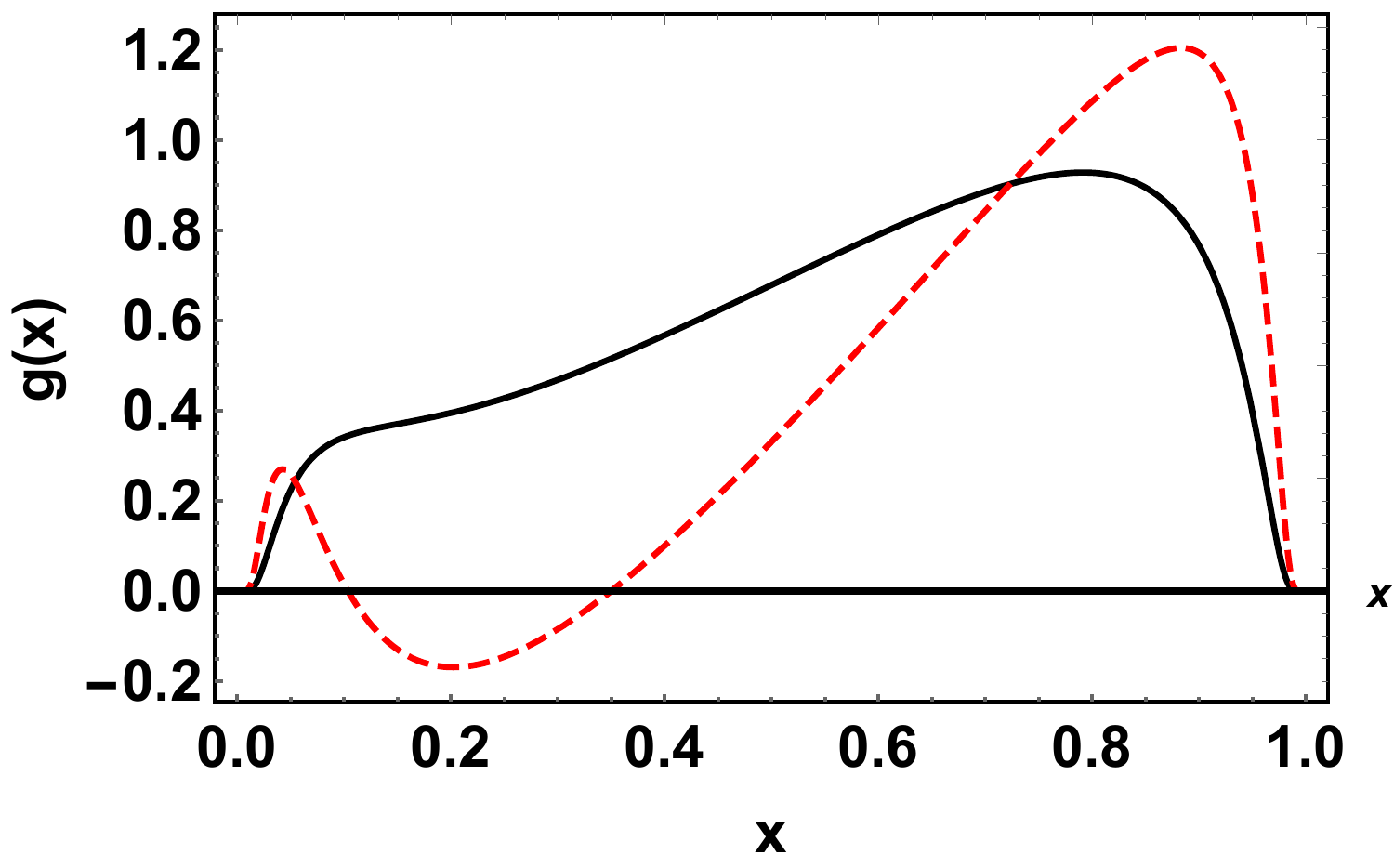}
\hspace{0.03cm}	
(d)\includegraphics[width=7.5cm]{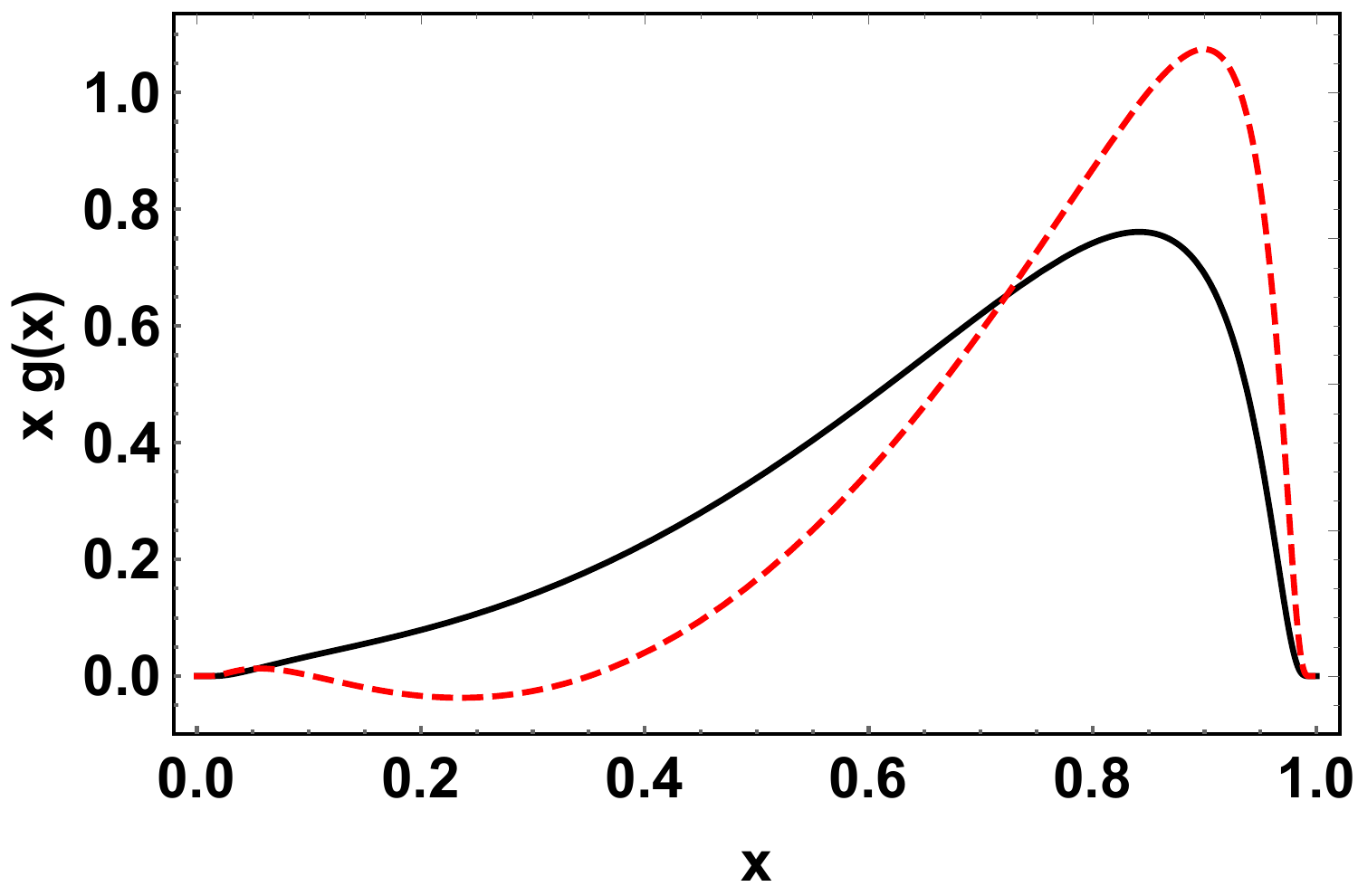} 
\hspace{0.03cm}
\end{minipage}
\centering
\begin{minipage}[c]{0.98\textwidth}
(e)\includegraphics[width=7.5cm]{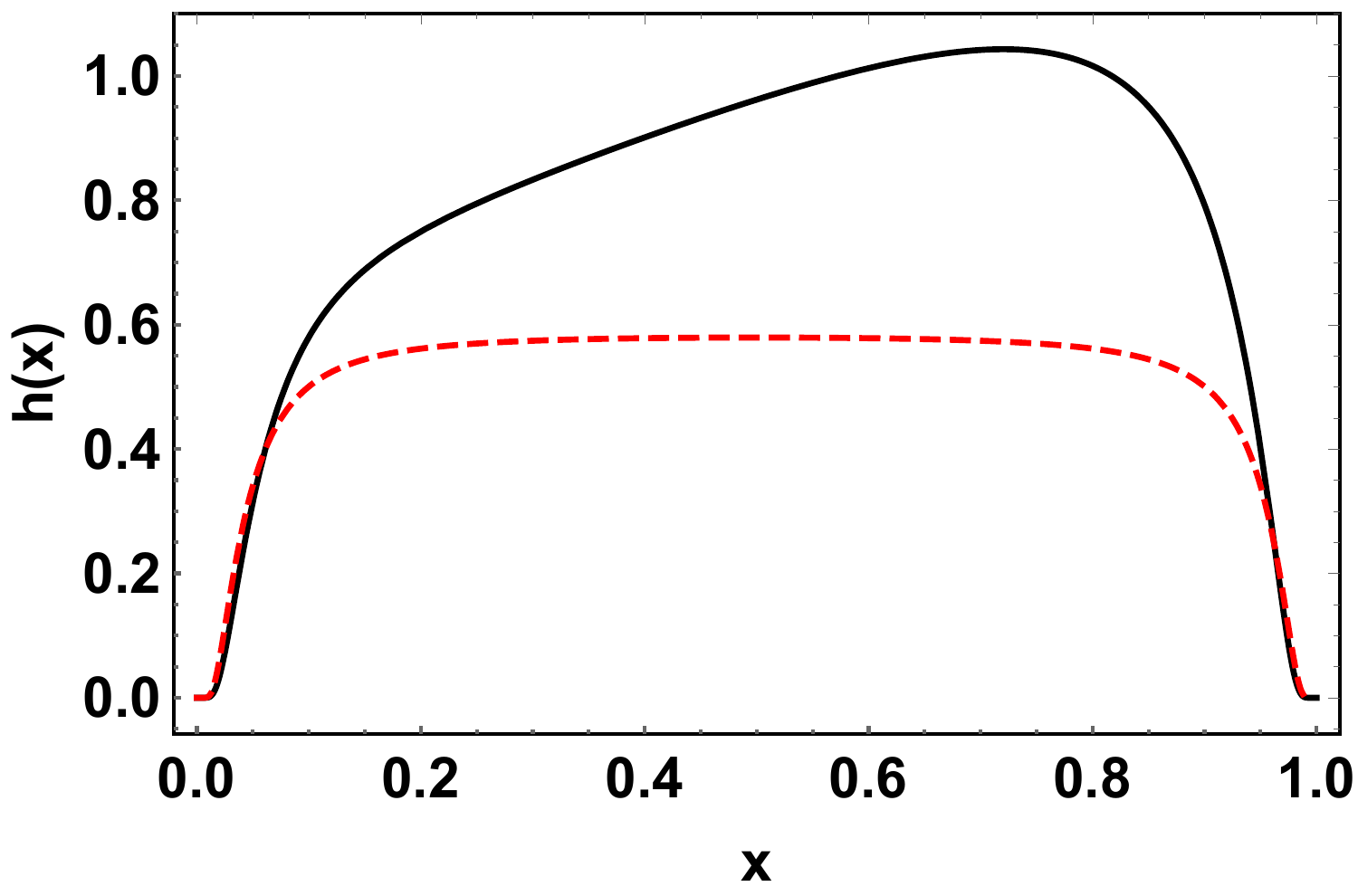}
\hspace{0.03cm}	
(f)\includegraphics[width=7.5cm]{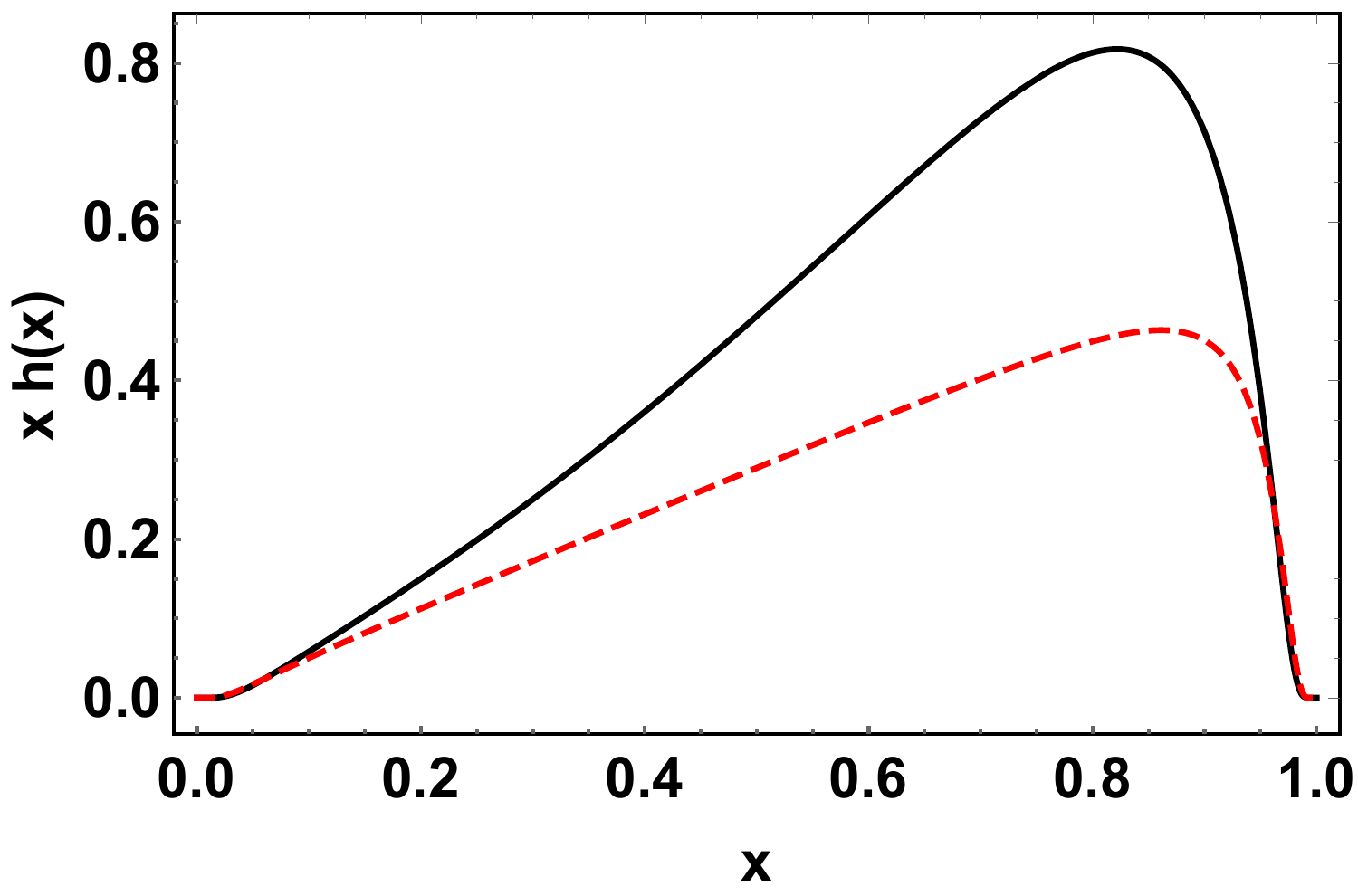} 
\hspace{0.03cm}
\end{minipage}
\centering
\begin{minipage}[c]{0.98\textwidth}
(g)\includegraphics[width=7.5cm]{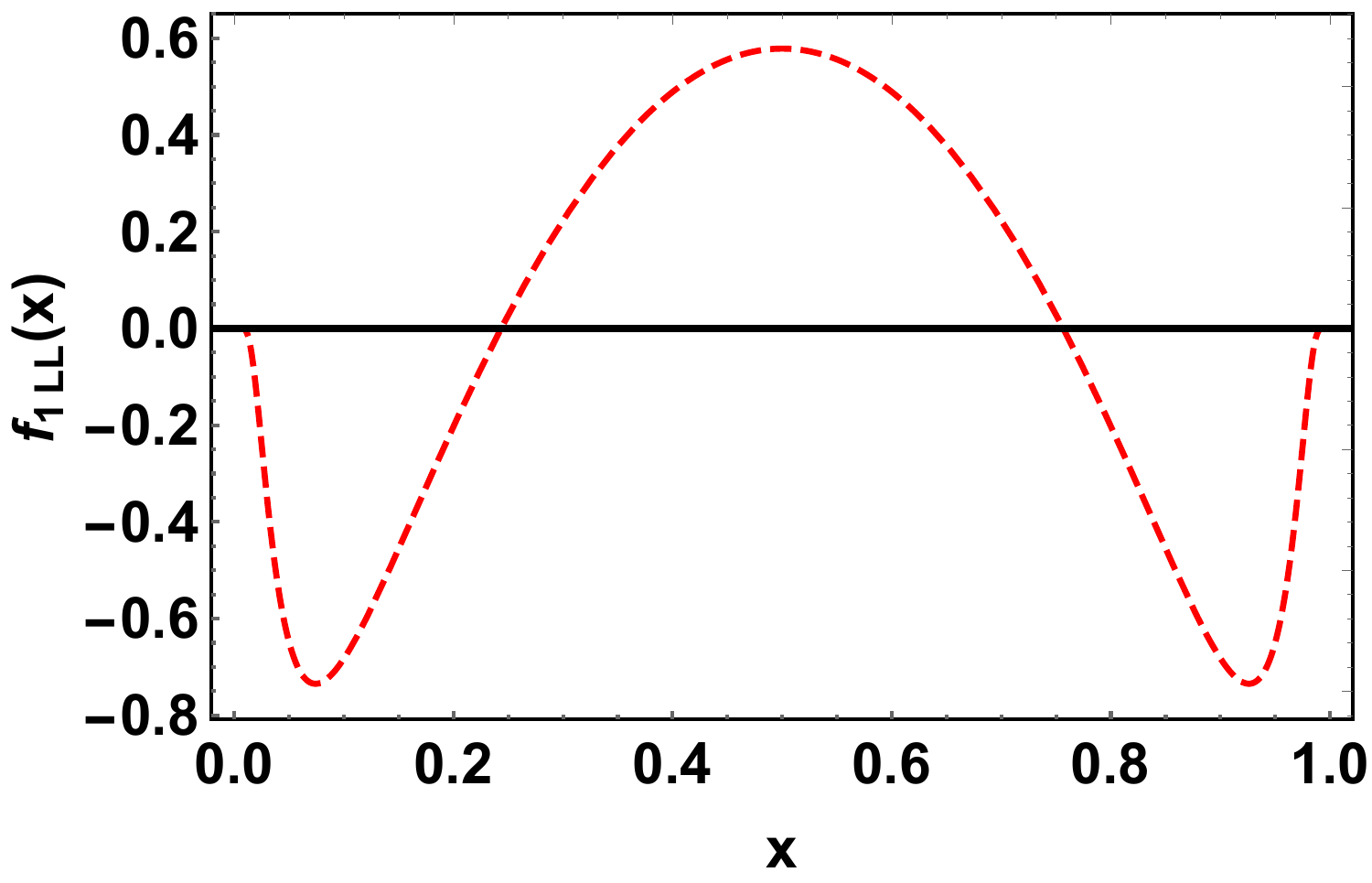}
\hspace{0.03cm}	
(h)\includegraphics[width=7.5cm]{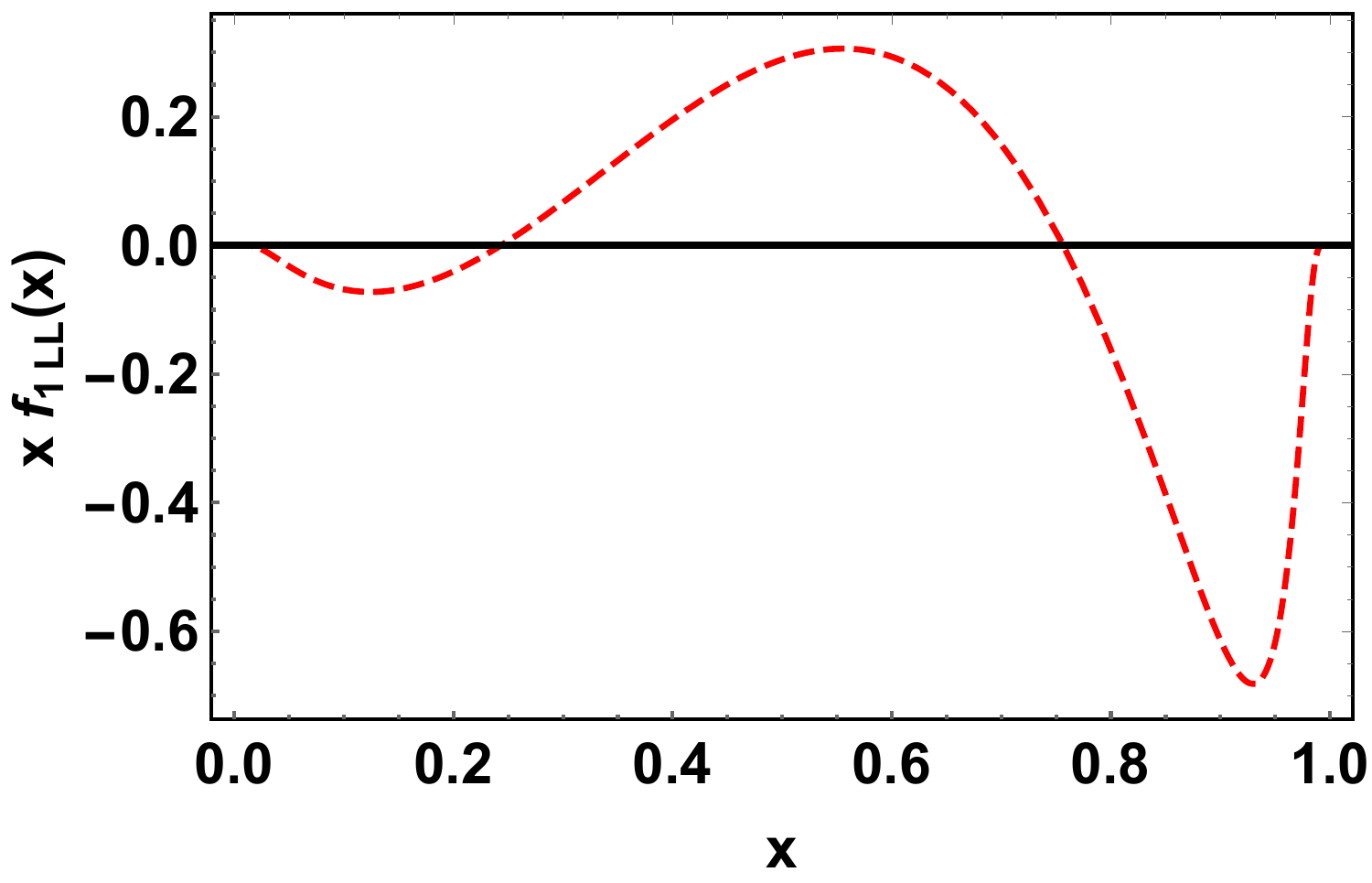} 
\hspace{0.03cm}
\end{minipage}
\caption{Quark PDFs for S-1 and S-2 type spin wave functions at the model scale $Q^2 = 0.20~\mathrm{GeV}^2$ as functions of the longitudinal momentum fraction $x$: 
(a)-(b) unpolarized $f_1(x)$, 
(c)-(d) helicity $g(x)$, 
(e)-(f) transversity $h(x)$, and 
(g)-(h) tensor-polarized $f_{1LL}(x)$.
%
\label{fig1}
}
\end{figure*}

\begin{figure*}
\centering
\begin{minipage}[c]{0.98\textwidth}
(a)\includegraphics[width=7.5cm]{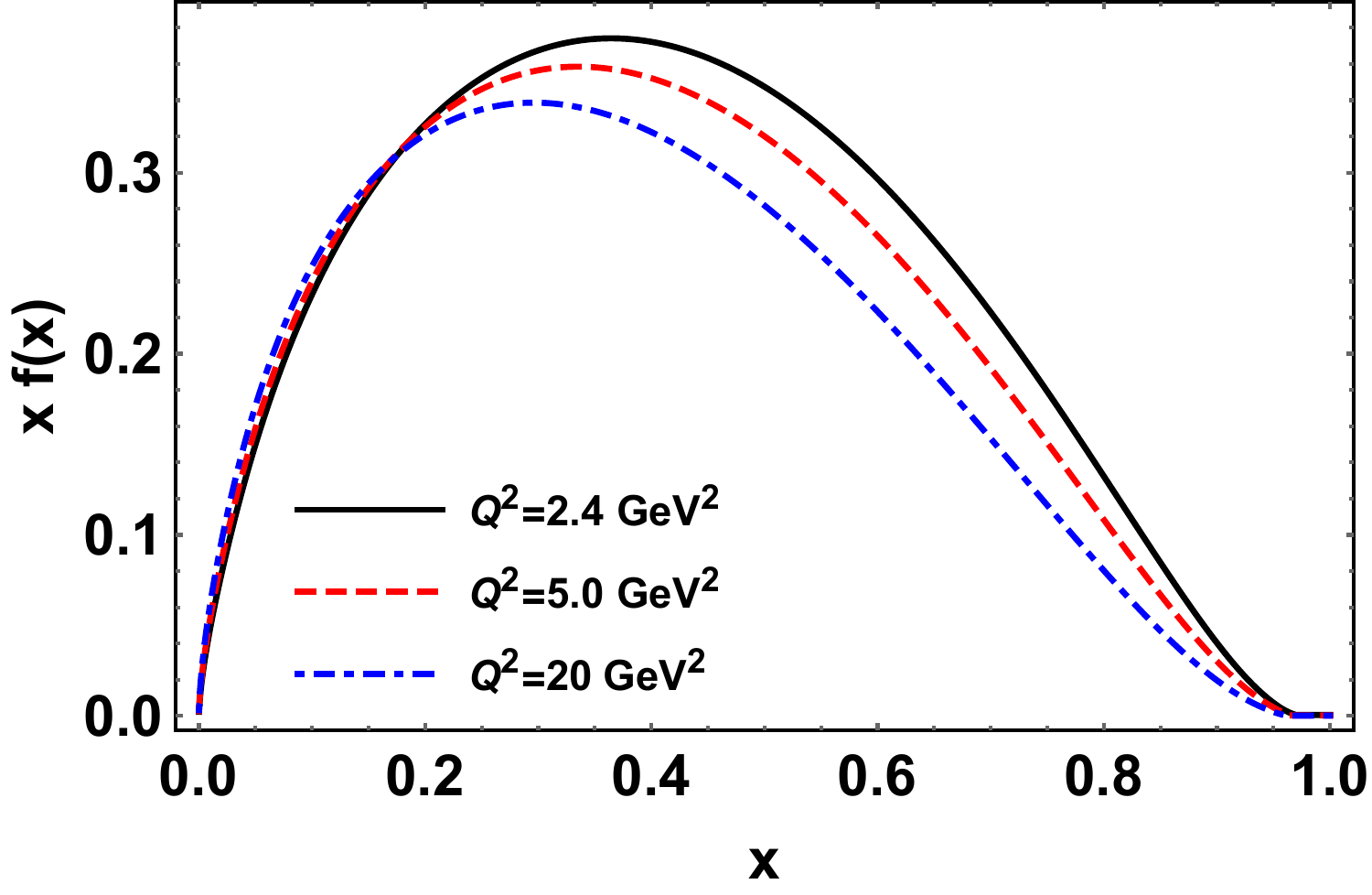}
\hspace{0.03cm}	
(b)\includegraphics[width=7.5cm]{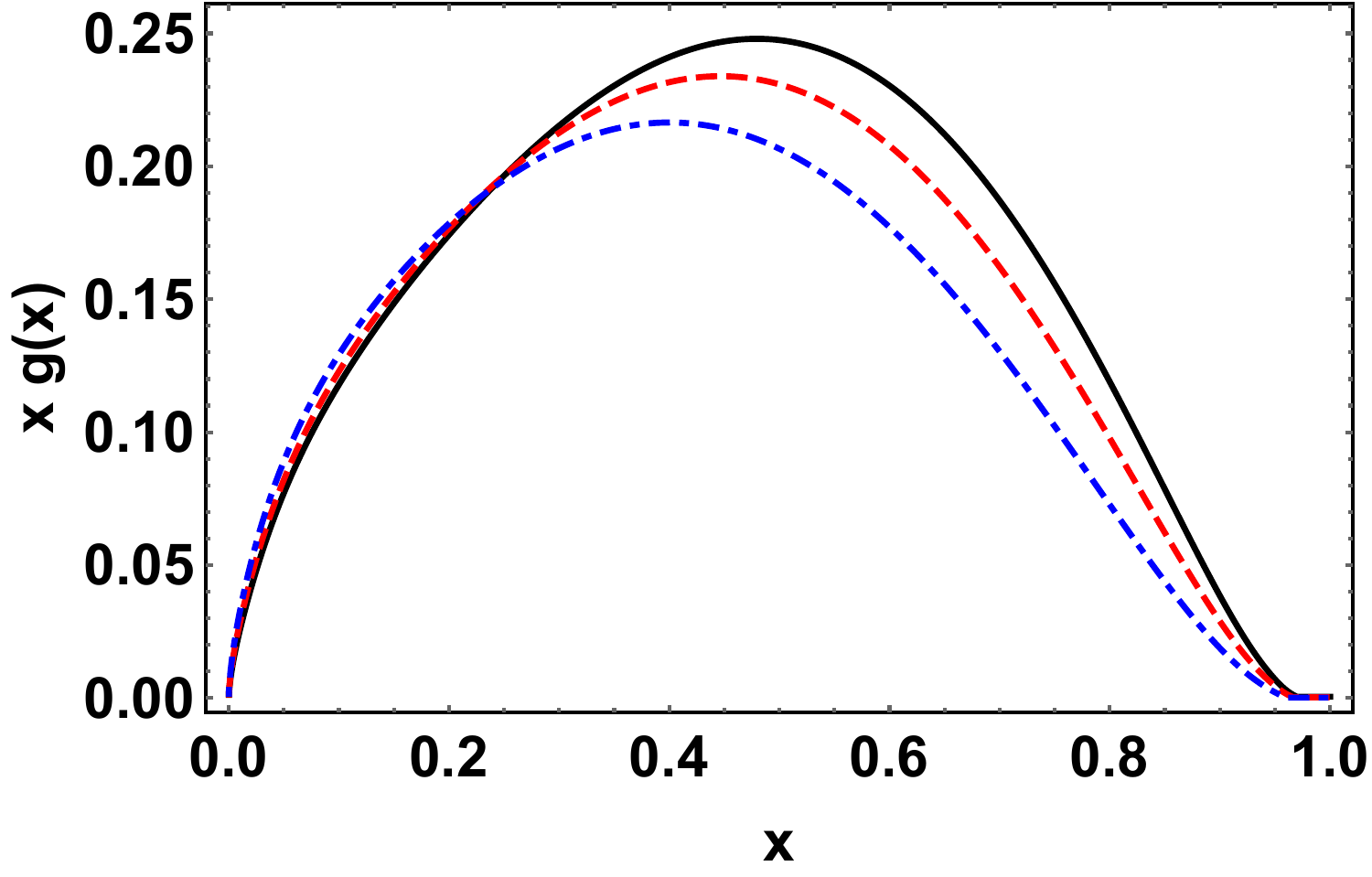} 
\hspace{0.03cm}
\end{minipage}
\centering
\begin{minipage}[c]{0.98\textwidth}
(c)\includegraphics[width=7.5cm]{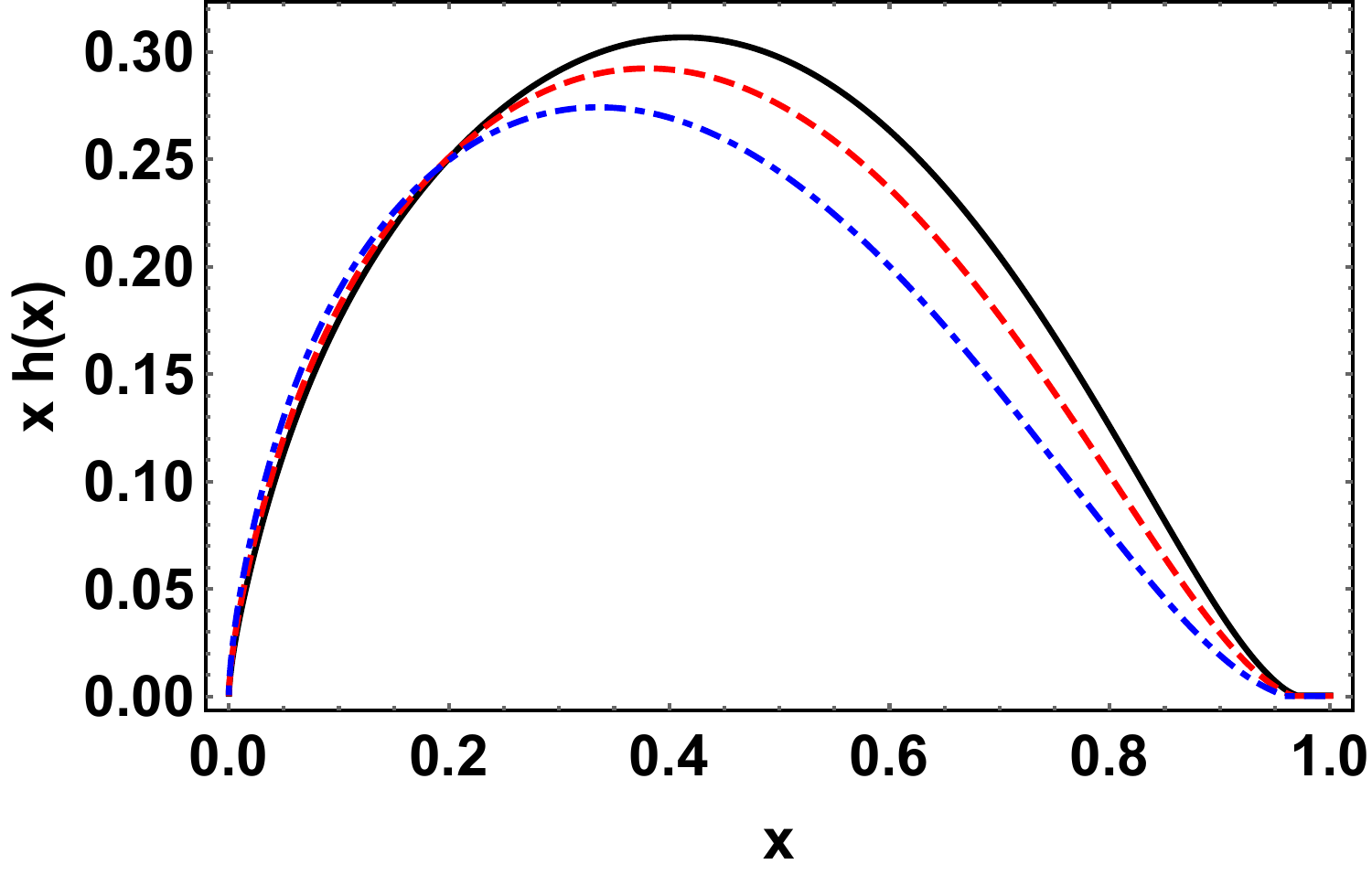}
\end{minipage}
\caption{Evolved quark PDFs of S-1 type spin wave function at $Q^2 = 2.4$, $5$, and $20~\mathrm{GeV}^2$ obtained through NLO DGLAP evolution: (a) $x f(x)$, (b) $x g(x)$, and (c) $x h(x)$ as functions of $x$.\label{fig2}}
%
\end{figure*}
%
\begin{figure*}
\centering
\begin{minipage}[c]{0.98\textwidth}
(a)\includegraphics[width=7.5cm]{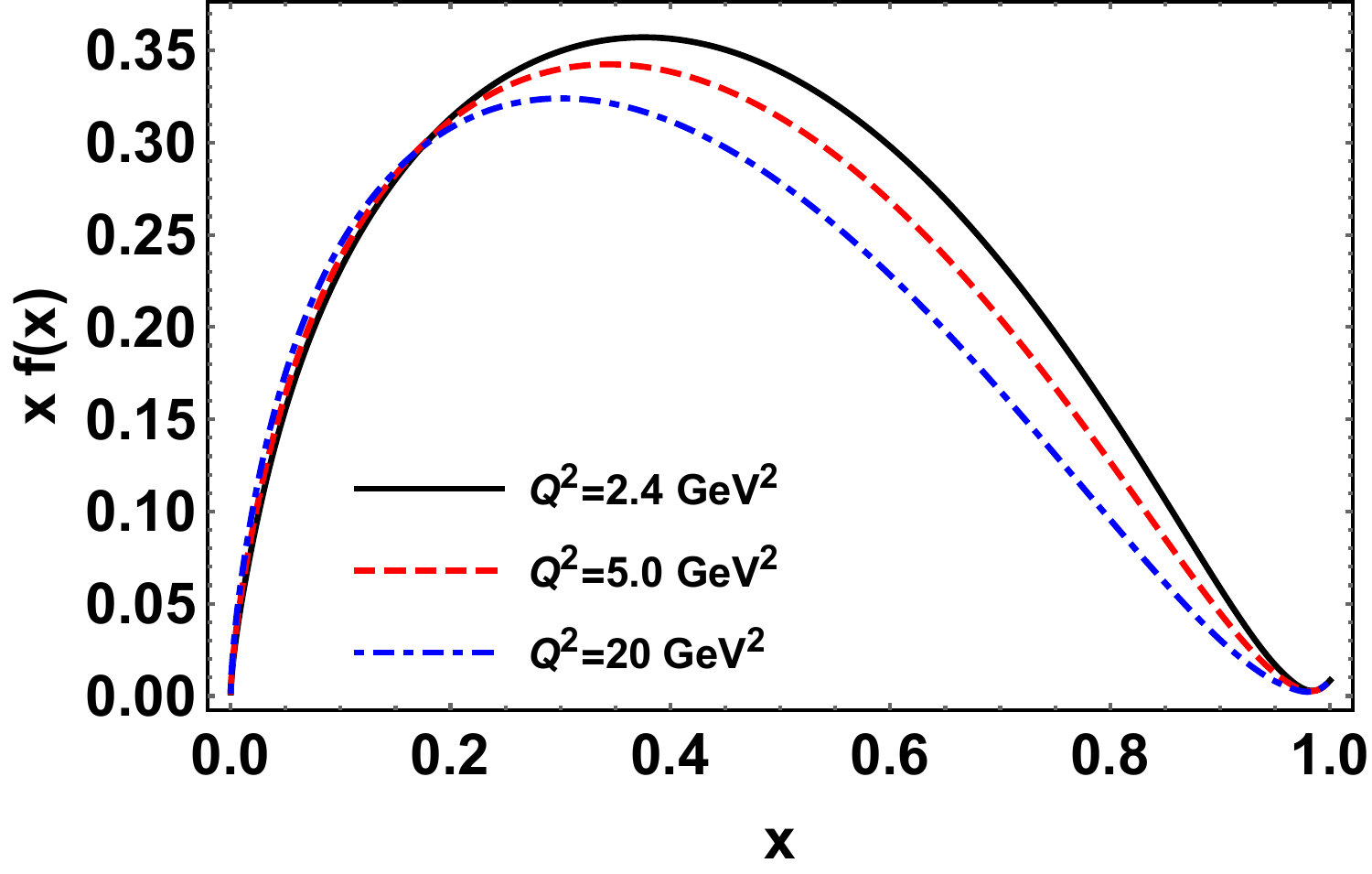}
\hspace{0.03cm}	
(b)\includegraphics[width=7.5cm]{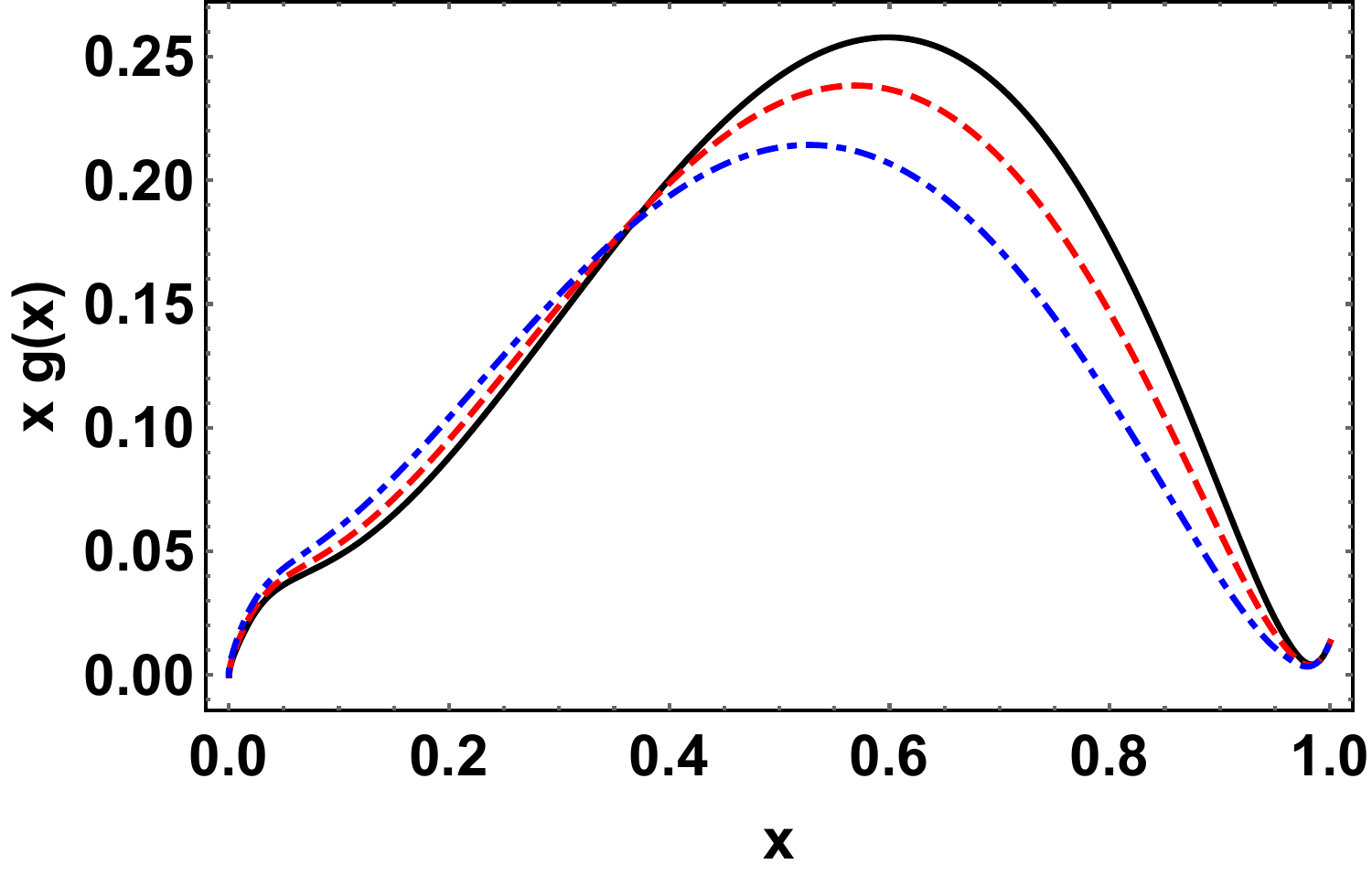} 
\hspace{0.03cm}
\end{minipage}
\centering
\begin{minipage}[c]{0.98\textwidth}
(c)\includegraphics[width=7.5cm]{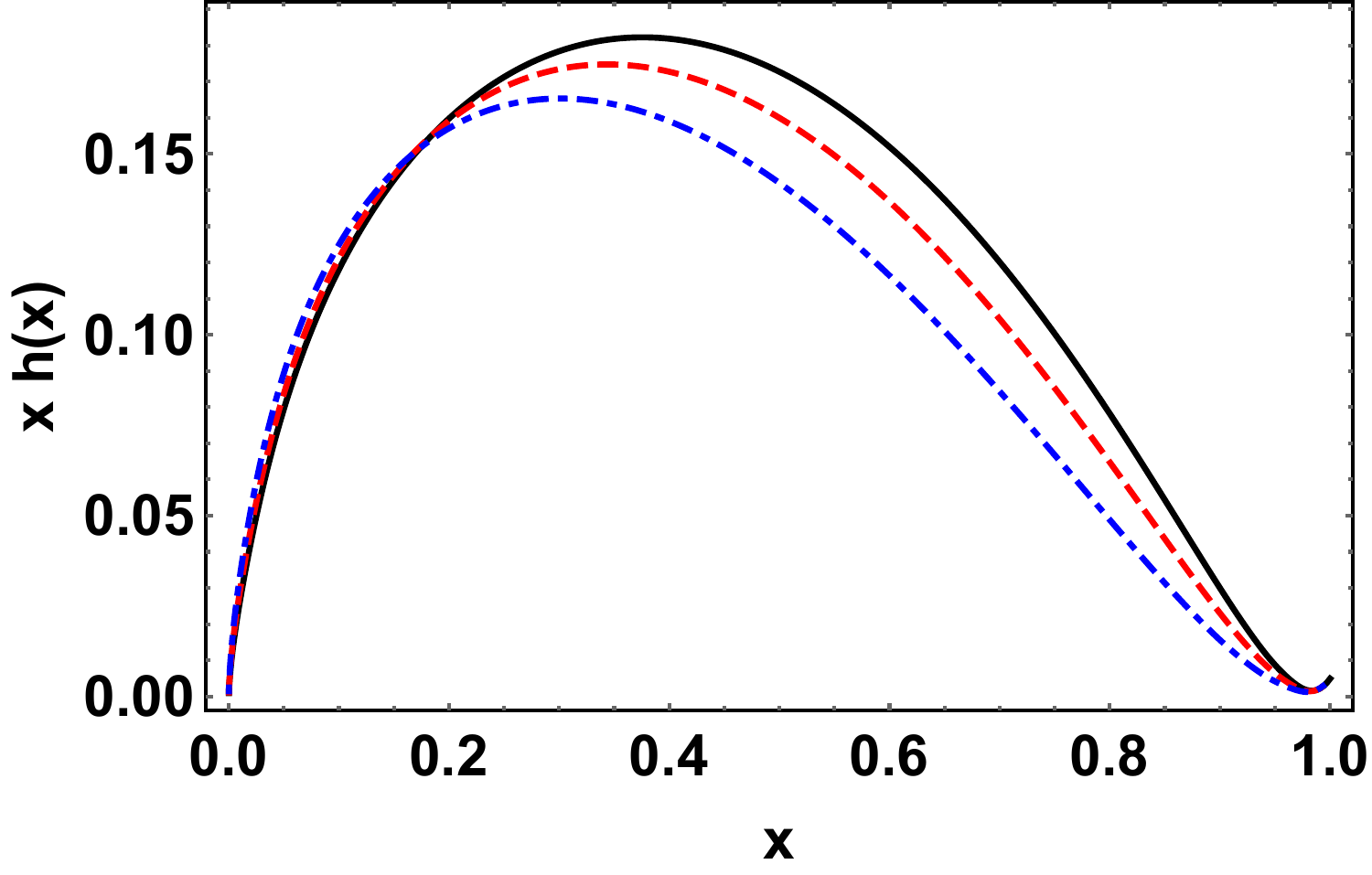}
\hspace{0.03cm}	
(d)\includegraphics[width=7.5cm]{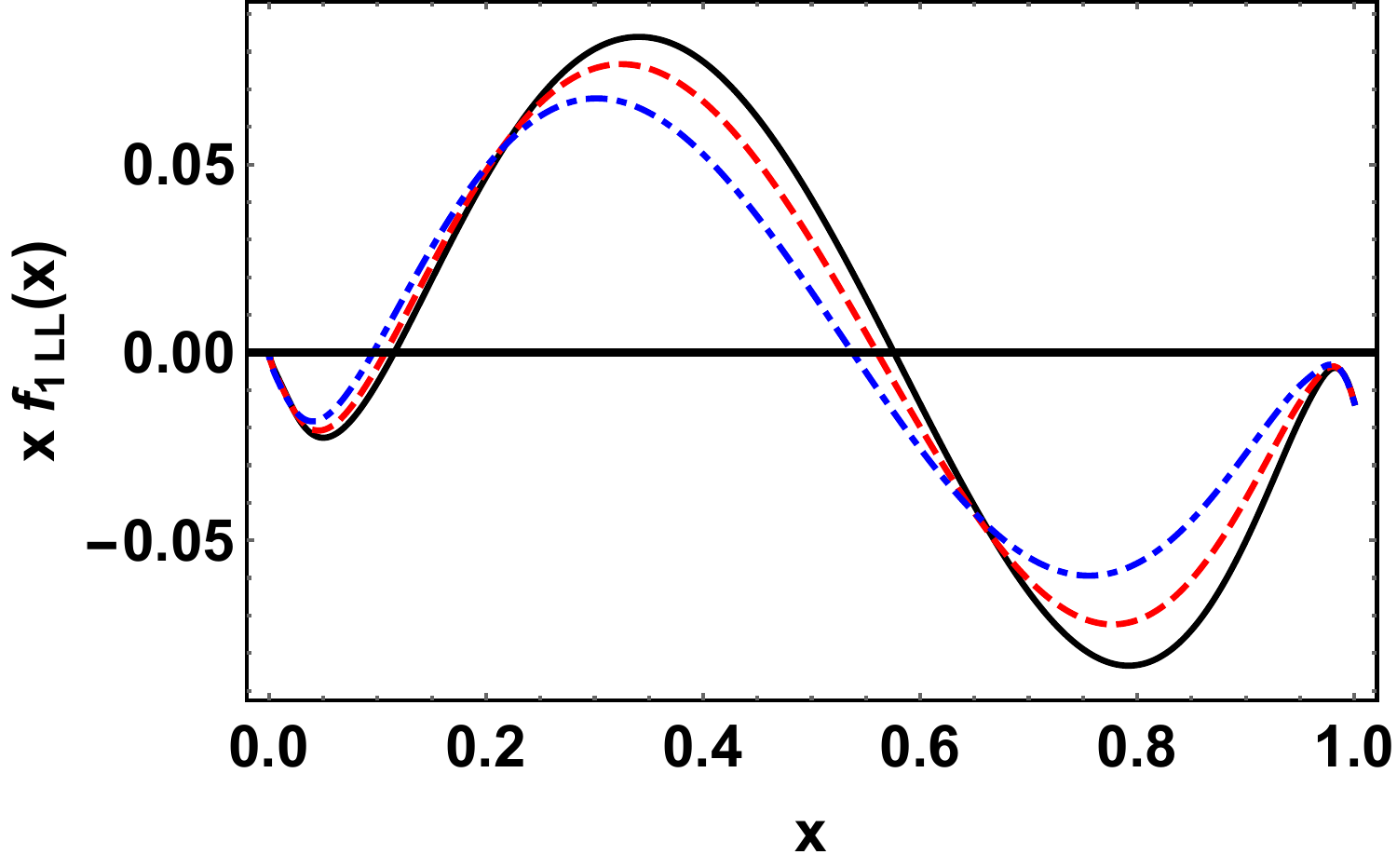} 
\hspace{0.03cm}
\end{minipage}
\caption{Evolved quark PDFs of the S-2 type spin wave function at $Q^2 = 2.4$, $5$, and $20~\mathrm{GeV}^2$ obtained through NLO DGLAP evolution: 
(a) $x f(x)$, 
(b) $x g(x)$, 
(c) $x h(x)$, and 
(d) $x f_{1LL}(x)$ as functions of $x$.
%
\label{fig3}}
\end{figure*}
%
\begin{figure*}[htbp]
\centering
\begin{subfigure}{0.32\textwidth}
    \includegraphics[width=\linewidth]{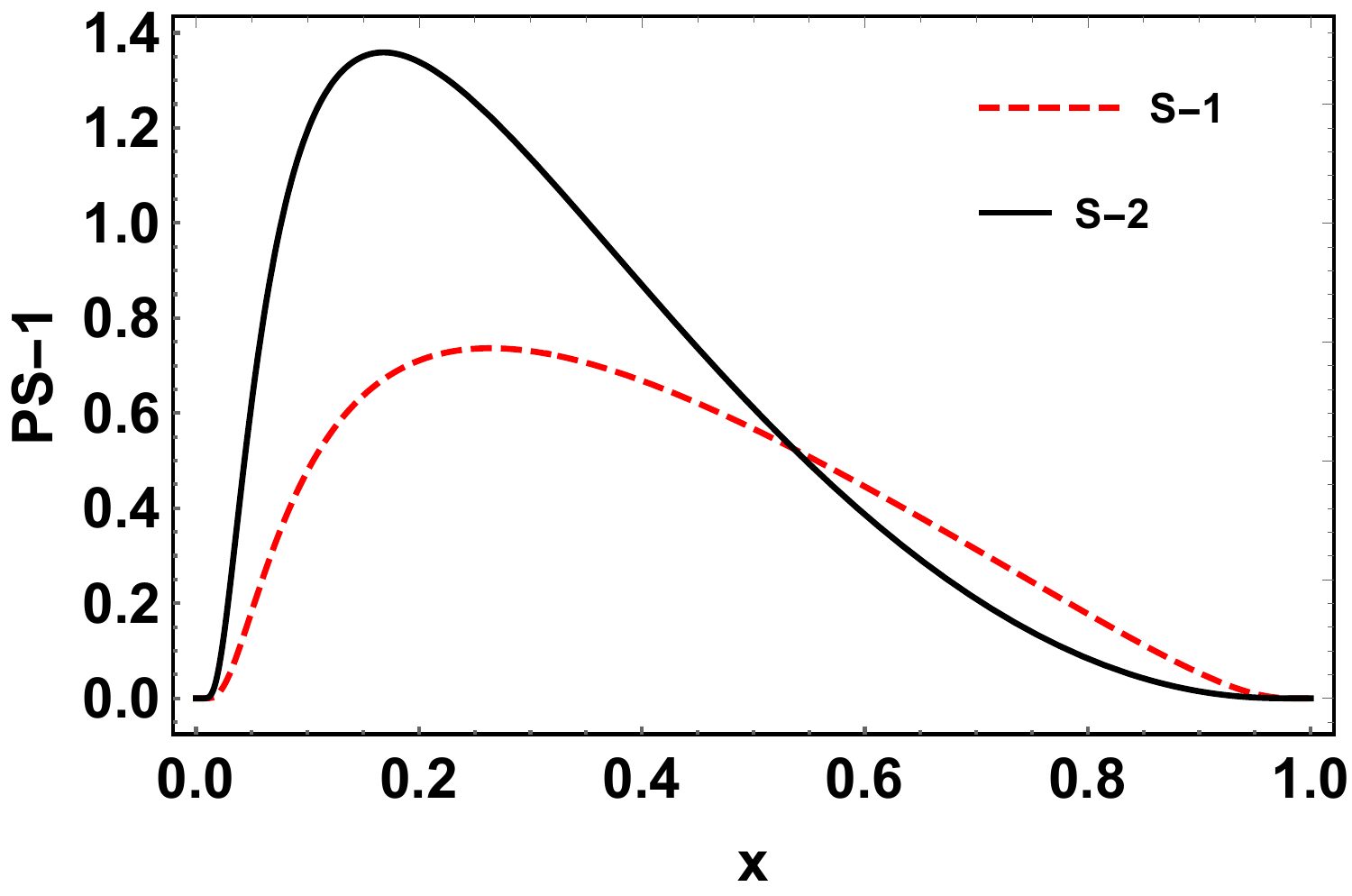}
    \subcaption{}
\end{subfigure}
\hfill
\begin{subfigure}{0.32\textwidth}
    \includegraphics[width=\linewidth]{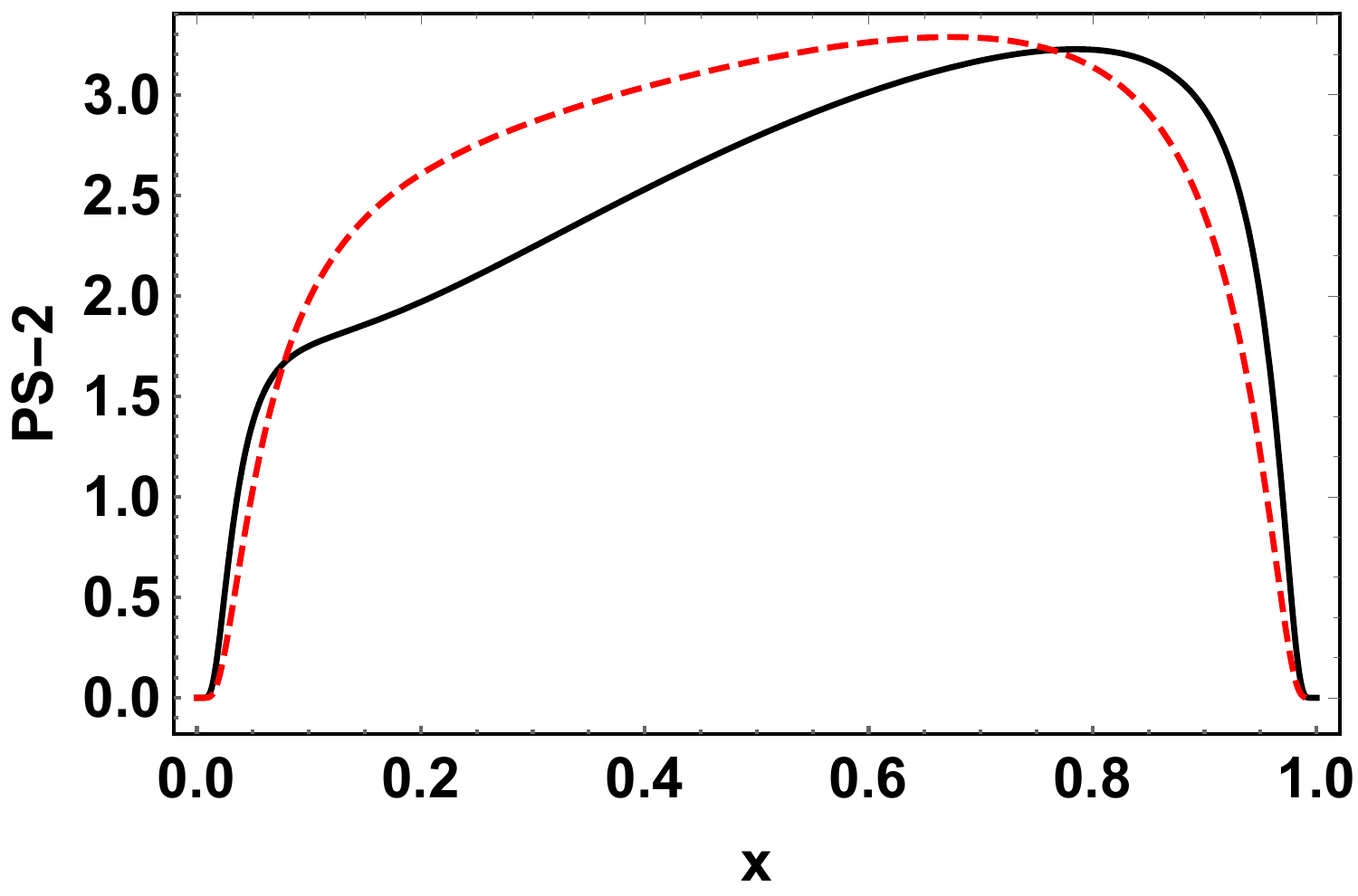}
    \subcaption{}
\end{subfigure}
\hfill
\begin{subfigure}{0.32\textwidth}
    \includegraphics[width=\linewidth]{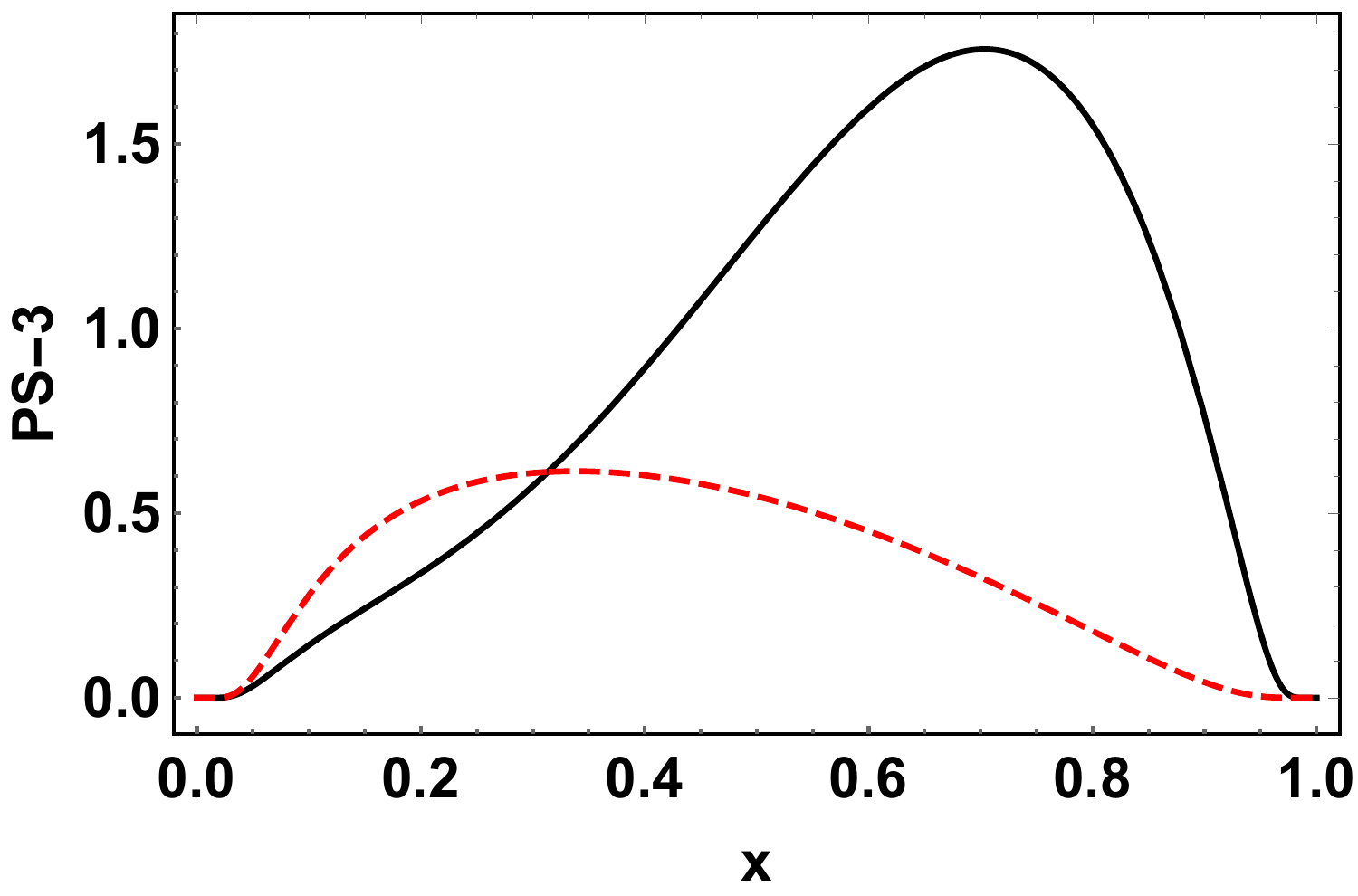}
    \subcaption{}
\end{subfigure}
\hfill
\begin{subfigure}{0.32\textwidth}
    \includegraphics[width=\linewidth]{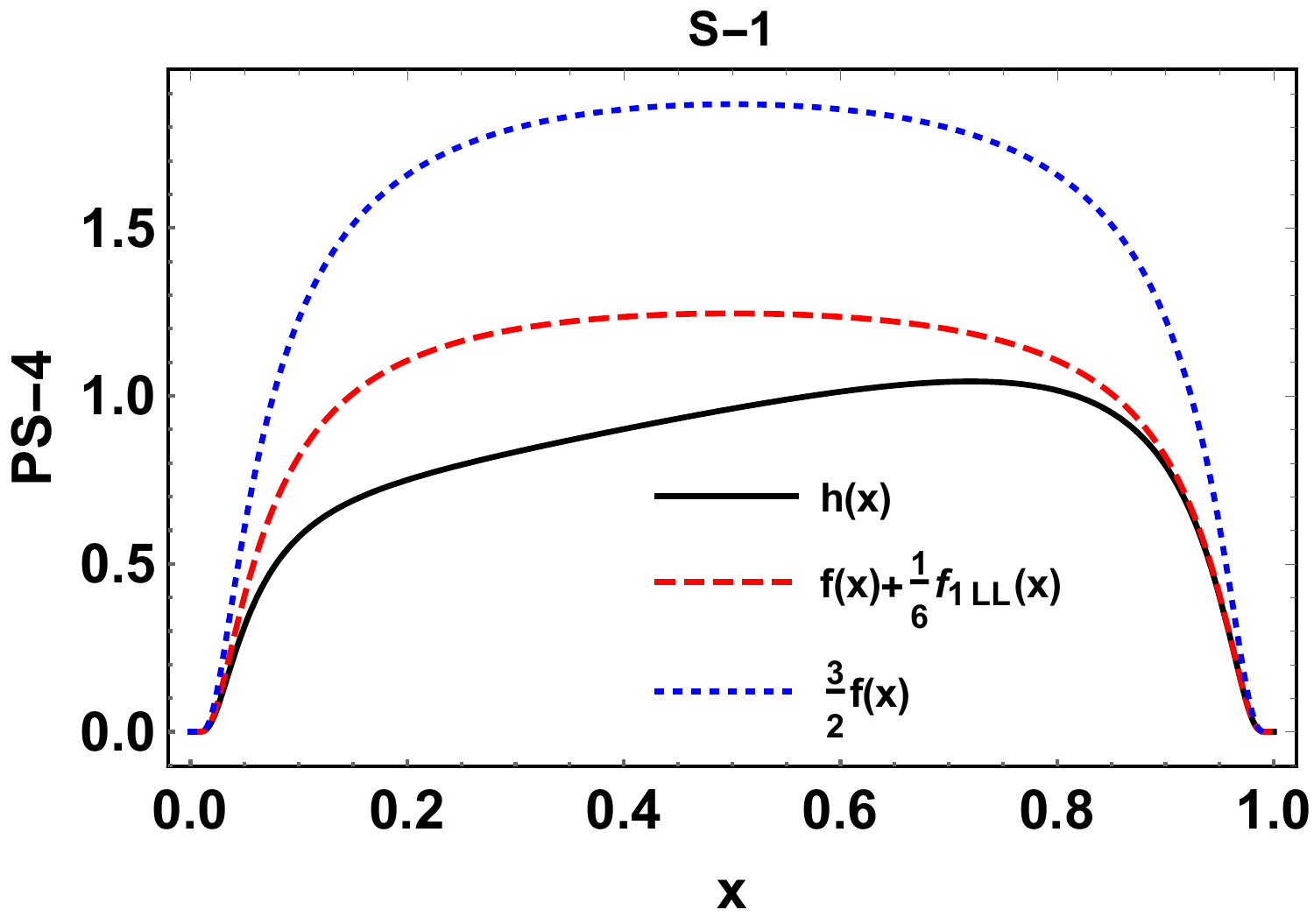}
    \subcaption{The PS of PDFs using the S-1 type spin wave functions.}
\end{subfigure}
\hfill
\begin{subfigure}{0.32\textwidth}
    \includegraphics[width=\linewidth]{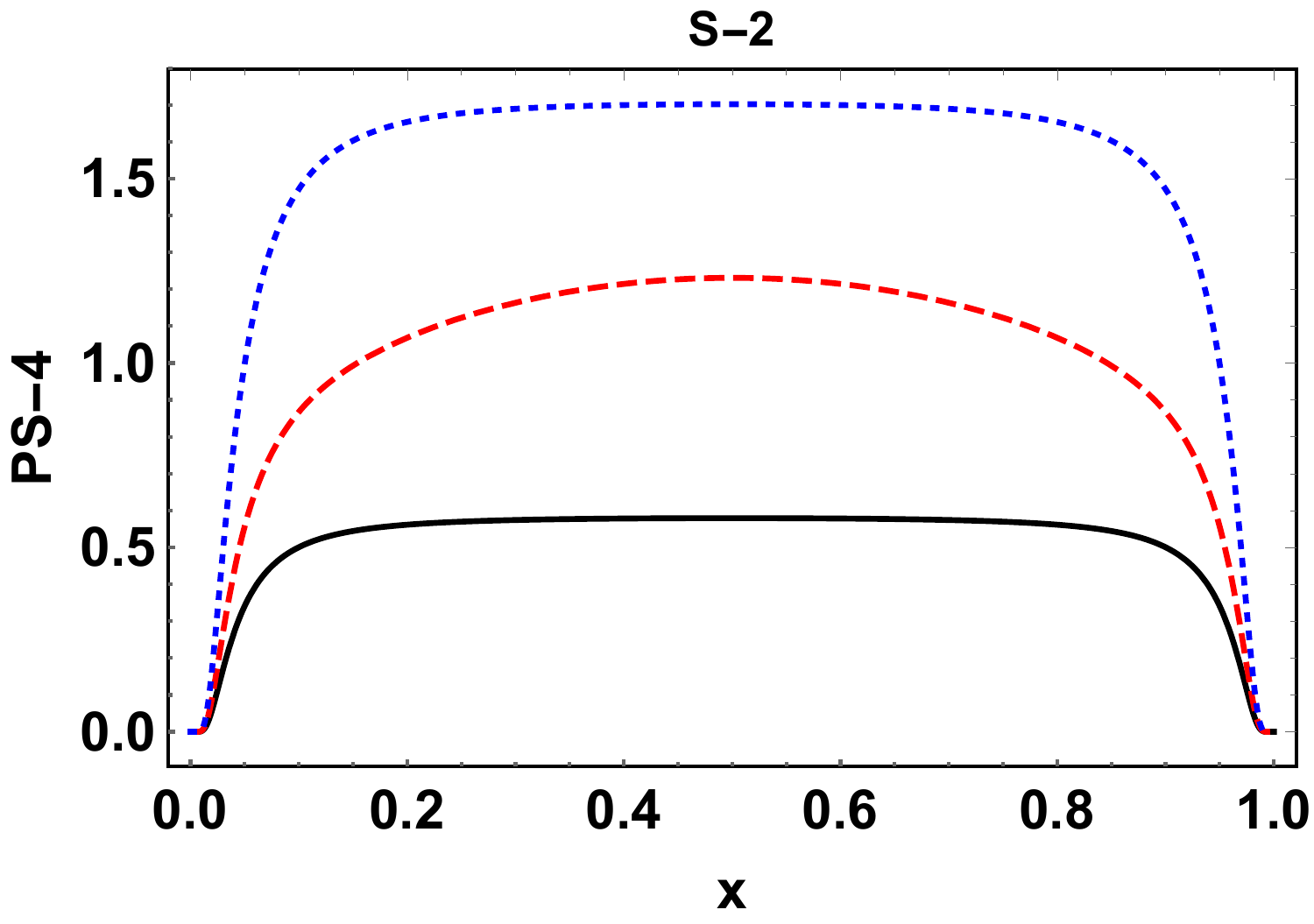}
    \subcaption{The PS of PDFs using the S-2 type spin wave functions.}
\end{subfigure}
\caption{\label{pcpdf1}
Positivity constraints (PS) of the PDFs
for the S-1 and S-2 type spin wave function.}
\end{figure*}
\begin{table*}[t]
\renewcommand{\arraystretch}{1.5}
    \centering
    \begin{tabular}{|c| c|c |c|c|c|c|c|c|c|c|c|}
    \hline
       &  & $f(x)$ &$g(x)$ & $ h(x)$   & $ f_{1LL}(x) $ & $e(x)$ & $g_T(x)$ & $h_L(x)$ & $e_{LL}(x)$ & $f_{LT}(x)$\\
       \hline
         $S-2$  \\
         \hline
         &$\langle x^0 \rangle$ & 1 & 0.41256& 0.510215  & 0& -&- & -& 0& -\\
          &$\langle x \rangle$ & 0.5 & 0.31638& 0.255108  &0& 0.285714& 0.455707& 0.39064& 0& -\\
         
        $Q^2=0.20$ & $\langle x^2 \rangle$ & 0.317 & 0.25 & 0.161815  & 0&0.142857 & 0.227853&0.19532 &0 & 0.21662\\
        & $\langle x^3 \rangle$ & 0.225683 & 0.201608 & 0.115168  & 0& 0.0906064& 0.137254& 0.123891&0 & 0.21662\\
           & $\langle x^4 \rangle$ & 0.171511 & 0.138866 & 0.0875344  & 0&0.064481 & 0.0919547&0.088177 &0 &0.182896 \\
        \hline
         
         &$\langle x \rangle$ & 0.236699 &0.148762 & 0.120677  & 0.001531& & & & & \\
         
        $Q^2=2.4$ & $\langle x^2 \rangle$ & 0.101811 &0.080224 &  0.052195 &$\approx 0$ & & & & & \\
        & $\langle x^3 \rangle$ & 0.055388 & 0.049722& 0.028626  & $\approx 0$ & & & & & \\
        & $\langle x^4 \rangle$ & 0.034226 &0.033419 & 0.017868  & $\approx 0$& & & & & \\
         \hline
         
         &$\langle x \rangle$ & 0.223957 &0.138764 & 0.113528 & 0.001334& & & & & \\
         
        $Q^2=5$ & $\langle x^2 \rangle$ & 0.093155 & 0.072187& 0.047162 &$\approx 0$& & & & & \\
        & $\langle x^3 \rangle$ & 0.049524 &0.043597 & 0.025044  & $\approx 0$& & & & & \\
        & $\langle x^4 \rangle$ & 0.030060 & 0.028700&0.015191 &$\approx 0$& & & & & \\
        \hline
         
         &$\langle x \rangle$ & 0.202116 &0.126463 & 0.103665 & 0.001010& & & & & \\
         
        $Q^2=20$ & $\langle x^2 \rangle$ & 0.079374 & 0.062411&0.040940 & $\approx 0$& & & & & \\
        & $\langle x^3 \rangle$ & 0.040454 &0.036311 &0.020969 & $\approx 0$& & & & & \\
        & $\langle x^4 \rangle$ & 0.023739 &0.023219 & 0.012361  & $\approx 0$& & & & & \\
         \hline
         $S-1$  \\
         \hline
         &$\langle x^0 \rangle$ & 1 & 0.58635 & 0.793175  & 0&- & -&- & -& -\\
          &$\langle x \rangle$ & 0.5 & 0.336676& 0.418338  &0 & 0.220275&0.266317 &0.31236&0 & 0\\
         
        $Q^2=0.20$ & $\langle x^2 \rangle$ & 0.309504 & 0.227842 & 0.268673  & 0& 0.110137& 0.133159&0.15618 &0 &0 \\
        & $\langle x^3 \rangle$ & 0.214257 & 0.167112 & 0.190684  & 0&0.0715785 & 0.0852643&0.09895 & 0&0 \\
           & $\langle x^4 \rangle$ & 0.158548 & 0.128662 & 0.143605  &0 &0.0522991 & 0.0613171&0.0703351 &0 & 0\\
        \hline
        &$\langle x \rangle$ & 0.235988 & 0.158875& 0.197774  & 0& & & & & \\
         
        $Q^2=2.4$ & $\langle x^2 \rangle$ & 0.098805 &0.072937 &0.086191& 0& & & & & \\
        & $\langle x^3 \rangle$ & 0.052037 & 0.040858&0.046754 & 0& & & & & \\
        & $\langle x^4 \rangle$ & 0.031116 & 0.025526&0.028605& 0& & & & & \\
         \hline
        &$\langle x \rangle$ & 0.221210 & 0.148560& 0.189440 &0 & & & & & \\
         
        $Q^2=5$ & $\langle x^2 \rangle$ & 0.089419 &0.065659 & 0.080770  & 0& & & & & \\
        & $\langle x^3 \rangle$ & 0.045938 & 0.035780& 0.043333 & 0& & & & & \\
        & $\langle x^4 \rangle$ & 0.026942 &0.021866 &0.026373&0 & & & & & \\
         \hline
        &$\langle x \rangle$ & 0.204570 & 0.135748&0.169819 & 0& & & & & \\
         
        $Q^2=20$ & $\langle x^2 \rangle$ & 0.079364 &0.057033 & 0.067819  & 0& & & & & \\
        & $\langle x^3 \rangle$ & 0.039730 &0.030023 &0.034547 & 0& & & & & \\
        & $\langle x^4 \rangle$ & 0.022882 & 0.017876&0.020106 & 0& & & & & \\
         \hline
    \end{tabular}
\caption{Mellin moments $\langle x^n \rangle$ of twist-2 and twist-3 quark PDFs up to $n=4$ for S-1 and S-2 spin wave functions, evaluated at $Q^2 = 0.20$ (model scale), $2.4$, $5$, and $20~\mathrm{GeV}^2$ using NLO DGLAP evolution.}
\label{tab1}
\end{table*}

\begin{table*}
    \begin{tabular}{|c|c|c|c|c|c|c|}
    \hline
    & \multicolumn{2}{c|}{$g(x)$} & \multicolumn{2}{c|}{$h(x)$}&\multicolumn{2}{c|}{$f_{1LL}(x)$} \\ \cline{2-7}
    & $\langle x^0 \rangle$ & $\langle x \rangle$ &$\langle x^0 \rangle$ &$\langle x \rangle$ & $\langle x^0 \rangle$ &$\langle x \rangle$\\ \hline
    S-1 (model scale) & 0.59 & 0.34 &0.79 &0.41 &0 & 0 \\ \hline
    $Q^2=2.4$ GeV$^2$ & - &  0.24 & - & 0.16 & 0 &  0\\ \hline
    S-2 (model scale) & 0.41 & 0.32 &0.51 & 0.26 & 0& 0 \\ \hline
    $Q^2=2.4$ GeV$^2$ & - &0.15  &- & 0.12& 0& 0.002 \\ \hline
    ILM results \cite{Liu:2025fuf} & 0.97 &-  &  0.704& -& 0& 0 \\ \hline
    BSE \cite{Shi:2022erw} & 0.57 & 0.212 & 0.79&- & 0&  0\\ \hline
    $Q^2=2.4$ GeV$^2$ & 0.66 & 0.227 & -&- &- &  -\\ \hline
    M-NJL \cite{Zhang:2024plq} &0.505 &  0.279 & 0.803&- &0 & 0 \\ \hline
    NJL \cite{Ninomiya:2017ggn} & 0.56 &  0.369 &  0.94&  & &  \\ \hline
    \end{tabular}
\caption{We have compared our Mellin moment of $g(x)$, $h(x)$ and $f_{1LL}(x)$ quark PDFs with available theoretical model predictions \cite{Shi:2022erw,Zhang:2024plq,Ninomiya:2017ggn,Liu:2025fuf} at model scale as well as at $Q^2=2.4$ GeV$^2$.}
    \label{tab3}
\end{table*}
\begin{table*}
    \centering
    \begin{tabular}{|c|c|c|c|c|c|c|c|c|}
        \hline
        & Our work (S-1) & Our work (S-2)
        & M-NJL \cite{Zhang:2024plq}
        & ILM \cite{Liu:2025fuf}
        & NJL \cite{Ninomiya:2017ggn} & LFHM \cite{Kaur:2020emh} & BHL \cite{Kaur:2020emh} & BSE \cite{Shi:2022erw}\\
        \hline
        $f_1(x,\bfk)$  & 0.32 & 0.44 & 0.368 &0.0810 &  0.32 &0.238  &0.328 & 0.399\\
        \hline
        $g_{1L}(x,\bfk)$  & 0.27 & 0.29 &  0.327 &  0.0751& 0.08& 0.204& 0.269&0.318\\
        \hline
        $g_{1T}(x,\bfk)$  & 0.27 & 0.41 &  0.346 &0.0785 &  0.34& 0.229&0.269 &0.358\\
        \hline
        $h_1(x,\bfk)$  & 0.31 & 0.41 &  0.346 &  0.0785& 0.34&  0.229&0.307 & 0.367\\
        \hline
        $h_{1L}(x,\bfk)$  & 0.27 & 0.29 & 0.327 &  0.0751& 0.33 & 0.204& 0.269& 0.368\\
        \hline
        $h_{1T}(x,\bfk)$  & 0.24 & - & - & -& -&- & 0.237&0.365\\
        \hline
        $f_{1LL}(x,\bfk)$  & - & - & - &-&- & -&- &-\\
        \hline
        $f_{1LT}(x,\bfk)$  & - & 0.37 & - & -& -& -& -&-\\
        \hline
        $f_{1TT}(x,\bfk)$  & - & 0.32 & 0.322 & 0.0734& 0.32&  0.211& -& 0.338\\
        \hline
    \end{tabular}
    \caption{Average transverse momenta $\langle \mathbf{k}_\perp \rangle$ of quark TMDs for S-1 and S-2 type spin wave functions compared with available theoretical models \cite{Shi:2022erw,Kaur:2020emh,Ninomiya:2017ggn,Liu:2025fuf,Zhang:2024plq}.}
    \label{transverse}
\end{table*}
\section{Parton Distribution Functions}
\label{dis}
\subsection{Leading twist}
Quark PDFs play a crucial role in describing hadron structure in terms of the longitudinal momentum fraction ($x$). At leading twist, spin-$1$ mesons possess four independent quark PDFs, compared to one for spin-$0$ mesons and three for spin-$\tfrac{1}{2}$ nucleons. These are the unpolarized distribution $f(x)$, the helicity distribution $h(x)$, the transversity distribution $g(x)$, and the tensor distribution $f_{1LL}(x)$. Among them, $f(x)$, $g(x)$, and $h(x)$ are common to spin-$\tfrac{1}{2}$ nucleons \cite{Meissner:2009ww}, while the tensor PDF $f_{1LL}(x)$ arises uniquely from the tensor polarization of spin-$1$ hadrons \cite{Kumano:2020ijt}. These quark PDFs can be obtained by evaluating the quark–quark correlator for spin-$1$ mesons, while taking their polarization states into consideration, as
\begin{eqnarray}
\Phi_{ij}(x) &=& \int \frac{dz^-}{2\pi} e^{i k \cdot z}%
\allowbreak \nonumber\\
&\times& \langle M_{\lambda}(P) | 
\overline{\psi}_j(0) \psi_i(z) | M_{\lambda}(P) \rangle_{z^+ = z_\perp = 0}.
\label{pdf}
\end{eqnarray}
For the spin-$1$ case, the quark PDFs can be calculated using the trace with different $\Gamma$ matrices as \cite{Ninomiya:2017ggn,Bacchetta:2001rb}
\begin{eqnarray}
    \langle \Gamma\rangle^{\lambda}_{ij} (x)&=&\frac{1}{2}\text{Tr} \Big[\Gamma \Phi_{ij}(x)\Big]\nonumber\\
    &&
    =\epsilon^{* \mu}_{\lambda}(P^*)\langle \Gamma \rangle^{\mu \nu}(x)\epsilon^{\nu}_{\lambda}(P).\label{a21}
\end{eqnarray}
The spin-$1$ quark PDFs corresponding to different choices of $\Gamma$ for the unpolarized, polarized, and tensor-polarized hadron cases are expressed as
\begin{eqnarray}
    \langle \gamma^+ \rangle^{\lambda}_{ij} (x)&=& f(x) + S_{LL} f_{1LL},\label{a11}\\
     \langle \gamma^+ \gamma_5 \rangle^{\lambda}_{ij} (x)&=& \lambda S_{L} g(x),\label{a221}\\
     \langle \gamma^+ \gamma^{i_1} \gamma_5 \rangle^{\lambda}_{ij} (x)&=& S^{i_1}_{\perp} h(x).\label{a31}
\end{eqnarray}
Here, $S_{LL}$ denotes the longitudinal tensor polarization of the hadron, written as \begin{eqnarray}
    S_{LL}=(3\lambda^2-2)\Big(\frac{1}{6}-\frac{S_L^2}{2}\Big),
\end{eqnarray} 
with $S=(S_L,S_T^x,S_T^y)$ representing the hadron polarization vector. 

\par For a given polarization direction $S$, the hadron can have three possible spin projections $\lambda=0,\pm 1$. 
For longitudinal polarization, where $|S_T|=0$ and $|S_L|=1$, the spin projections are $\lambda=\pm1,0$ along the momentum direction. 
For transverse polarization, where $|S_L|=0$ and $|S_T|=1$, the three projections are along the direction perpendicular to the momentum. Here, $i_1$ denotes the transverse components $x$ and $y$. The LF wave function (LFWF) overlap representations of these quark PDFs can be obtained by equating Eqs.~(\ref{a11})--(\ref{a31}) with Eq.~(\ref{a21}). The explicit derivation of the LFWF overlap form is provided in Appendix~\ref{appendixspin}. 
The resulting overlap representations of the quark PDFs are given as
\begin{widetext}
\begin{eqnarray}
    f(x)&=&\int \frac{d^2 \textbf{k}_\perp}{6 (2\pi)^3}\sum_{i,j}\Bigg[|\Psi^{+1}_{i,j}(x,\textbf{k}^2_\perp)|^2 +|\Psi^{0}_{i,j}(x,\textbf{k}^2_\perp)|^2+|\Psi^{-1}_{i,j}(x,\textbf{k}^2_\perp)|^2\Bigg],\\
    g(x)&=& \int \frac{d^2 \textbf{k}_\perp}{4 (2\pi)^3}\sum_{j}\Bigg[|\Psi^{+1}_{\uparrow,j}(x,\textbf{k}^2_\perp)|^2+|\Psi^{-1}_{\downarrow,j}(x,\textbf{k}^2_\perp)|^2-|\Psi^{+1}_{\downarrow,j}(x,\textbf{k}^2_\perp)|^2 -|\Psi^{-1}_{\uparrow,j}(x,\textbf{k}^2_\perp)|^2\Bigg], \\
    h(x)&=& \int \frac{d^2 \textbf{k}_\perp}{4 \sqrt{2} (2\pi)^3}\sum_{j}\Bigg[ \Psi^{+1*}_{\uparrow,j}(x,\textbf{k}^2_\perp)\Psi^{0}_{\downarrow,j}(x,\textbf{k}^2_\perp) +\Psi^{0*}_{\downarrow,j}(x,\textbf{k}^2_\perp)\Psi^{+1}_{\uparrow,j}(x,\textbf{k}^2_\perp)+\Psi^{0*}_{\uparrow,j}(x,\textbf{k}^2_\perp)\Psi^{-1}_{\downarrow,j}(x,\textbf{k}^2_\perp)\nonumber\\
    &&
    +\Psi^{-1*}_{\downarrow,j}(x,\textbf{k}^2_\perp)\Psi^{0}_{\uparrow,j}(x,\textbf{k}^2_\perp)\Bigg],\\
    f_{1LL}(x)&=&\int \frac{d^2 \textbf{k}_\perp}{2 (2\pi)^3}\sum_{i,j}\Bigg[\frac{-1}{2}\bigg(|\Psi^{+1}_{i,j}(x,\textbf{k}^2_\perp)|^2 +|\Psi^{-1}_{i,j}(x,\textbf{k}^2_\perp)|^2\bigg)+|\Psi^{0}_{i,j}(x,\textbf{k}^2_\perp)|^2\Bigg].   
\end{eqnarray}
\end{widetext}
The explicit expressions of these quark PDFs, obtained using the S-1 spin wave function, are given by
\begin{widetext}
  \begin{eqnarray}
      f(x)&=&\int \frac{d^2 \textbf{k}_\perp}{2 (2 \pi)^3}|\phi(x,\textbf{k}^2_\perp)|^2,\\
       g(x)&=&\int \frac{d^2 \textbf{k}_\perp}{4 (2 \pi)^3}\Bigg(\frac{2 \left(\bfk (2 x-1)+m^2\right) M_{q\bar q}+4 m \left(\bfk+m^2\right)}{\left(\bfk+m^2\right) (M_{q\bar q}+2 m)}\Bigg)|\phi(x,\textbf{k}^2_\perp)|^2,\\
       h(x)&=&\int \frac{d^2 \textbf{k}_\perp}{2(2 \pi)^3}\Bigg(\frac{ \left(\bfk \left(2 m (x+1) M_{q \bar q}+x (2 x-1) M_{q \bar q}^2+6 m^2\right)+m^2 (M_{q \bar q}+2 m)^2+2 \textbf{k}_\perp^4\right)}{\left(\bfk+m^2\right) (M_{q \bar q}+2 m)^2}\Bigg)|\phi(x,\textbf{k}^2_\perp)|^2,\\
        f_{1LL}(x)&=&0.
  \end{eqnarray}  
\end{widetext}
Similarly, the explicit expressions of the leading-twist quark PDFs using the S-2 type spin wave function are obtained as
\begin{widetext}
  \begin{eqnarray}
      f(x)&=&\int \frac{d^2 \textbf{k}_\perp}{6 (2 \pi)^3}\Bigg(\frac{2 \left(\text{A}_L^2 \left(\bfk+m^2-\text{M}_{q \bar q}^2 (x-1) x\right)^2+2 \text{A}_T^2 \left(\bfk (2 (x-1) x+1)+m^2\right)\right)}{(1-x) x}\Bigg)|\phi(x,\textbf{k}^2_\perp)|^2,\\
       g(x)&=&\int \frac{d^2 \textbf{k}_\perp}{4 (2 \pi)^3}\Bigg(\frac{4 \text{A}_T^2 \left(\bfk (2 x-1)+m^2\right)}{(1-x) x}\Bigg)|\phi(x,\textbf{k}^2_\perp)|^2,\\
       h(x)&=&\int \frac{d^2 \textbf{k}_\perp}{(2 \pi)^3}\Bigg(\frac{\text{A}_L \text{A}_T m \left(\bfk+m^2-\text{M}_{q \bar q}^2 (x-1) x\right)}{(1-x) x}\Bigg)|\phi(x,\textbf{k}^2_\perp)|^2,\\
        f_{1LL}(x)&=&\int \frac{d^2 \textbf{k}_\perp}{2 (2\pi)^3}\Bigg(\frac{-2 \text{A}_T^2 \left(\bfk (2 (x-1) x+1)+m^2\right)+2 \text{A}_L^2 \left(\bfk+m^2-\text{M}_{q \bar q}^2 (x-1) x\right)^2}{(1-x) x}\Bigg)|\phi(x,\textbf{k}^2_\perp)|^2.
  \end{eqnarray}  
\end{widetext}
The LFWFs satisfy angular momentum conservation projected along the $z$-axis, i.e., $J_z = s_q + s_{\bar q} + L_z$, where $s_q$ and $s_{\bar q}$ denote the spin contributions of the quark and antiquark, respectively, and $L_z$ is the orbital angular momentum of the $\rho$ meson, taking values $L_z = 0, \pm 1, \pm 2$ depending on the specific LFWF. The unpolarized $f(x)$, helicity $g(x)$, and tensor-polarized $f_{1LL}(x)$ quark PDFs arise from the diagonal matrix elements of the $6 \times 6$ helicity amplitude matrix of Eq.~(\ref{maatrix}), whereas the transversity $h(x)$ corresponds to non-diagonal elements. The PDFs $f(x)$, $g(x)$, and $f_{1LL}(x)$ are associated with non-flip hadron and quark polarizations (i.e., zero orbital angular momentum transfer between the initial and final hadron), while $h(x)$ involves polarization flips. For the S-1 type spin wave functions, the tensor-polarized PDF $f_{1LL}(x)$ vanishes, whereas it is non-zero for the S-2 type.
In Fig.~\ref{fig1}, we plot all leading-twist quark PDFs as functions of the longitudinal momentum fraction $x$ carried by the active quark at the model scale $Q^2 = 0.20~\mathrm{GeV}^2$. The unpolarized $f(x)$ and tensor-polarized $f_{1LL}(x)$ PDFs for both spin wave function types are found to satisfy the PDF sum rule
\begin{eqnarray}
    \int _{0}^{1} dx f(x)=1, \ \ \ \ \int _{0}^{1} dx f_{1LL}(x)=0.
\end{eqnarray}
This implies that the valence quark and antiquark carry $100\%$ of the light-front momentum, independent of the hadron polarization. Both $f(x)$ and $f_{1LL}(x)$ quark PDFs exhibit symmetry under the exchange $x \longleftrightarrow (1-x)$ around $x=1/2$. For the $S$-2 type spin wave function, the $g(x)$ and $f_{1LL}(x)$ PDFs show two nodes. 

At the model scale, the quark PDFs computed using the $S$-1 method are strictly positive, whereas in the $S$-2 case, $g(x)$ and $f_{1LL}(x)$ take both positive and negative values. This kind of mixed distribution has also been reported in the NJL model \cite{Ninomiya:2017ggn}. 
Our $f_{1LL}(x)$ and $x f_{1LL}(x)$ distributions obtained with the $S$-2 method show qualitative agreement with results from the LFHM \cite{Kaur:2020emh}, NJL model \cite{Ninomiya:2017ggn,Zhang:2024plq}, and ILM \cite{Liu:2025fuf}. 
The tensor PDF $f_{1LL}(x)$ is related to the structure function $b_1(x)$ via
\begin{eqnarray}
    b_1(x) = \frac{1}{2} f_{1LL}(x),
\end{eqnarray}
which measures the difference in the spin projection of the $\rho$ meson, depending solely on the quark spin-average distribution. 

\par In Figs.~\ref{fig2} and \ref{fig3}, we show the evolution of the quark PDFs obtained from both methods at $Q^2=2.4$, $5$, and $20$ GeV$^2$ using NLO DGLAP evolution \cite{Miyama:1995bd,Hirai:1997gb,Hirai:1997mm}. 
With increasing $Q^2$, the quark PDFs decrease in magnitude for both methods. 
After evolution, all quark PDFs remain positive except $f_{1LL}(x)$. 

\par The average longitudinal momentum fraction, or $x$-moments, of the PDFs are defined as
\begin{eqnarray}
    \langle x^n \rangle_{Q^2} = \int_{0}^{1} dx \, x^n \, \text{PDF}(x,Q^2).
\end{eqnarray}

In this work, we have also calculated $\langle x^n \rangle$ moments up to $n=4$ for different PDFs at various scales by fitting the evolved PDFs using a neural network (NN) architecture of $(1,4,1)$, i.e., one input layer, four hidden layers (each with 128 neurons), and one output layer. 
We tested several activation functions and chose the Tanh function for the hidden layers, defined as
\begin{eqnarray}
    \sigma(z) = \frac{e^z - e^{-z}}{e^z + e^{-z}}.
\end{eqnarray}
The resulting $\langle x^n \rangle$ values are summarized in Table~\ref{tab1}. 
\par At $Q^2=2.4$ GeV$^2$, the mean longitudinal momentum fraction $\langle x \rangle$ of the unpolarized PDF $f(x)$ is $0.24$ for both methods, indicating that the quark and antiquark together carry only $48\%$ of the hadron momentum at this scale. This fraction decreases to $40\%$ at $Q^2=20$ GeV$^2$. For comparison, the ILM \cite{Liu:2025fuf} and BSE \cite{Shi:2022erw} approaches give $\langle x \rangle = 0.201$ and $0.316$, respectively, at a similar scale.
\par The transversity PDF $g(x)$ encodes information about the total spin of the $\rho$ meson. 
The spin sum rule, $\int_0^1 dx \, g(x)$, yields
\[
\int_0^1 dx \, g(x) =
\begin{cases}
0.59, & \text{S-1}, \\
0.41, & \text{S-2}.
\end{cases}
\]
This indicates that, at the model scale, valence quarks contribute $59\%$ to the $\rho$ meson spin in the S-1 method, and $42\%$ in the S-2 method. 
Consequently, in the S-2 approach, the quark orbital angular momentum (OAM) accounts for $58\%$ of the meson spin at the model scale, increasing to $74\%$ at $Q^2 = 20$ GeV$^2$. 
A similar trend is observed for the S-1 method at higher scales. 
Table~\ref{tab3} compares our spin sum rule results at the model scale and at $Q^2=2.4$ GeV$^2$ with other theoretical predictions \cite{Shi:2022erw,Liu:2025fuf,Ninomiya:2017ggn,Zhang:2024plq}. 

\par We also find that the valence quark and antiquark contribute $80\%$ and $51\%$ to the total tensor charge, $\int_0^1 dx \, h(x)$, of the $\rho$ meson in the S-1 and S-2 methods at the model scale, respectively. 
These contributions decrease to $34\%$ and $20\%$ at $Q^2=20$ GeV$^2$. 
Examining the tensor-polarized PDF $f_{1LL}(x)$, we observe that $\int_0^1 dx \, x f_{1LL}(x)$ remains small but finite at high $Q^2$; for example, at $Q^2=5$ GeV$^2$, we find $\int_0^1 dx \, x f_{1LL}(x) = 0.001$, compared to $0.03$ in the NJL model \cite{Ninomiya:2017ggn}. 
\par Finally, we have studied the positivity constraints for spin-1 meson leading twist PDFs, which can be expressed as 
\begin{widetext}
  \begin{align}
f(x) &\ge 0, \\[4pt]
f(x) - g(x) - \tfrac{1}{3} f_{1LL}(x) &\ge 0, \quad (\text{PS-1}) \\[4pt]
2 f(x) + g(x) + \tfrac{1}{3} f_{1LL}(x) &\ge 0, \quad (\text{PS-2}) \\[4pt]
\Big(f(x) + \tfrac{2}{3} f_{1LL}(x)\Big)
\Big(f(x) + g(x) - \tfrac{1}{3} f_{1LL}(x)\Big)
&\ge 2\,|h(x)|^2, \quad (\text{PS-3})\\[4pt]
|h(x)| \le f(x)+\frac{1}{6}f_{1LL}(x)\le \frac{3}{2}f(x). \quad (\text{PS-4})
\end{align}  
\end{widetext}
For positivity constraints, we have used the notation ``PS" throughout this work. The positivity constraints of both S-1 and S-2 type spin wave functions have been plotted in Fig. \ref{pcpdf1}. We observed that all the positivity constraints are obeyed by both the spin wave functions. 
\subsection{Sub-leading twist}
Similarly, at subleading twist, spin-$1$ mesons possess a total of five quark PDFs, in contrast to the three found in spin-$\tfrac{1}{2}$ nucleons. The two additional distributions correspond to tensor quark PDFs that are unique 
to spin-$1$ systems. The twist-3 quark PDFs can be expressed in the form
%
%
%
\begin{eqnarray}
    \langle \textbf{1} \rangle^{\lambda}_{ij} (x)&=& \frac{M_\rho}{P^+}(e(x) + S_{LL} e_{LL}(x)),\label{a5}\\
     \langle \gamma^{i_1} \gamma_5 \rangle^{\lambda}_{ij} (x)&=&\frac{M_\rho}{P^+} ( S^{i_1}_{\perp} g_T(x)),\label{a6}\\
     \langle i \sigma^{+-}\gamma_5 \rangle^{\lambda}_{ij} (x)&=&\frac{M_\rho}{P^+} (\lambda S_{L} h_L(x)),\label{a7} \\
     \langle \gamma^{i_1} \rangle^{\lambda}_{ij} (x)&=& \frac{M_\rho}{P^+} ( S^{i_1}_{LT} f_{LT}(x)).\label{a8}
\end{eqnarray}
The overlap representations of these quark PDFs, which incorporate all possible helicity and polarization states of the hadron, are obtained by solving the $6 \times 6$ matrix presented in the Appendix \ref{appendixspin}. The resulting overlap expressions for these quark PDFs are
\begin{widetext}
    \begin{eqnarray}
    e(x)&=&\int \frac{d^2 \textbf{k}_\perp}{6 (2\pi)^3}\sum_{i,j}\frac{ m}{xM_\rho}\Bigg[|\Psi^{+1}_{i,j}(x,\textbf{k}^2_\perp)|^2 +|\Psi^{0}_{i,j}(x,\textbf{k}^2_\perp)|^2+|\Psi^{-1}_{i,j}(x,\textbf{k}^2_\perp)|^2\Bigg]=\frac{m}{xM_\rho}f(x),\\
    g_T(x)&=& \int \frac{d^2 \textbf{k}_\perp}{4\sqrt{2} (2\pi)^3}\frac{1}{x M_\rho}\sum_{j}\Bigg[\frac{\textbf{k}^R_\perp}{2}\Bigg(\Psi^{+1*}_{\uparrow,j}(x,\textbf{k}^2_\perp)\Psi^{0}_{\uparrow,j}(x,\textbf{k}^2_\perp)-\Psi^{+1*}_{\downarrow,j}(x,\textbf{k}^2_\perp)\Psi^{0}_{\downarrow,j}(x,\textbf{k}^2_\perp)\nonumber\\
    &&+\Psi^{0*}_{\uparrow,j}(x,\textbf{k}^2_\perp)\Psi^{-1}_{\uparrow,j}(x,\textbf{k}^2_\perp)-\Psi^{0*}_{\downarrow,j}(x,\textbf{k}^2_\perp)\Psi^{-1}_{\downarrow,j}(x,\textbf{k}^2_\perp)\Bigg)+\frac{\textbf{k}^L_\perp}{2}\Bigg(\Psi^{0*}_{\uparrow,j}(x,\textbf{k}^2_\perp)\Psi^{+1}_{\uparrow,j}(x,\textbf{k}^2_\perp)-\Psi^{0*}_{\downarrow,j}(x,\textbf{k}^2_\perp)\Psi^{+}_{\downarrow,j}(x,\textbf{k}^2_\perp)\nonumber\\
    &&+\Psi^{-1*}_{\uparrow,j}(x,\textbf{k}^2_\perp)\Psi^{0}_{\uparrow,j}(x,\textbf{k}^2_\perp)-\Psi^{-1*}_{\downarrow,j}(x,\textbf{k}^2_\perp)\Psi^{0}_{\downarrow,j}(x,\textbf{k}^2_\perp)\Bigg)+m\Bigg(\Psi^{+1*}_{\uparrow,j}(x,\textbf{k}^2_\perp)\Psi^{0}_{\downarrow,j}(x,\textbf{k}^2_\perp)+\Psi^{0*}_{\downarrow,j}(x,\textbf{k}^2_\perp)\Psi^{+1}_{\uparrow,j}(x,\textbf{k}^2_\perp)\nonumber\\
    &&+\Psi^{0*}_{\uparrow,j}(x,\textbf{k}^2_\perp)\Psi^{-1}_{\downarrow,j}(x,\textbf{k}^2_\perp)+\Psi^{-1*}_{\downarrow,j}(x,\textbf{k}^2_\perp)\Psi^{0}_{\uparrow,j}(x,\textbf{k}^2_\perp)\Bigg)\Bigg], \\
    h(x)&=& \int \frac{d^2 \textbf{k}_\perp}{4  (2\pi)^3}\sum_{j}\frac{1}{x M_\rho}\Bigg[m\Bigg(|\Psi^{+1*}_{\uparrow,j}(x,\textbf{k}^2_\perp)|^2-|\Psi^{+1*}_{\downarrow,j}(x,\textbf{k}^2_\perp)|^2+|\Psi^{-1*}_{\downarrow,j}(x,\textbf{k}^2_\perp)|^2-|\Psi^{-1*}_{\uparrow,j}(x,\textbf{k}^2_\perp)|^2\Bigg)\nonumber\\
    &&-\textbf{k}^R_\perp\Bigg(\Psi^{+1*}_{\downarrow,j}(x,\textbf{k}^2_\perp)\Psi^{+1}_{\uparrow,j}(x,\textbf{k}^2_\perp)-\Psi^{-1*}_{\downarrow,j}(x,\textbf{k}^2_\perp)\Psi^{-1}_{\uparrow,j}(x,\textbf{k}^2_\perp)\Bigg)\nonumber\\
&&+\textbf{k}^L_\perp\Bigg(\Psi^{+1*}_{\uparrow,j}(x,\textbf{k}^2_\perp)\Psi^{+1}_{\downarrow,j}(x,\textbf{k}^2_\perp)-\Psi^{-1*}_{\uparrow,j}(x,\textbf{k}^2_\perp)\Psi^{-1}_{\downarrow,j}(x,\textbf{k}^2_\perp)\Bigg)\Bigg],\\
    e_{LL}(x)&=&\int \frac{d^2 \textbf{k}_\perp}{2 (2\pi)^3}\sum_{i,j}\frac{m}{xM_\rho}\Bigg[\frac{-1}{2}(|\Psi^{+1}_{i,j}(x,\textbf{k}^2_\perp)|^2 +|\Psi^{-1}_{i,j}(x,\textbf{k}^2_\perp)|^2)+|\Psi^{0}_{i,j}(x,\textbf{k}^2_\perp)|^2\Bigg]=\frac{m}{xM_\rho}f_{1LL}(x),\\ 
    f_{LT}(x)&=&\int \frac{d^2 \textbf{k}_\perp}{2 (2\pi)^3}\sum_{i,j}\frac{1}{4\sqrt{2}x M_\rho}\Bigg[\textbf{k}^R_\perp\Bigg(\Psi^{+1*}_{i,j}(x,\textbf{k}^2_\perp)\Psi^{0}_{i,j}(x,\textbf{k}^2_\perp) - \Psi^{0*}_{i,j}(x,\textbf{k}^2_\perp)\Psi^{-1}_{i,j}(x,\textbf{k}^2_\perp)\Bigg)\nonumber\\
    &&+\textbf{k}^L_\perp\Bigg(\Psi^{0*}_{i,j}(x,\textbf{k}^2_\perp)\Psi^{+1}_{i,j}(x,\textbf{k}^2_\perp) - \Psi^{-1*}_{i,j}(x,\textbf{k}^2_\perp)\Psi^{0}_{i,j}(x,\textbf{k}^2_\perp)\Bigg)\Bigg].
\end{eqnarray}
\end{widetext}
The explicit expressions of the quark PDFs, obtained using the S-1 type spin wave functions, are given by
\begin{widetext}
    \begin{eqnarray}
        e(x)&=& \int \frac{d^2 \textbf{k}_\perp}{2 (2 \pi)^3}\Bigg(\frac{m}{xM_\rho}\Bigg)|\phi(x,\bfk)|^2,\\
        g_T(x)&=&\int \frac{ d^2 \textbf{k}_\perp}{2 (2\pi)^3}\Bigg(\frac{M_{q\bar q} \left(m \left(\bfk (2 (x-1) x+1)+m^2\right) M_{q \bar q}+\textbf{k}_\perp^4+5 \bfk m^2+4 m^4\right)+4 m \left(\bfk+m^2\right)^2}{\text{M}_\rho x \left(\bfk+m^2\right) (M_{q \bar q}+2 m)^2}\Bigg)\nonumber\\
        && \times|\phi(x,\bfk)|^2,\\
        h(x)&=&\int \frac{d^2 \textbf{k}_\perp}{2 (2 \pi)^3}\Bigg(\frac{ \left(m M_{q \bar q}+2 \left(\bfk+m^2\right)\right)}{ \text{M}_\rho x (M_{q\bar q}+2 m)}\Bigg)|\phi(x,\bfk)|^2,\\
        e_{LL}(x)&=&0,\\
        f_{LT}(x)&=&0.
    \end{eqnarray}
\end{widetext}
The explicit expressions of the twist-3 quark PDFs, calculated with the S-2 type spin wave function, are
\begin{widetext}
\begin{eqnarray}
        e(x)&=& \int \frac{d^2 \textbf{k}_\perp}{6 (2 \pi)^3}\Bigg(\frac{m}{xM_\rho}\Bigg)\Bigg(\frac{2 \left(\text{A}_L^2 \left(\bfk+m^2-\text{M}_{q \bar q}^2 (x-1) x\right)^2+2 \text{A}_T^2 \left(\bfk (2 (x-1) x+1)+m^2\right)\right)}{(1-x) x}\Bigg)\nonumber\\
        && \times|\phi(x,\textbf{k}^2_\perp)|^2,\\
        g_T(x)&=&\int \frac{ d^2 \textbf{k}_\perp}{2 (2\pi)^3}\Bigg(\frac{\text{A}_L \text{A}_T \left(\bfk+2 m^2\right) \left(\bfk+m^2+\text{M}_{q \bar q}^2 (1-x) x\right)}{\text{M}_\rho (1-x) x^2}\Bigg)|\phi(x,\bfk)|^2,\\
        h(x)&=&\int \frac{d^2 \textbf{k}_\perp}{2 (2 \pi)^3}\Bigg(\frac{2 \text{A}_T^2 m \left(\bfk+m^2\right)}{\text{M}_\rho (1-x) x^2}\Bigg)|\phi(x,\bfk)|^2,\\
        e_{LL}(x)&=&\int \frac{d^2 \textbf{k}_\perp}{2 (2 \pi)^3}\Bigg(\frac{m}{xM_\rho}\Bigg)\Bigg(\frac{-2 \text{A}_T^2 \left(\bfk (2 (x-1) x+1)+m^2\right)+2 \text{A}_L^2 \left(\bfk+m^2-\text{M}_{q \bar q}^2 (x-1) x\right)^2}{(1-x) x}\Bigg)\nonumber\\
        &&\times|\phi(x,\bfk)|^2,\\
        f_{LT}(x)&=&\int \frac{d^2 \textbf{k}_\perp}{2 (2 \pi)^3}\Bigg(\frac{\text{A}_L \text{A}_T \bfk (2 x-1) \left(\bfk+m^2+\text{M}_{q\bar q}^2 (1-x) x\right)}{\text{M}_\rho (1-x) x^2}\Bigg)|\phi(x,\bfk)|^2.
\end{eqnarray}
\end{widetext}
\begin{figure*}
\centering
\begin{minipage}[c]{0.98\textwidth}
(a)\includegraphics[width=7.5cm]{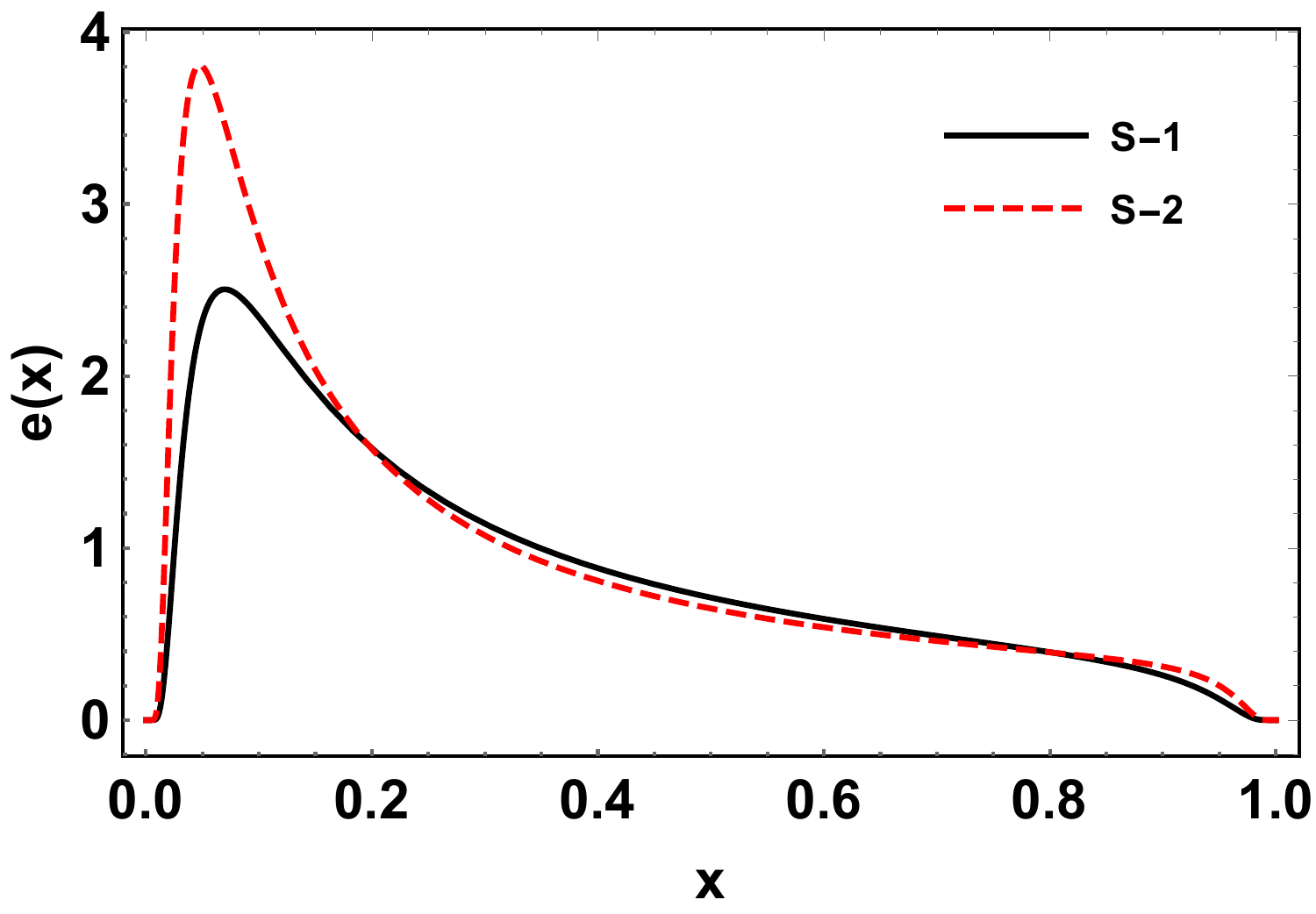}
\hspace{0.03cm}	
(b)\includegraphics[width=7.5cm]{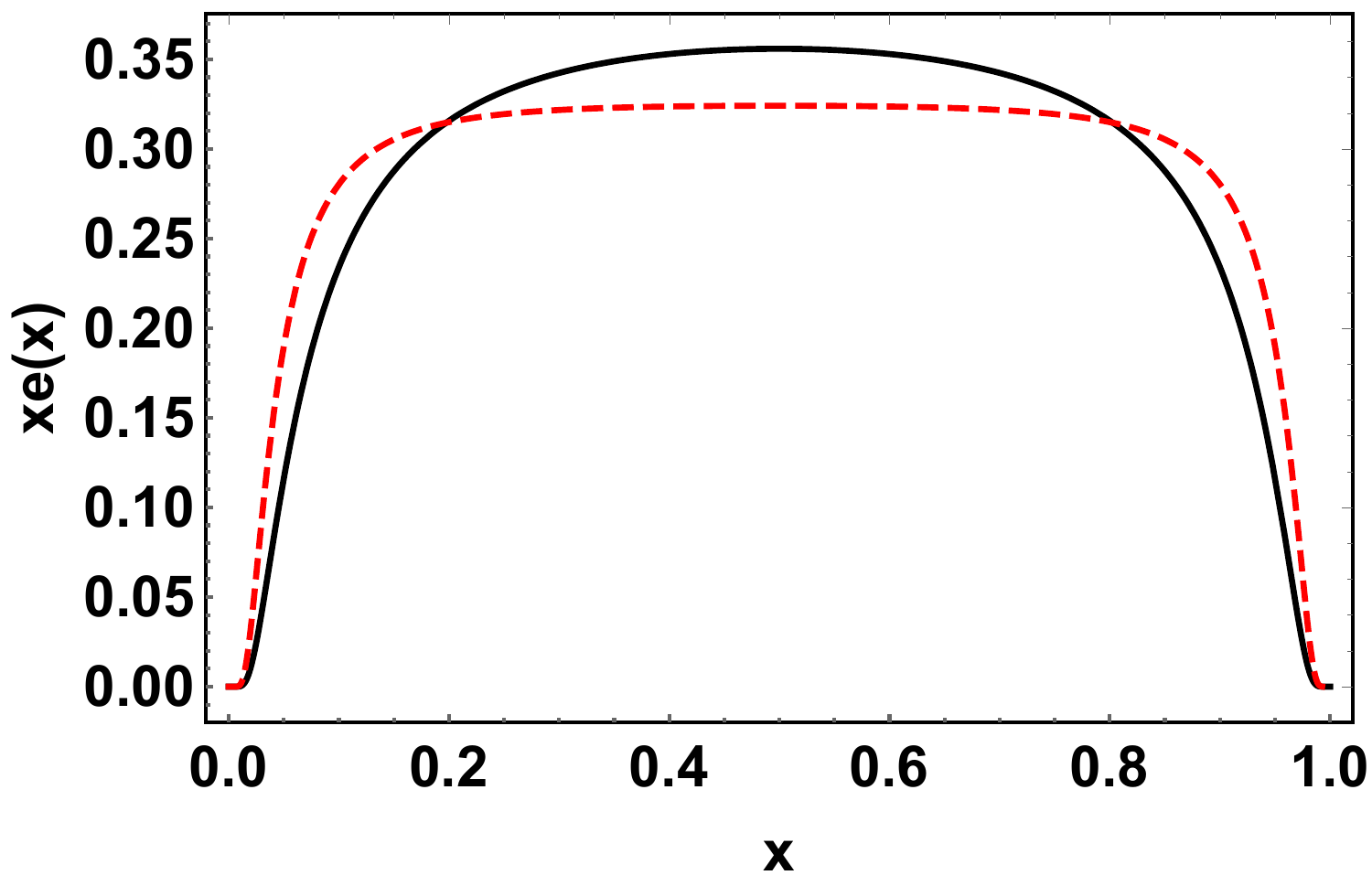} 
\hspace{0.03cm}
\end{minipage}
\centering
\begin{minipage}[c]{0.98\textwidth}
(c)\includegraphics[width=7.5cm]{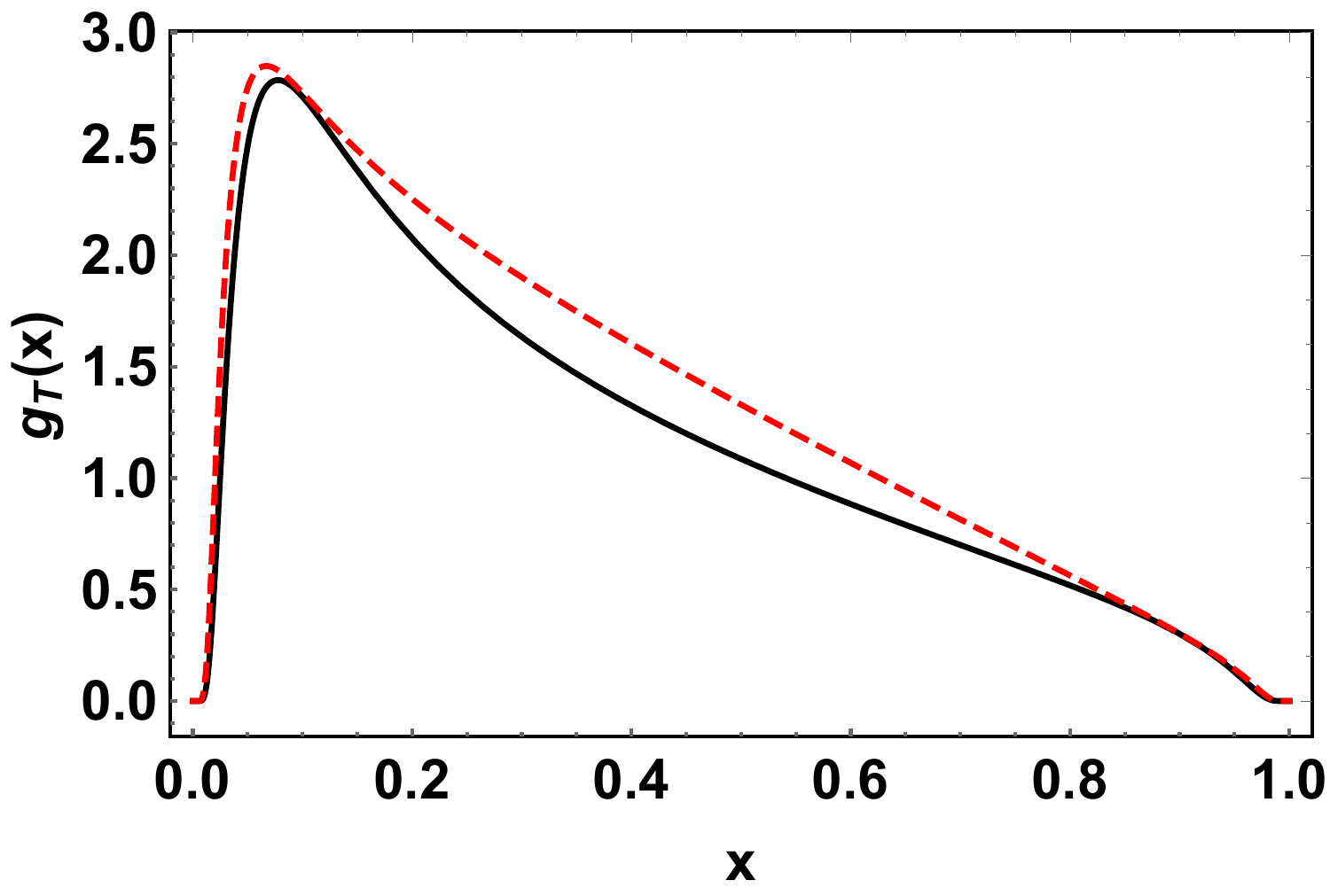}
\hspace{0.03cm}	
(d)\includegraphics[width=7.5cm]{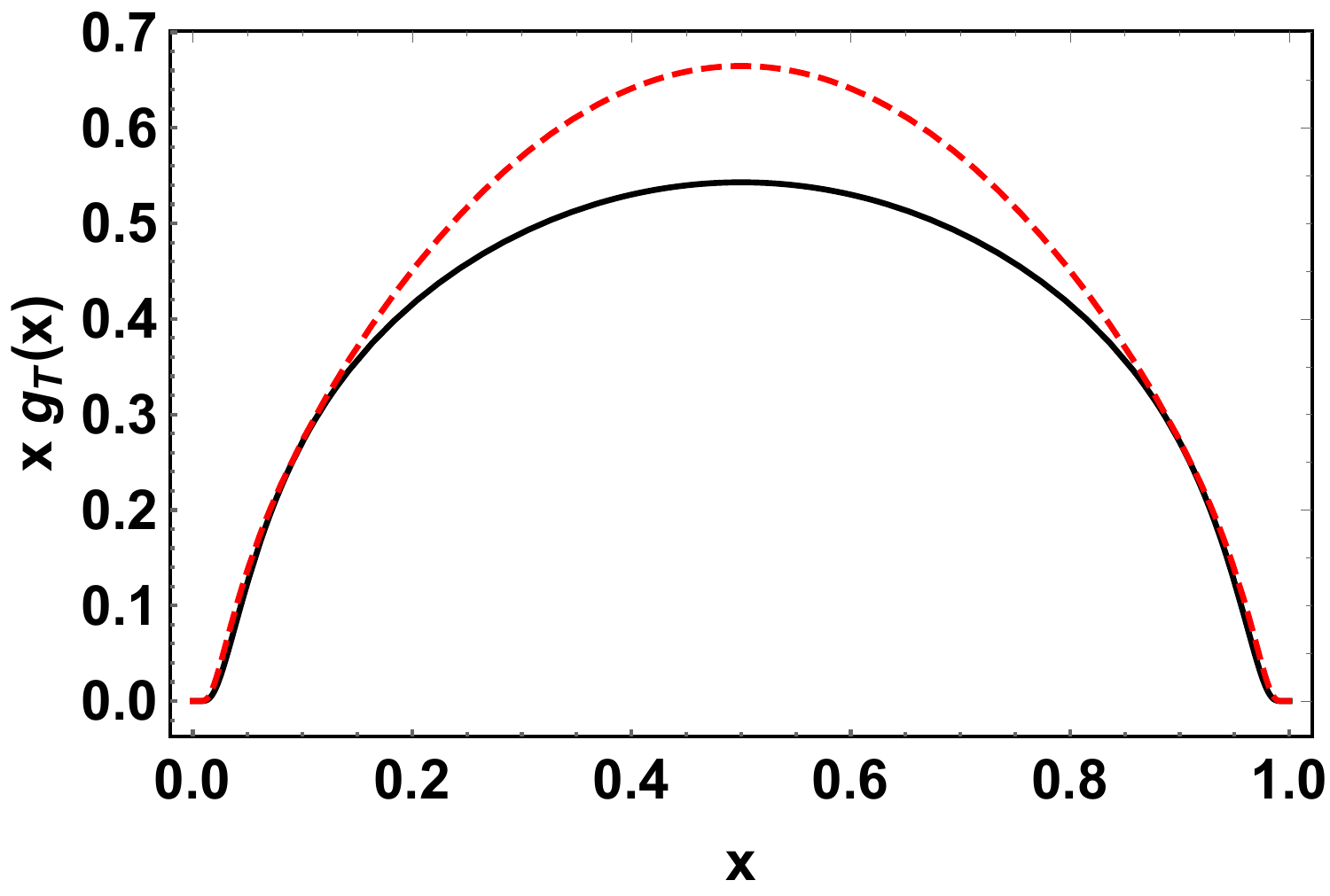} 
\hspace{0.03cm}
\end{minipage}
\centering
\begin{minipage}[c]{0.98\textwidth}
(e)\includegraphics[width=7.5cm]{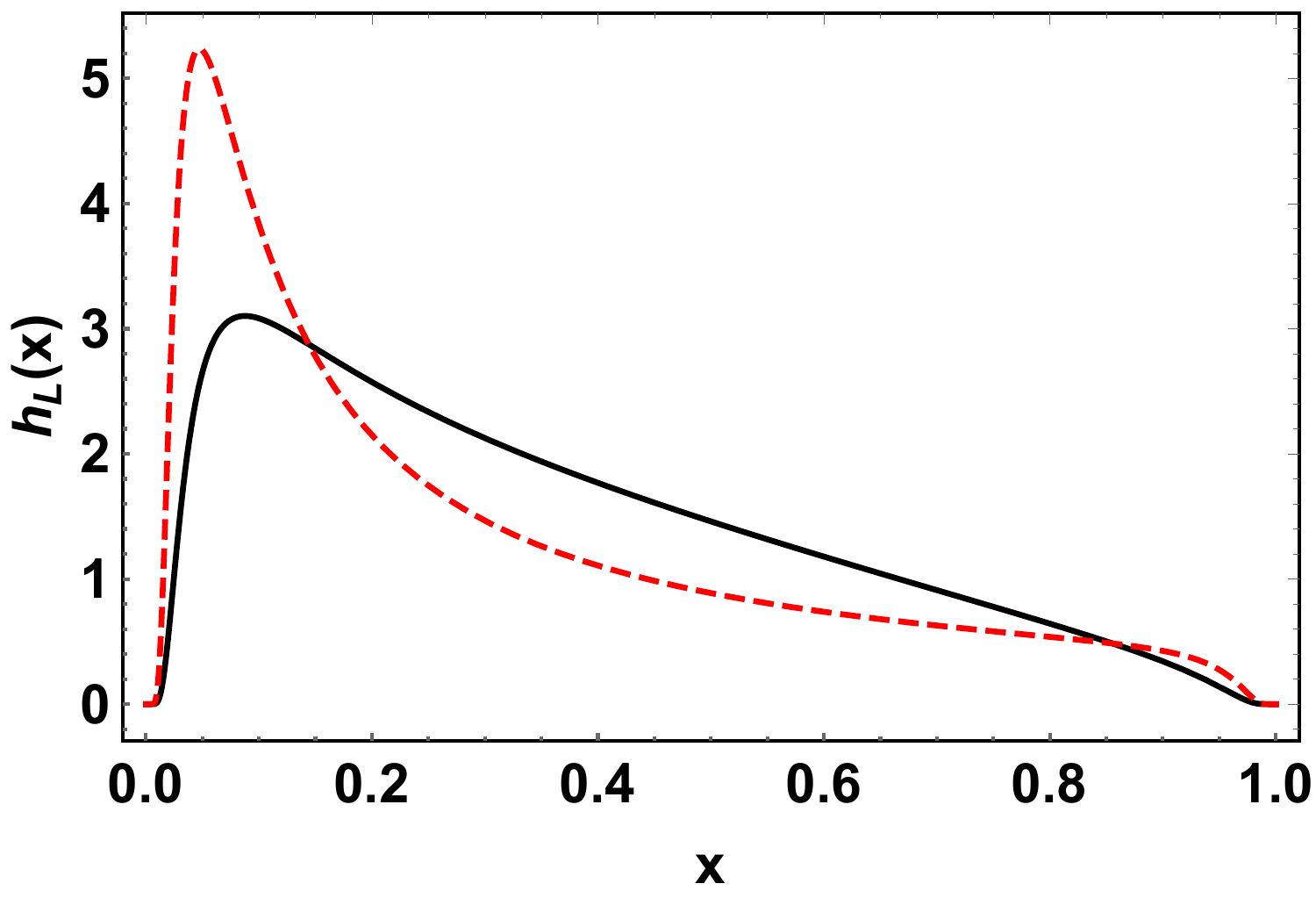}
\hspace{0.03cm}	
(f)\includegraphics[width=7.5cm]{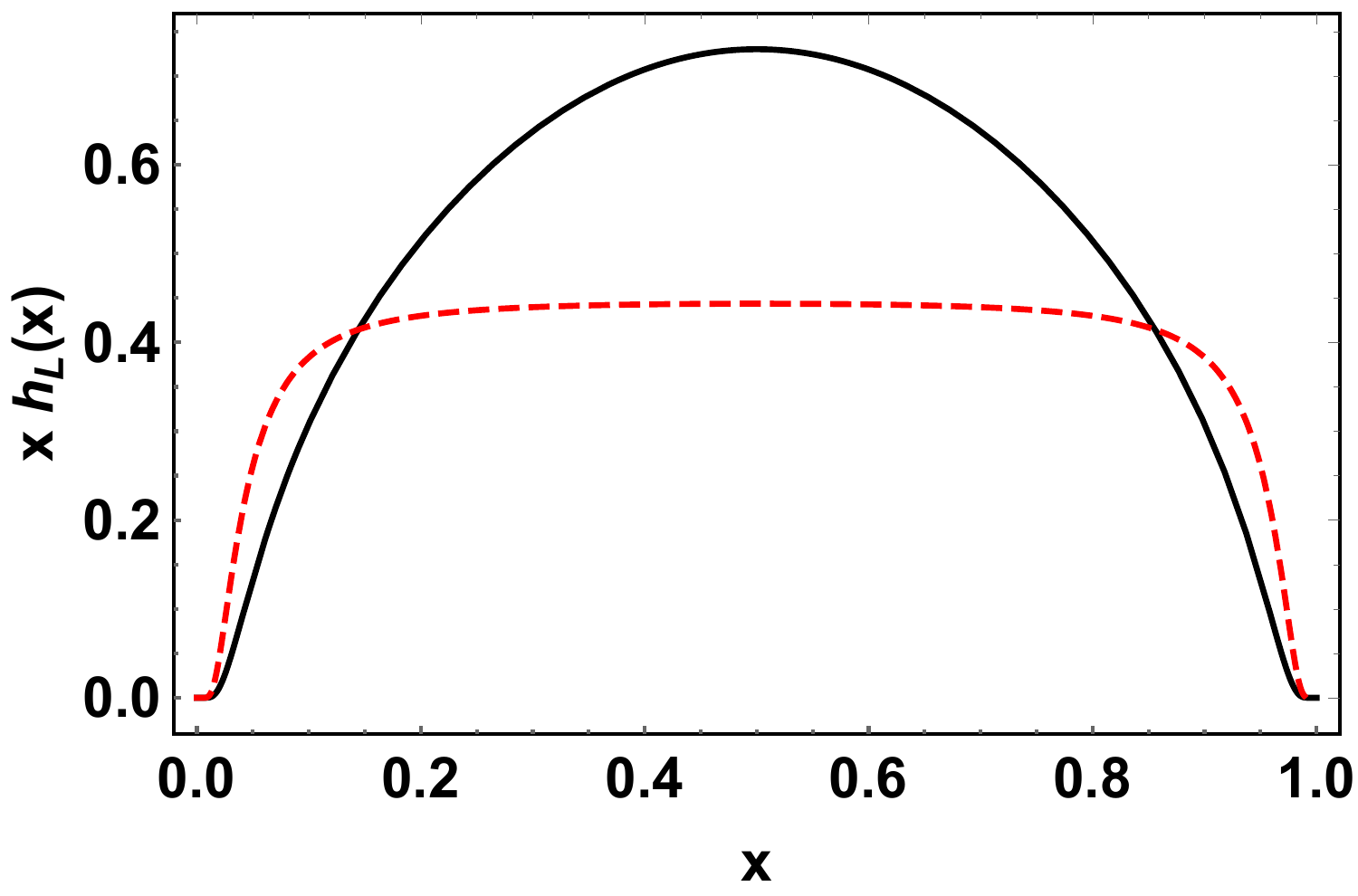} 
\hspace{0.03cm}
\end{minipage}
\centering
\begin{minipage}[c]{0.98\textwidth}
(g)\includegraphics[width=7.5cm]{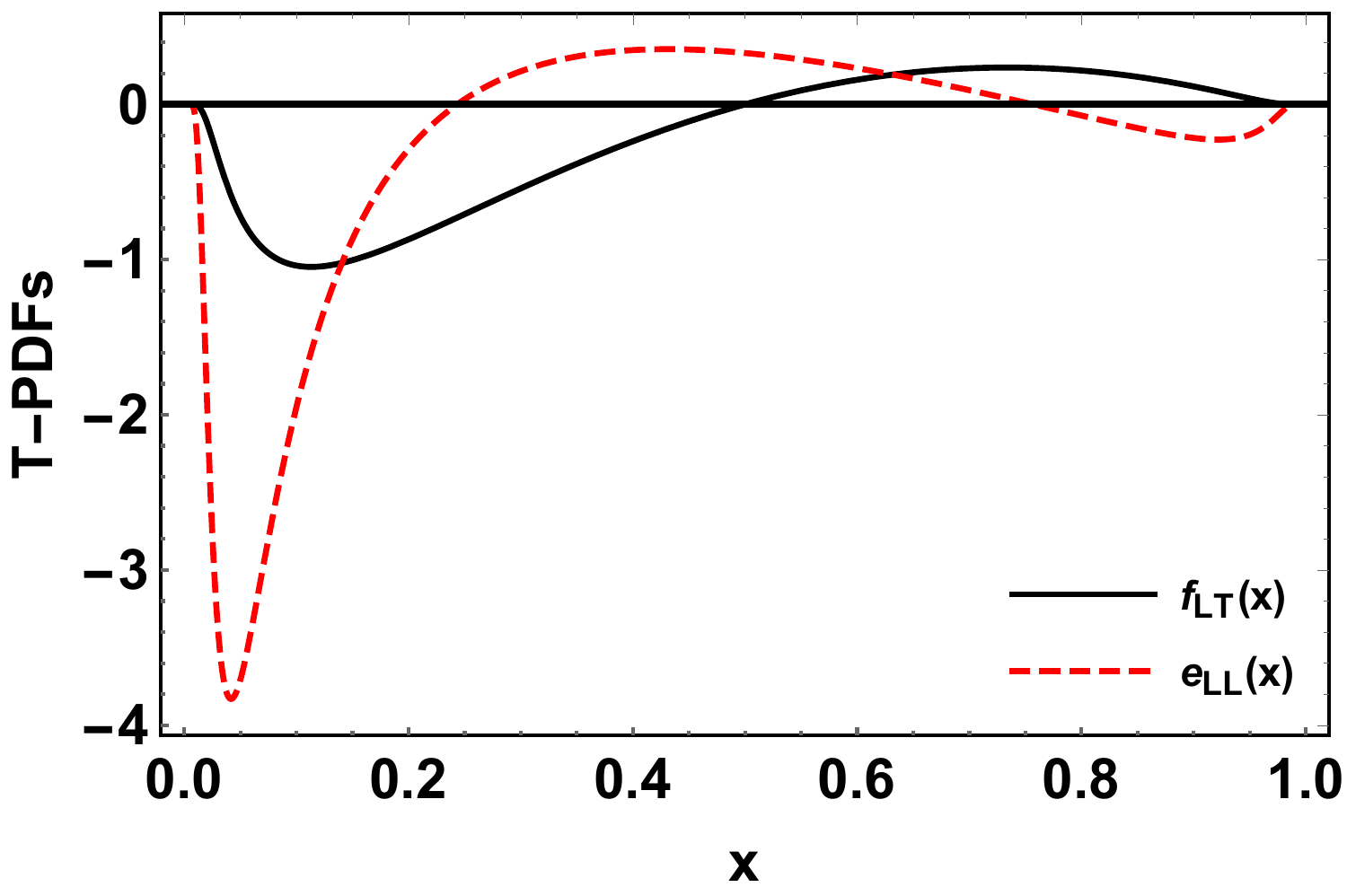}
\hspace{0.03cm}	
(h)\includegraphics[width=7.5cm]{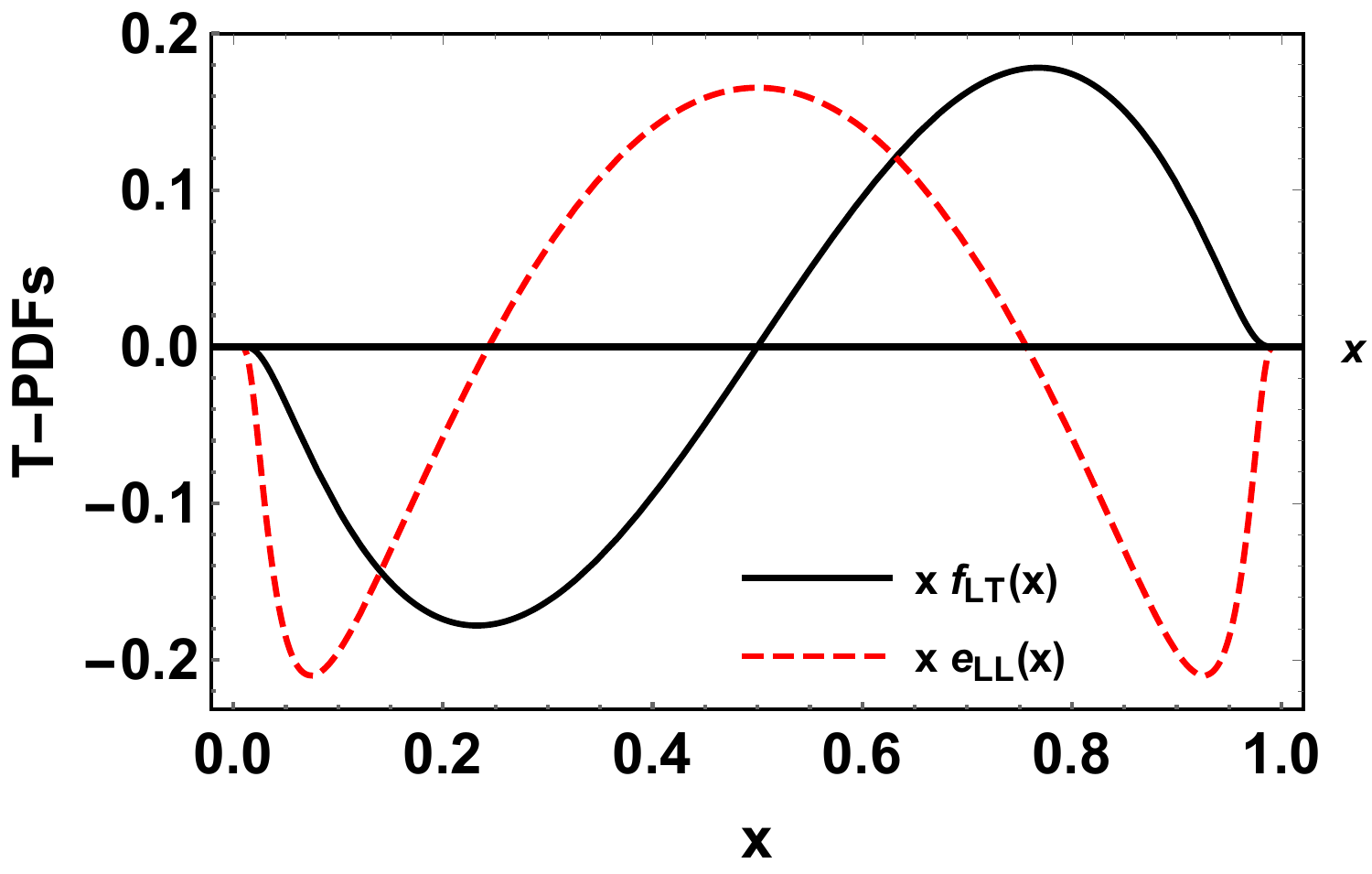} 
\hspace{0.03cm}
\end{minipage}
\caption{Sub-leading twist quark PDFs for S-1 and S-2 spin wave functions at the model scale $Q^2 = 0.20~\mathrm{GeV}^2$ as functions of $x$: 
(a) $e(x)$, (b) $x e(x)$, 
(c) $g_T(x)$, (d) $x g_T(x)$, 
(e) $h_L(x)$, (f) $x h_L(x)$, 
(g) $e_{LL}(x)$, and (h) $f_{LT}(x)$. The $e_{LL}(x)$, and $f_{LT}(x)$ PDFs are zero for the case of S-1 spin wave functions.
%
%
\label{fig4}}
\end{figure*}
Compared to $f_{1LL}(x)$ PDF at the leading twist, there are total $2$ tensor PDF $e_{LL}$ and $f_{LT}$ present at the twist-3 level. As we have calculated all the PDFs through the quark-quark correlator, the gluon contributions are coming to zero in this work. Similar to leading twist, the tensor PDFs are coming zero in the case of S-1 spin wave function. The twist-3 $e(x)$ and $e_{LL}(x)$ are the results of the diagonal matrix elements of the LFWFs as shown in Eq. \ref{twist3pdf}, while $h_L(x)$ quark PDFs are observed to have both diagonal and non-diagonal matrix elements of the LFWFs. All the quark PDFs have been plotted with respect to longitudinal momentum fraction $x$ in Fig. \ref{fig4}. All the quark PDFs except the tensor PDFs are found to have positive distributions in both the spin wave functions. However, the tensor PDFs are found to have both positive and negative distributions for S-2 spin wave function. All the quark distributions are found to have high at low $x$ region due to the $1/x$ behavior in the LFWFs form. Similar to $f_{1LL}$, the twist-3 $e_{LL}$ also have two nodes in the range $0 < x<1$. While the tensor $f_{LT}$ has only one node in the same region. As there is no theoretical prediction or experimental data available for the twist-3 $\rho$ meson case, we can not compare our results. However, one recent work on twist-3 $f_{LT}$ PDF has been reported for the case of deuteron \cite{Kumano:2025rai}. The tensor $f_{LT}(x)$ quark PDF of $\rho$ meson shows a similar kind of distribution behavior to deuteron results \cite{Kumano:2025rai}. All the twist-3 quark PDFs are found to have higher distributions compared to the leading twist PDFs for both spin wave functions. The twist-3 sum rules and the equation of motion for the spin-1 system are also satisfied in this work \cite{Kumano:2021xau,Kumano:2021fem}. The tensor $f_{LT}$ at the Wandzura-Wilczek limit also obeys the sum rule \cite{Kumano:2021fem}
\begin{eqnarray}
    \int_0^1 dx \Bigg( f_{LT}(x)-f_{LT}(-x)\Bigg)=0.
\end{eqnarray}
We have also computed the Mellin moment corresponding to the twist-3 quark PDFs, which are given in Table \ref{tab1}. The $\langle x\rangle$ of the twist-3 PDFs of S-1 type spin wave function are found to have lower values compared to the leading twist quark PDFs, which is not followed by the S-2 type spin wave function. 

\section{Transverse Momentum Parton Distribution Functions}\label{tmdss}
It is also of interest to investigate the transverse structure of quarks inside spin-$1$ hadrons. At leading twist, spin-$1$ mesons possess a total of $18$ quark TMDs, compared to $8$ for spin-$\tfrac{1}{2}$ nucleons and $2$ for spin-$0$ mesons. Imposing time-reversal symmetry reduces the number of independent TMDs to $9$ T-even distributions at leading twist. These quark TMDs can be extracted from the quark–quark correlation functions as \cite{Ninomiya:2017ggn,Kaur:2018ewq,Liu:2025fuf,Bacchetta:2001rb,Zhang:2024nxl}
\begin{widetext}
\begin{eqnarray}
    {\Phi}_{ij}^{\lambda}(x,\bm{k}^2_{\perp})
		&=&\int \frac{\mathrm{d}k^+\mathrm{d}k^-}{(2\pi)^4P^+}\delta\Bigg(x-\frac{k^+}{P^+}\Bigg) \int \mathrm{d}^4z e^{ik\cdot z} \, \langle M_\lambda(P)|\bar{\psi}_{j}(0)\Gamma\psi_{i}(z)|M_{\lambda}(P)\rangle \nonumber\\
		&&
        =\int \frac{\mathrm{d}z^-\mathrm{d}^2\bm{z}_{\perp}}{(2\pi)^3} e^{ik\cdot z} \langle M_\lambda(P)|\bar{\psi}_{j}(0)\Gamma\psi_{i}|M_\lambda(P)\rangle_{z^+=0} \nonumber\\
		&&
        =\varepsilon_{\lambda}^{*\mu}(P)\langle \Gamma\rangle_{ij}^{\mu \nu}\varepsilon_{\lambda}^{\nu}(P)=\langle \Gamma \rangle^{\lambda}_{ij}(x,\bfk).
\end{eqnarray}\label{TMD}
\end{widetext}
Here, $\Gamma = [\gamma^+, \gamma^+ \gamma_5, \gamma^+ \gamma^i \gamma_5]$ encodes the dependence on the hadron polarization. Analogous to the quark PDFs, the quark TMDs corresponding to different hadron polarizations can be obtained by selecting the appropriate Dirac matrix $\Gamma$ as
%
%

%
%
\begin{align}\label{a93}
&\langle\gamma^+\rangle_{ij}^{\lambda}(x,\bfk)=\frac{1}{2}\text{Tr}\left[\gamma^+ \Phi^{\lambda}_{ij}(x,\bfk)\right]\nonumber\\
		&\equiv \varepsilon_{\lambda}^{*\mu}(P^*)\langle\gamma^+\rangle_{ij}^{\mu\nu}(x,\bfk)\varepsilon_{\lambda}^{\nu}(P)\nonumber\\
        &\equiv f_1(x,\bfk)+S_{LL}f_{1LL}(x,\bfk)+\frac{S_{LT}.\textbf{k}_\perp}{M_\rho}f_{1LT}(x,\bfk)\nonumber\\
        &+\frac{\textbf{k}_\perp.S_{TT}.\textbf{k}_\perp}{M_\rho^2}f_{1TT}(x,\bfk),\\
    %
    \label{a94}
		&\langle\gamma^+\gamma_5\rangle_{ij}^{\lambda}(x,\bfk)=\frac{1}{2}\text{Tr}\left[\gamma^+\gamma_5\Phi^{\lambda}_{ij}(x,\bfk)\right]\nonumber\\
		&\equiv \varepsilon_{\lambda}^{*\mu}(P^*)\langle\gamma^+\gamma_5\rangle^{\mu\nu}_{ij}(x,\bfk)\varepsilon_{\lambda}^{\nu}(P)\nonumber\\
        &\equiv \lambda \Bigg[S_L g_{1L}(x,\bfk)+\frac{\textbf{k}_\perp.S_T}{M_\rho}g_{1T}(x,\bfk)\Bigg],\\
    %
    \label{a95}
		&\langle \gamma^+\gamma^{i_1}\gamma_5\rangle_{ij}^{\lambda}(x,\bfk)=\frac{1}{2}\text{Tr}\left[\gamma^+\gamma^{i_1}\gamma_5\Phi^{\lambda}_{ij}(x,\bfk)\right]\nonumber\\
		&\equiv \varepsilon_{\lambda}^{*\mu}(P^*)\langle \gamma^+\gamma^{i_1}\gamma^5\rangle^{\mu\nu}_{ij}(x,\bfk)\varepsilon_{\lambda}^{\nu}(P)\nonumber\\
        &\equiv \lambda \Bigg[ S_T^{i_1} h_1(x,\bfk)+\frac{S_L\textbf{k}^{i_1}_\perp}{M_\rho}h_{1L}(x,\bfk)\nonumber\\
        &+\frac{\Big(2\textbf{k}_\perp^{i_1}\textbf{k}_\perp.S_T-S_T^{i_1}\bfk\Big)}{2M_\rho^2}h_{1T}(x,\bfk)\Bigg].
\end{align}
Here, $S_{LT}$ and $S_{TT}$ denote the tensor polarization of the hadron, and are defined as
\begin{align}
S^{i_1}_{LT} &= (3\lambda^2 - 2)\, S_L S_T^{i_1}, \\
S_{TT}^{i_1 j_1} &= (3\lambda^2 - 2)\Big(S_T^{i_1} S_T^{j_1} - \frac{S_T^2}{2}\delta^{i_1 j_1}\Big).
\end{align}
The overlap representations of the T-even quark TMDs are then given by
\begin{widetext}
\begin{eqnarray}
f_1(x,\bfk)&=&\frac{1}{6(2\pi)^3}\sum_{i,j}\Bigg[|\Psi^{+1}_{i,j}(x,\bfk)|^2+|\Psi^{0}_{i,j}(x,\bfk)|^2+  |\Psi^{-1}_{i,j}(x,\bfk)|^2\Bigg],\nonumber\\
g_{1L}(x,\bfk)&=&\frac{1}{4(2\pi)^3}\sum_{j}\Bigg[|\Psi^{+1}_{\uparrow,j}(x,\bfk)|^2+|\Psi^{-1}_{\downarrow,j}(x,\bfk)|^2- |\Psi^{-1}_{\uparrow,j}(x,\bfk)|^2-|\Psi^{+1}_{\downarrow,j}(x,\bfk)|^2\Bigg],\\       
g_{1T}(x,\bfk)&=&\frac{M_\rho}{4\sqrt{2}(2\pi)^3\bfk}\sum_{j}\Bigg[\textbf{k}^L_\perp\Bigg(\Psi^{0*}_{\uparrow,j}(x,\bfk)\Psi^{+1}_{\uparrow,j}(x,\bfk)+\Psi^{-1*}_{\uparrow,j}(x,\bfk)\Psi^{0}_{\uparrow,j}(x,\bfk)-\Psi^{0*}_{\downarrow,j}(x,\bfk)\Psi^{+1}_{\downarrow,j}(x,\bfk)- \nonumber\\&&\Psi^{-1*}_{\downarrow,j}(x,\bfk)\Psi^{0}_{\downarrow,j}(x,\bfk)\Bigg)+\textbf{k}^R_\perp\Bigg(\Psi^{+1*}_{\uparrow,j}(x,\bfk)\Psi^{0}_{\uparrow,j}(x,\bfk)+\Psi^{0*}_{\uparrow,j}(x,\bfk)\Psi^{-1}_{\uparrow,j}(x,\bfk)-\Psi^{+1*}_{\downarrow,j}(x,\bfk)\nonumber\\
&& \Psi^{0}_{\downarrow,j}(x,\bfk)-\Psi^{0*}_{\downarrow,j}(x,\bfk)\Psi^{-1}_{\downarrow,j}(x,\bfk)\Bigg)\Bigg], \\
h_{1}(x,\bfk)&=&\frac{1}{4\sqrt{2}(2\pi)^3}\sum_{j}\Bigg[\Psi^{+1*}_{\uparrow,j}(x,\bfk)\Psi^{0}_{\downarrow,j}(x,\bfk)+\Psi^{0*}_{\downarrow,j}(x,\bfk)\Psi^{+1}_{\uparrow,j}(x,\bfk)+\Psi^{0*}_{\uparrow,j}(x,\bfk)\Psi^{-1}_{\downarrow,j}(x,\bfk)\nonumber\\
&&+\Psi^{-1*}_{\downarrow,j}(x,\bfk)\Psi^{0}_{\uparrow,j}(x,\bfk)\Bigg],\\
          h_{1T}(x,\bfk)&=&\frac{M_\rho^2}{2\sqrt{2}(2\pi)^3\textbf{k}_\perp^4}\sum_{j}\Bigg[(\textbf{k}_\perp^R)^2\Bigg(\Psi^{+1*}_{\downarrow,j}(x,\bfk)\Psi^{0}_{\uparrow,j}(x,\bfk)+\Psi^{0*}_{\downarrow,j}(x,\bfk)\Psi^{-1}_{\uparrow,j}(x,\bfk)\Bigg)\nonumber\\
&&+(\textbf{k}_\perp^L)^2\Bigg(\Psi^{0*}_{\uparrow,j}(x,\bfk)\Psi^{+1}_{\downarrow,j}(x,\bfk)+\Psi^{-1*}_{\uparrow,j}(x,\bfk)\Psi^{0}_{\downarrow,j}(x,\bfk)\Bigg)\Bigg],\\
         h_{1L}(x,\bfk)&=&\frac{M_\rho}{4(2\pi)^3\bfk}\sum_{j}\Bigg[\textbf{k}_\perp^L\Bigg(\Psi^{+1*}_{\uparrow,j}(x,\bfk)\Psi^{+}_{\downarrow,j}(x,\bfk)-\Psi^{-1*}_{\uparrow,j}(x,\bfk)\Psi^{-1}_{\downarrow,j}(x,\bfk)\Bigg)\nonumber\\
         &&+\textbf{k}_\perp^R\Bigg(\Psi^{+1*}_{\downarrow,j}(x,\bfk)\Psi^{+1}_{\uparrow,j}(x,\bfk)-\Psi^{-1*}_{\downarrow,j}(x,\bfk)\Psi^{-1}_{\uparrow,j}(x,\bfk)\Bigg],\\
    f_{1LL}(x,\bfk)&=&\frac{1}{2(2\pi)^3}\sum_{i,j}\Bigg[\frac{-1}{2}(|\Psi^{+1}_{i,j}(x,\textbf{k}^2_\perp)|^2 +|\Psi^{-1}_{i,j}(x,\textbf{k}^2_\perp)|^2)+|\Psi^{0}_{i,j}(x,\textbf{k}^2_\perp)|^2\Bigg],\\
    f_{1LT}(x,\bfk)&=&\frac{M_\rho}{4\sqrt{2}(2\pi)^3\bfk}\sum_{i,j}\Bigg[\textbf{k}_\perp^L\Bigg(\Psi^{0*}_{i,j}(x,\bfk)\Psi^{+1}_{i,j}(x,\bfk)-\Psi^{-1*}_{i,j}(x,\bfk)\Psi^{0}_{i,j}(x,\bfk)\Bigg)\nonumber\\
&&+\textbf{k}_\perp^R\Bigg(\Psi^{+1*}_{i,j}(x,\bfk)\Psi^{0}_{i,j}(x,\bfk)-\Psi^{0*}_{i,j}(x,\bfk)\Psi^{-1}_{i,j}(x,\bfk)\Bigg)\Bigg],\\
f_{1TT}(x,\bfk)&=&\frac{M_\rho^2}{4(2\pi)^3\textbf{k}_\perp^4}\sum_{i,j}\Bigg[(\textbf{k}_\perp^L)^2\Bigg(\Psi^{-1*}_{i,j}(x,\bfk)\Psi^{+1}_{i,j}(x,\bfk)\Bigg)+(\textbf{k}_\perp^R)^2\Bigg(\Psi^{+1*}_{i,j}(x,\bfk)\Psi^{-1}_{i,j}(x,\bfk)\Bigg)\Bigg].
        \end{eqnarray}
\end{widetext}
The explicit expressions of these TMDs, obtained using the S-1 type spin wave functions, are given by
\begin{widetext}
    \begin{eqnarray}
        f_1(x,\bfk)&=&\frac{1}{2(2\pi)^3}|\phi(x,\bfk)|^2,\\
        g_{1L}(x,\bfk)&=&\frac{1}{2(2\pi)^3}\Bigg(\frac{2 \left(\bfk (2 x-1)+m^2\right) M_{q \bar q}+4 m \left(\bfk+m^2\right)}{2 \left(\bfk+m^2\right) (M_{q \bar q}+2 m)}\Bigg)|\phi(x,\bfk)|^2,\\
        g_{1T}(x,\bfk)&=&\frac{1}{2(2\pi)^3}\Bigg(\frac{2 \text{M}_\rho \left((m-m x) M_{q \bar q}+\bfk+m^2\right)}{\left(\bfk+m^2\right) (M_{q \bar q}+2 m)}\Bigg)|\phi(x,\bfk)|^2,\\
        h_1(x,\bfk)&=&\frac{1}{2(2\pi)^3}\Bigg(\frac{\bfk \left(2 m (x+1) M_{q \bar q}+x (2 x-1) M_{q \bar q}^2+6 m^2\right)+m^2 (M_{q \bar q}+2 m)^2+2 \textbf{k}_\perp^4}{\left(\bfk+m^2\right) (M_{q \bar q}+2 m)^2}\Bigg)|\phi(x,\bfk)|^2,\\
        h_{1L}(x,\bfk)&=&\frac{-1}{2(2\pi)^3}\Bigg(\frac{2 \text{M}_\rho \left((m-m x) M_{q \bar q}+\bfk+m^2\right)}{\left(\bfk+m^2\right) (M_{q \bar q}+2 m)}\Bigg)|\phi(x,\bfk)|^2,\\
        h_{1T}(x,\bfk)&=&\frac{-1}{2(2\pi)^3}\Bigg(\frac{2 \text{M}_\rho^2 \left(2 \left(\bfk+m^2\right)-(x-1) M_{q \bar q} ((1-2 x) M_{q \bar q}+2 m)\right)}{\left(\bfk+m^2\right) (M_{q \bar q}+2 m)^2}\Bigg)|\phi(x,\bfk)|^2,\\
        f_{1LL}(x,\bfk)&=&0,\\
        f_{1LT}(x,\bfk)&=&0,\\
        f_{1TT}(x,\bfk)&=&0.
    \end{eqnarray}
\end{widetext}
%
\begin{figure*}
\centering
\begin{minipage}[c]{0.98\textwidth}
(a)\includegraphics[width=7.5cm]{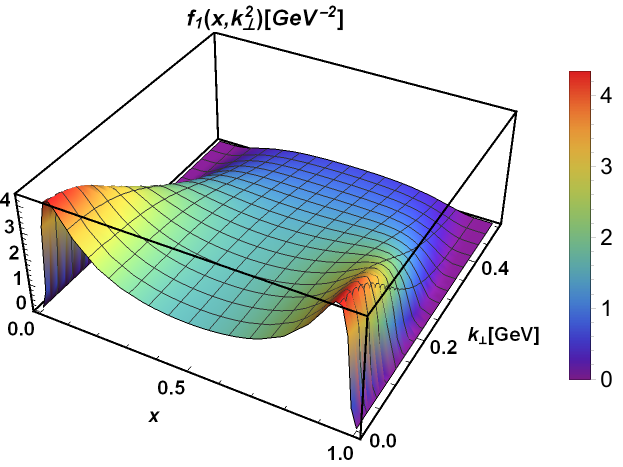}
\hspace{0.03cm}	
(b)\includegraphics[width=7.5cm]{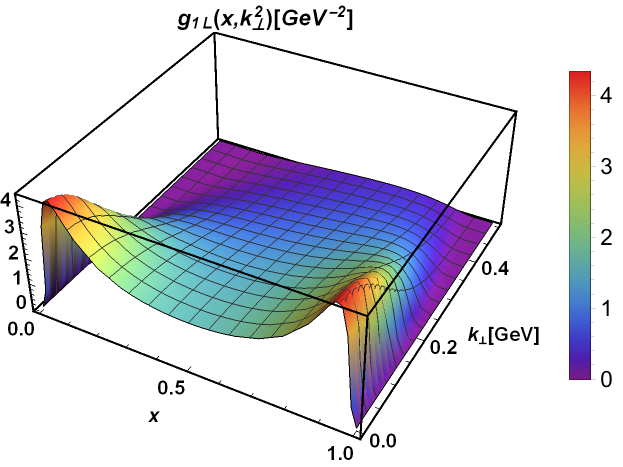} 
\hspace{0.03cm}
\end{minipage}
\centering
\begin{minipage}[c]{0.98\textwidth}
(c)\includegraphics[width=7.5cm]{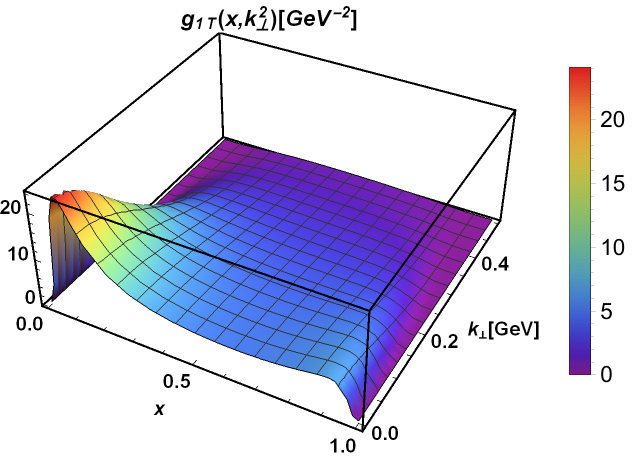}
\hspace{0.03cm}	
(d)\includegraphics[width=7.5cm]{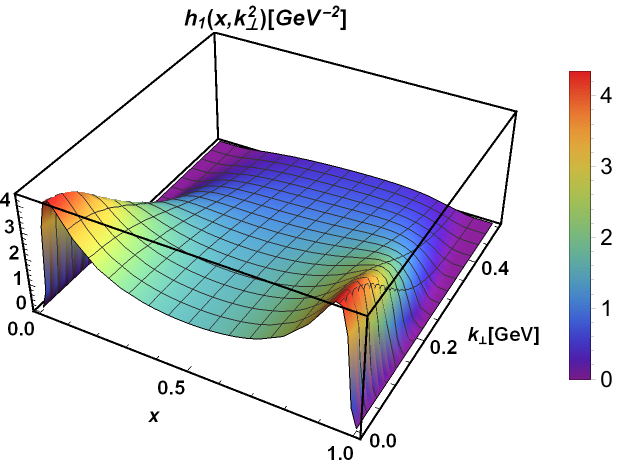} 
\hspace{0.03cm}
\end{minipage}
\centering
\begin{minipage}[c]{0.98\textwidth}
(e)\includegraphics[width=7.5cm]{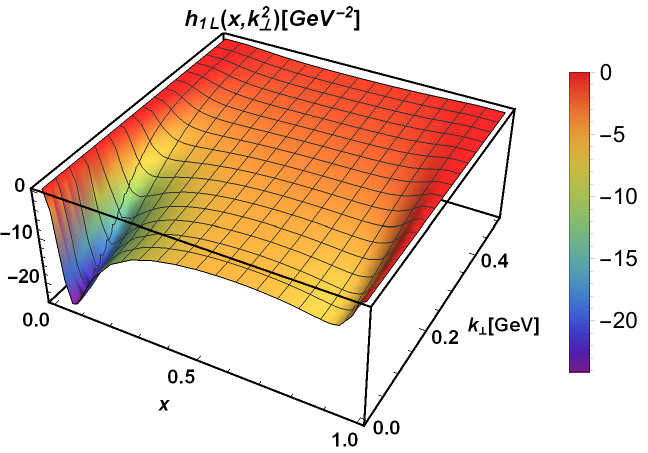}
\hspace{0.03cm}	
(f)\includegraphics[width=7.5cm]{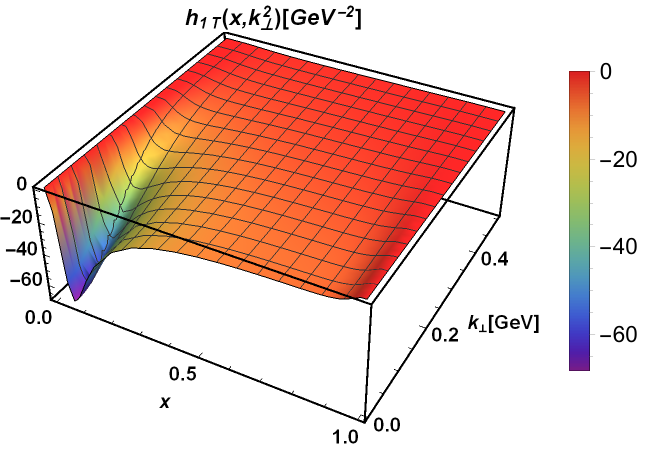} 
\hspace{0.03cm}
\end{minipage}
\caption{Quark TMDs for the S-1 type spin wave function as functions of the longitudinal momentum fraction $x$ and the transverse momentum $|\mathbf{k}_\perp|$ (GeV).
%
\label{fig5}}
\end{figure*}
\begin{figure*}
\centering
\begin{minipage}[c]{0.98\textwidth}
(a)\includegraphics[width=7.5cm]{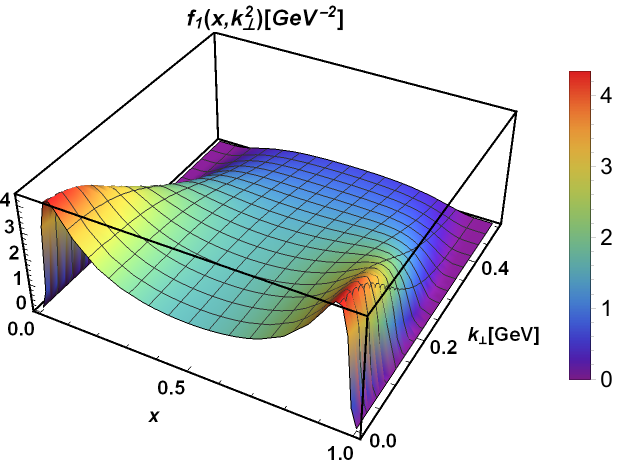}
\hspace{0.03cm}	
(b)\includegraphics[width=7.5cm]{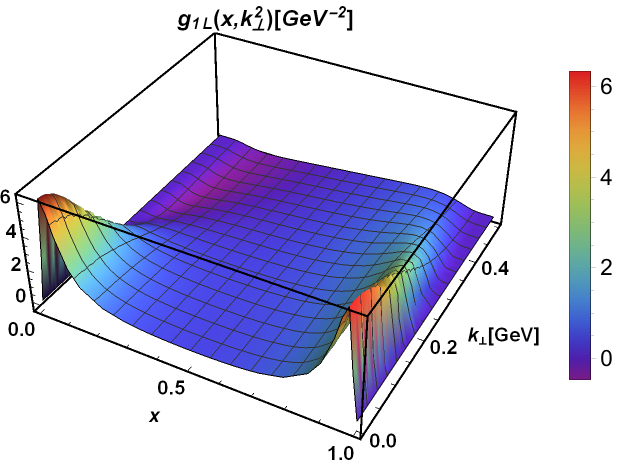} 
\hspace{0.03cm}
\end{minipage}
\centering
\begin{minipage}[c]{0.98\textwidth}
(c)\includegraphics[width=7.5cm]{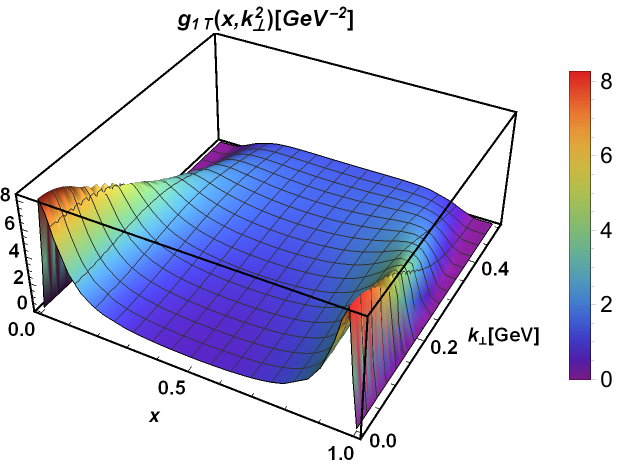}
\hspace{0.03cm}	
(d)\includegraphics[width=7.5cm]{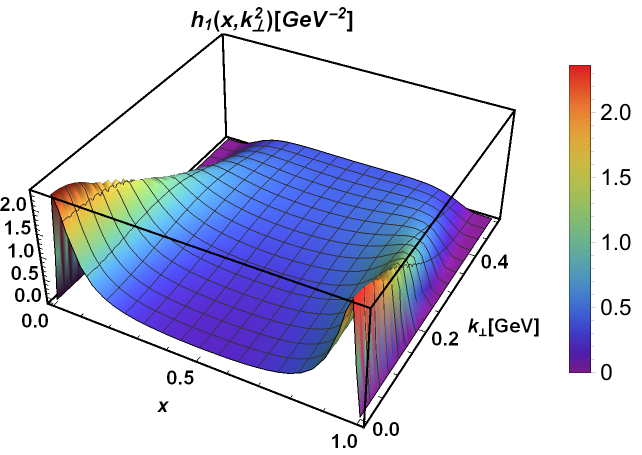} 
\hspace{0.03cm}
\end{minipage}
\centering
\begin{minipage}[c]{0.98\textwidth}
(e)\includegraphics[width=7.5cm]{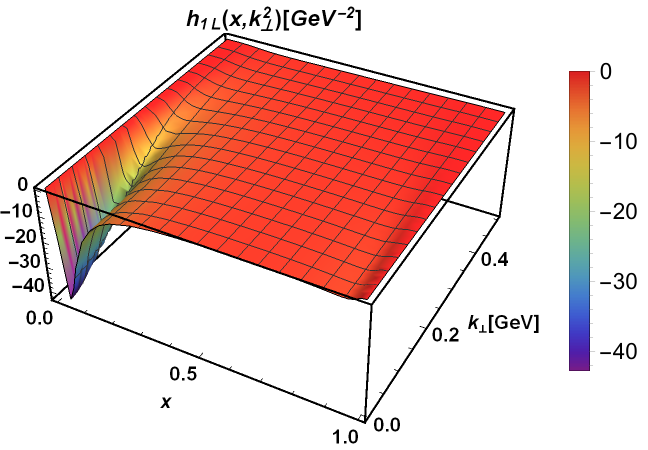}
\hspace{0.03cm}	
(f)\includegraphics[width=7.5cm]{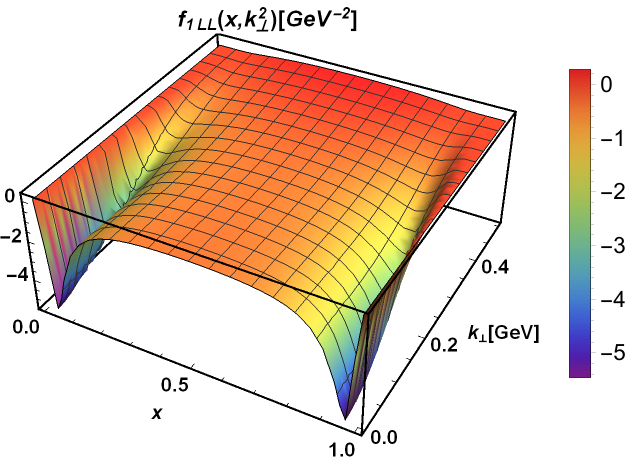} 
\hspace{0.03cm}
\end{minipage}
\centering
\begin{minipage}[c]{0.98\textwidth}
(g)\includegraphics[width=7.5cm]{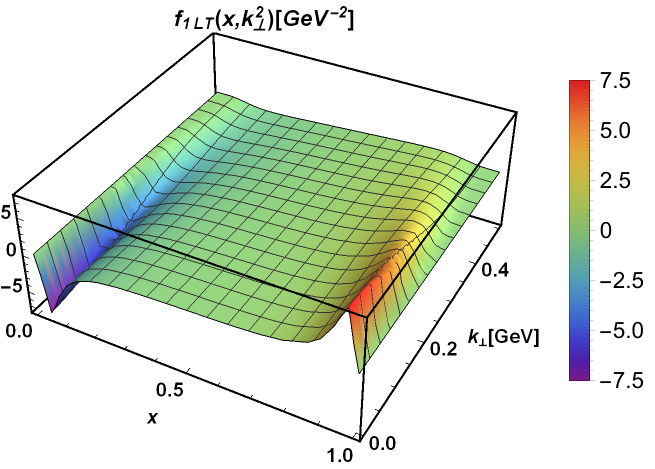}
\hspace{0.03cm}	
(h)\includegraphics[width=7.5cm]{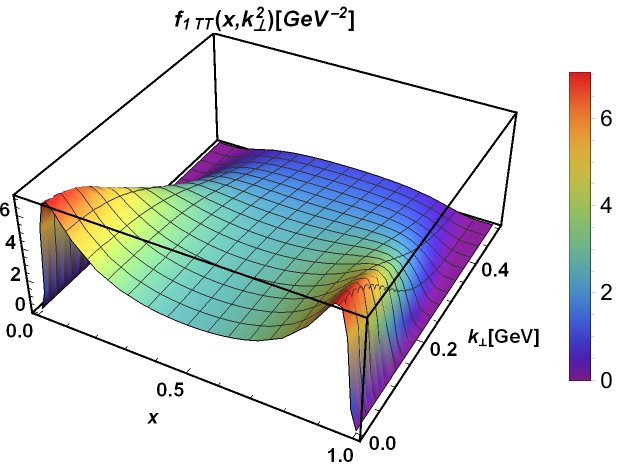} 
\hspace{0.03cm}
\end{minipage}
\caption{Quark TMDs for the S-2 type spin wave function as functions of the longitudinal momentum fraction $x$ and the transverse momentum $|\mathbf{k}_\perp|$ (GeV).
%
\label{fig6}}
\end{figure*}

\begin{figure*}[htbp]
	\centering
	\begin{subfigure}{0.32\textwidth}
		\centering
		\includegraphics[width=\linewidth]{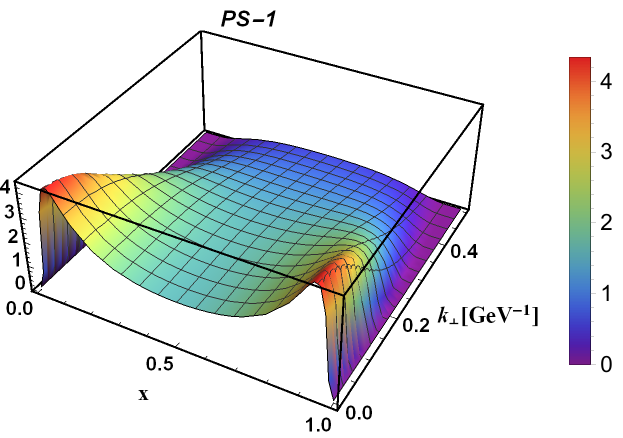}
	\end{subfigure}
	\hfill
	\begin{subfigure}{0.32\textwidth}
		\centering
		\includegraphics[width=\linewidth]{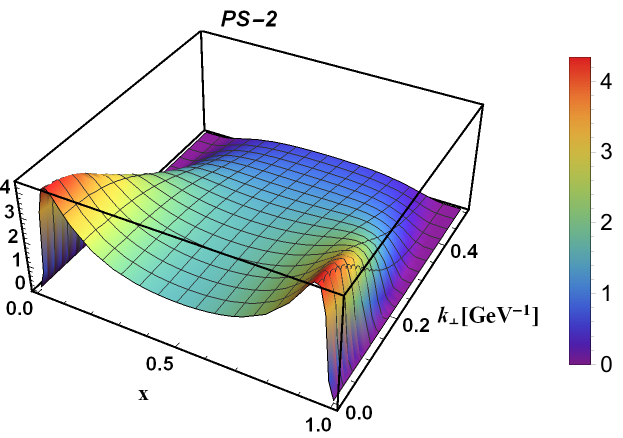}
	\end{subfigure}
	\hfill
	\begin{subfigure}{0.32\textwidth}
		\centering
		\includegraphics[width=\linewidth]{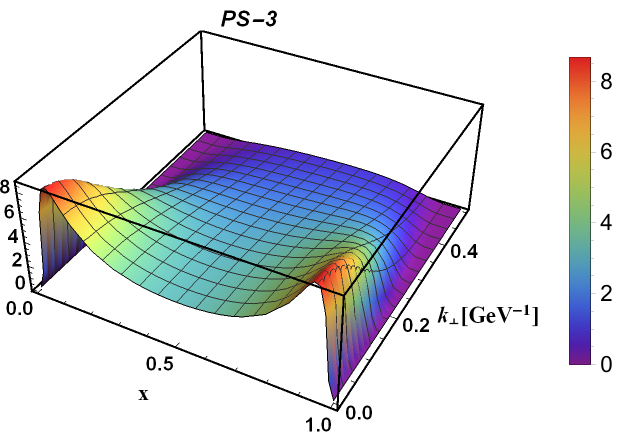}
	\end{subfigure}
	
	\vspace{0.2cm} 
	
	\begin{subfigure}{0.32\textwidth}
		\centering
		\includegraphics[width=\linewidth]{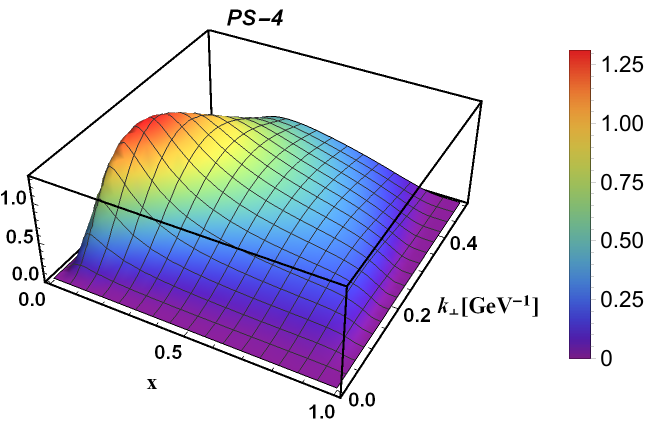}
	\end{subfigure}
	\hfill
	\begin{subfigure}{0.32\textwidth}
		\centering
		\includegraphics[width=\linewidth]{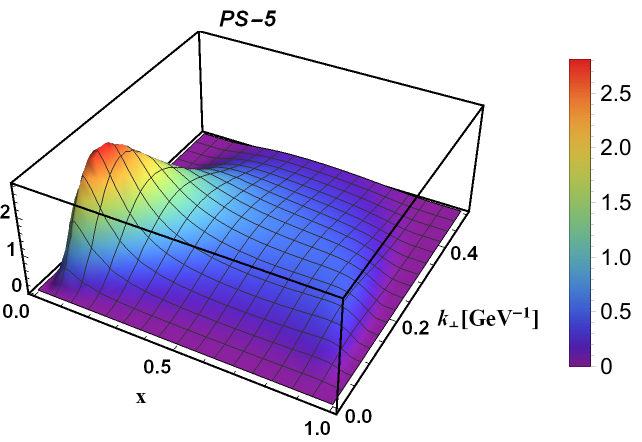}
	\end{subfigure}
	\hfill
	\begin{subfigure}{0.32\textwidth}
		\centering
		\includegraphics[width=\linewidth]{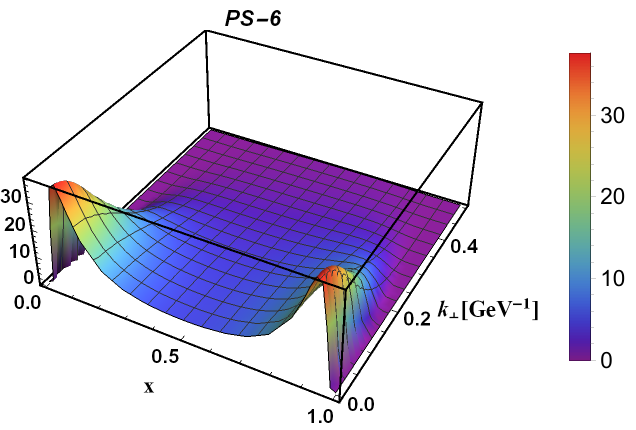}
	\end{subfigure}
	
	\vspace{0.2cm}
	
	\begin{subfigure}{0.32\textwidth}
		\centering
		\includegraphics[width=\linewidth]{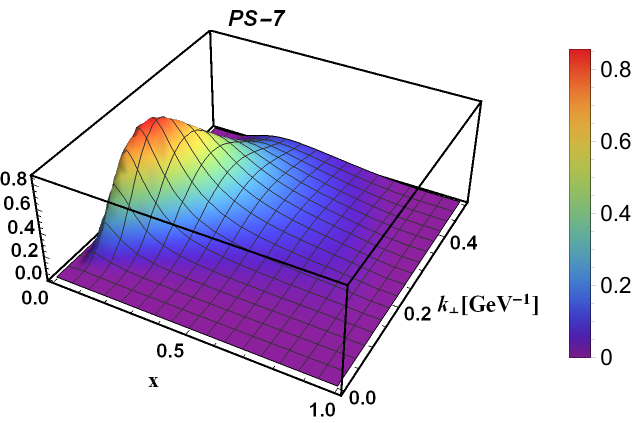}
	\end{subfigure}
	\hfill
	\begin{subfigure}{0.32\textwidth}
		\centering
		\includegraphics[width=\linewidth]{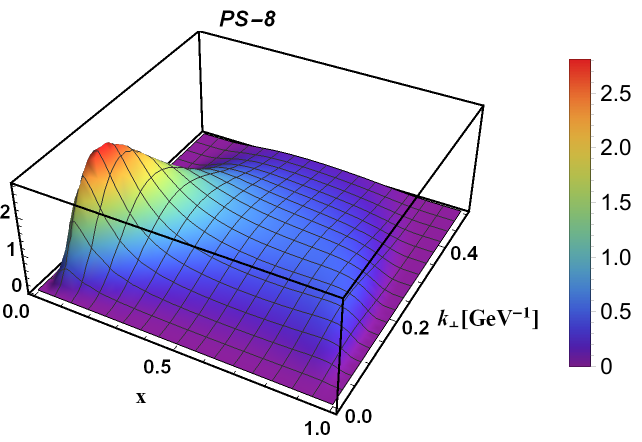}
	\end{subfigure}
	\hfill
	\begin{subfigure}{0.32\textwidth}
		\centering
		\includegraphics[width=\linewidth]{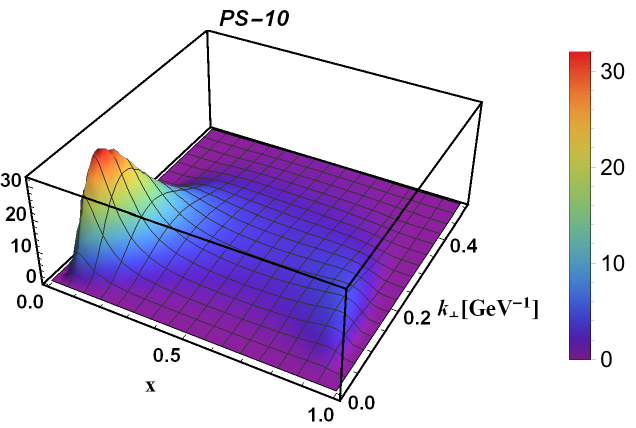}
	\end{subfigure}
	\caption{\label{tmdpcs1} All the positivity constraints on the spin-1 quark TMDs have been plotted with respect to x and $\textbf{k}_\perp$ for S-1 type spin wave function.}
\end{figure*}
\begin{figure*}[htbp]
	\centering
	\begin{subfigure}{0.32\textwidth}
		\centering
		\includegraphics[width=\linewidth]{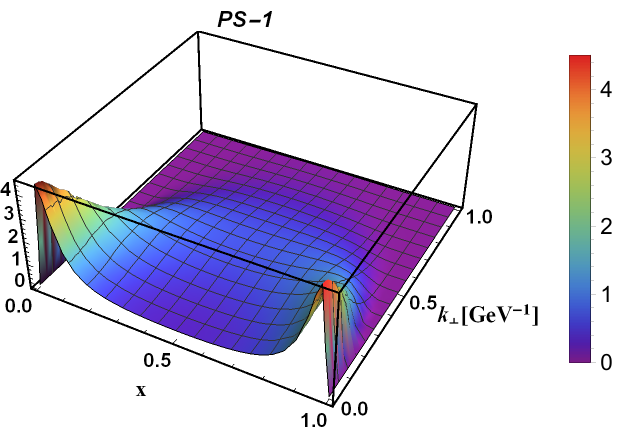}
	\end{subfigure}
	\hfill
	\begin{subfigure}{0.32\textwidth}
		\centering
		\includegraphics[width=\linewidth]{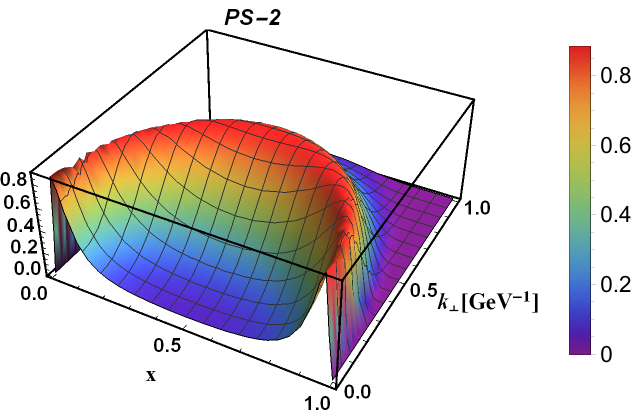}
	\end{subfigure}
	\hfill
	\begin{subfigure}{0.32\textwidth}
		\centering
		\includegraphics[width=\linewidth]{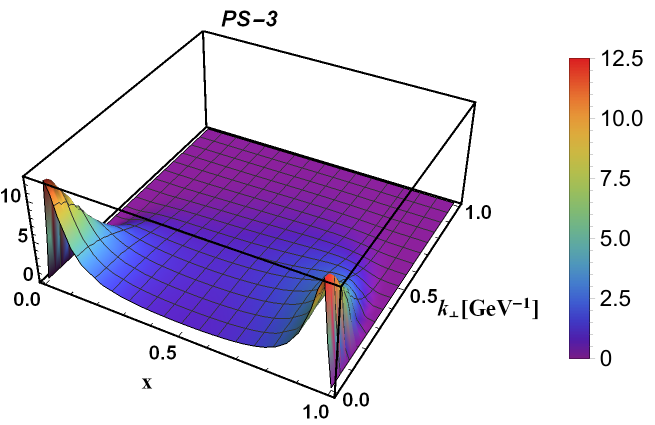}
	\end{subfigure}
	
	\vspace{0.2cm} 
	
	\begin{subfigure}{0.32\textwidth}
		\centering
		\includegraphics[width=\linewidth]{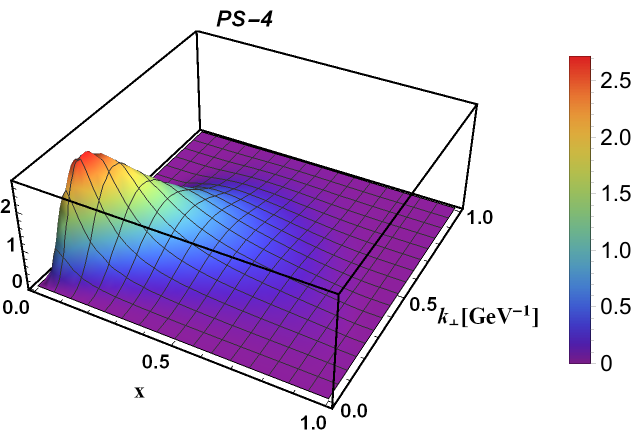}
	\end{subfigure}
	\hfill
	\begin{subfigure}{0.32\textwidth}
		\centering
		\includegraphics[width=\linewidth]{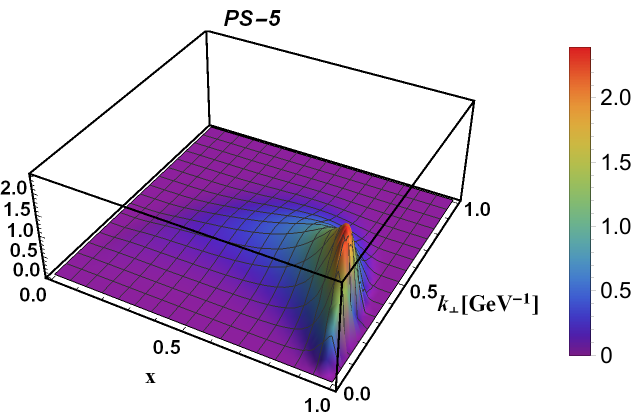}
	\end{subfigure}
	\hfill
	\begin{subfigure}{0.32\textwidth}
		\centering
		\includegraphics[width=\linewidth]{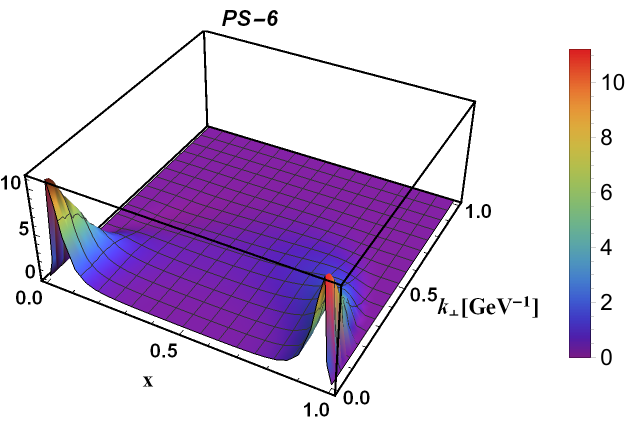}
	\end{subfigure}
	
	\vspace{0.2cm}
	
	\begin{subfigure}{0.32\textwidth}
		\centering
		\includegraphics[width=\linewidth]{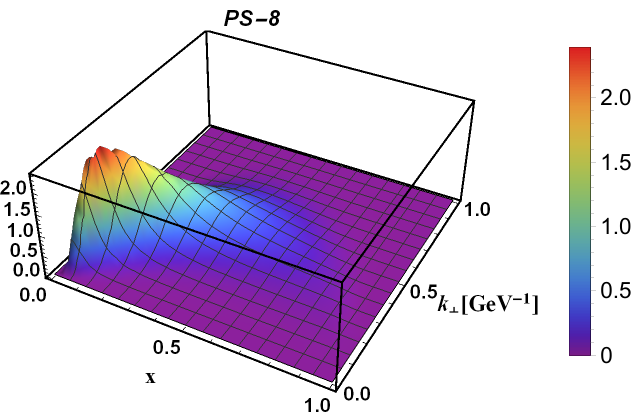}
	\end{subfigure}
	\hfill
	\begin{subfigure}{0.32\textwidth}
		\centering
		\includegraphics[width=\linewidth]{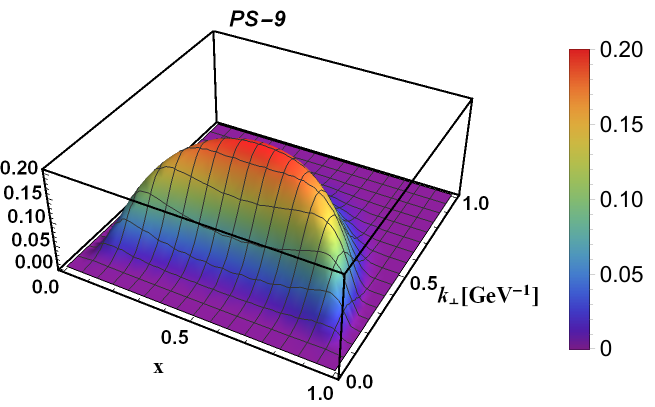}
	\end{subfigure}
	\hfill
	\begin{subfigure}{0.32\textwidth}
		\centering
		\includegraphics[width=\linewidth]{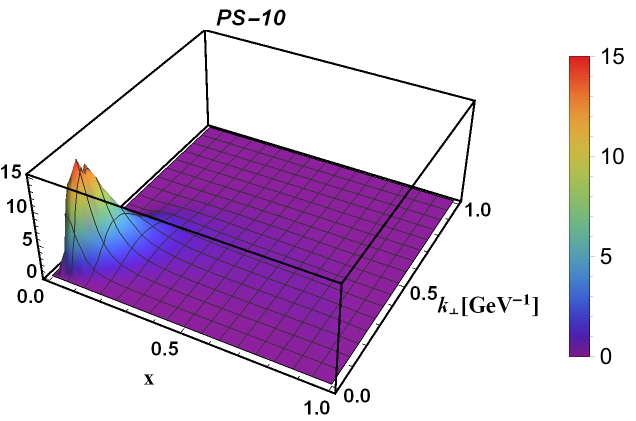}
	\end{subfigure}
	\caption{\label{tmdpc} All the positivity constraints on the spin-1 quark TMDs have been plotted with respect to x and $\textbf{k}_\perp$ for S-2 type spin wave function.}
\end{figure*}
Using the S-2 type spin wave functions, the corresponding TMDs are obtained as
\begin{widetext}
    \begin{eqnarray}
        f_1(x,\bfk)&=& \frac{1}{6 (2 \pi)^3}\Bigg(\frac{2 \left(\text{A}_L^2 \left(\bfk+m^2-\text{M}_{q \bar q}^2 (x-1) x\right)^2+2 \text{A}_T^2 \left(\bfk (2 (x-1) x+1)+m^2\right)\right)}{(1-x) x}\Bigg)|\phi(x,\textbf{k}^2_\perp)|^2,\\
        g_{1L}(x,\bfk)&=& \frac{1}{4 (2 \pi)^3}\Bigg(\frac{4 \text{A}_T^2 \left(\bfk (2 x-1)+m^2\right)}{(1-x) x}\Bigg)|\phi(x,\textbf{k}^2_\perp)|^2,\\
        g_{1T}(x,\bfk)&=&\frac{1}{2(2\pi)^3}\Bigg(\frac{2 \text{A}_L \text{A}_T \text{M}_\rho \left(\bfk+m^2-\text{M}_{q\bar q}^2 (x-1) x\right)}{(1-x) x}\Bigg)|\phi(x,\bfk)|^2,\\
        h_1(x,\bfk)&=&\frac{1}{2(2\pi)^3}\Bigg(\frac{2 \text{A}_L \text{A}_T m \left(\bfk+m^2-\text{M}_{q \bar q}^2 (1-x) x\right)}{(x-1) x}\Bigg)|\phi(x,\bfk)|^2,\\
        h_{1L}(x,\bfk)&=&\frac{-1}{2(2\pi)^3}\Bigg(\frac{4 \text{A}_T^2 m \text{M}_\rho}{x}\Bigg)|\phi(x,\bfk)|^2,\\
        h_{1T}(x,\bfk)&=&0,\\
        f_{1LL}(x,\bfk)&=&\frac{1}{2(2\pi)^3}\Bigg(\frac{2 \left(\text{A}_T^2 \left(\bfk (2 (x-1) x+1)+m^2\right)-\text{A}_L^2 \left(\bfk+m^2-\text{M}_{q \bar q}^2 (x-1) x\right)^2\right)}{(x-1) x}\Bigg)|\phi(x,\bfk)|^2,\\
        f_{1LT}(x,\bfk)&=&\frac{1}{2(2\pi)^3}\Bigg(\frac{2 \text{A}_L \text{A}_T \text{M}_\rho (2 x-1) \left(\bfk+m^2-M_{q \bar q}^2 (x-1) x\right)}{(1-x) x}\Bigg)|\phi(x,\bfk)|^2,\\
        f_{1TT}(x,\bfk)&=&\frac{1}{2(2\pi)^3}\Bigg(4 \text{A}_T^2 \text{M}_\rho^2\Bigg)|\phi(x,\bfk)|^2.
    \end{eqnarray}
\end{widetext}
All the tensor quark TMDs are found to be zero for the case of S-1 spin wave function, while the $h_{1T}(x,\bfk)$ quark TMD are coming to be zero for the S-2 type spin wave functions. We observed that the unpolarized $f_1(x,\bfk)$, helicity $g_{1L}(x,\bfk)$, transversity $h_1(x,\bfk)$, worm-gear-2 $h_{1L}(x,\bfk)$ and tensor $f_{1LL}(x,\bfk)$ quark TMDs have zero OAM transferred between the initial and final state hadron. The worm-gear-1 $g_{1T}(x,\bfk)$ and pretzelosity $h_{1T}(x,\bfk)$ quark TMDs have one unit of OAM transferred from the initial hadron to the final hadron, while the tensor $f_{1LT}(x,\bfk)$ and $f_{1TT}(x,\bfk)$ quark TMDs have two units of OAM transferred.
\par In Figs. \ref{fig5} and \ref{fig6}, we have plotted all the quark TMDs of S-1 and S-2 type spin wave function with respect to longitudinal momentum fraction $x$ and transverse momenta $\textbf{k}_\perp$, respectively. The unpolarized $f_1(x,\bfk)$ quark TMD, which describes the momentum distribution while both quark and hadron have no polarizations, is found to show symmetry under the transformation $x\longleftrightarrow(1-x)$. For both the spin wave function case, the $f_1(x,\bfk)$ is found to have a double peak distribution at $x \approx 0$ and 1, which can also be seen in the case of LFHM \cite{Kaur:2020emh}. However, in NJL \cite{Zhang:2024plq,Ninomiya:2017ggn}, ILM \cite{Liu:2025fuf} and BSE \cite{Shi:2022erw} models, the $f_1(x,\bfk)$ quark TMD found to have only one peak distributions. The $g_{1L}(x,\bfk)$ and $g_{1T}(x,\bfk)$ quark TMDs correspond to the longitudinally polarized quark in a longitudinally and transversely polarized hadron, respectively. While the $h_1(x,\bfk)$ and $h_{1T}(x,\bfk)$ quark TMDs describe the momentum distribution of a transversely polarized quark in a transversely polarized hadron. The remaining worm-gear-2 $h_{1L}(x,\bfk)$ TMD describes the transversely polarized quark in a longitudinally polarized hadron and all tensor TMDs describe the momentum distribution of an unpolarized quark in a tensor polarized hadron. Like $f_1(x,\bfk)$ quark TMD, the $g_{1L}(x,\bfk)$ and $h_1(x,\bfk)$ quark TMDs also show symmetry about $x \longleftrightarrow (1-x)$ along with double peak distributions for both the spin wave function. While the $g_{1T}(x,\bfk)$ quark TMDs shows single peak distribution at low $x$ for the S-1 type spin wave function and two peak distribution for the S-2 type spin wave function. The $f_1(x,\bfk)$, $g_{1L}(x,\bfk)$, $g_{1T}(x,\bfk)$ and $h_1(x,\bfk)$ quark TMDs shows only positive distributions, while the $h_{1L}(x,\bfk)$ and $h_{1T}(x,\bfk)$ shows negative distributions for both the spin wave functions. Similar kinds of observations have also been reported in NJL model \cite{Ninomiya:2017ggn,Zhang:2024plq}, BSE model \cite{Shi:2022erw}, LFHM \cite{Kaur:2020emh} and ILM \cite{Liu:2025fuf}. For the case of the S-1 spin wave function, the $h_{1T}(x,\bfk)$ quark TMDs survive and vanish for the case of S-2 spin wave function. This TMD is found to have a higher distribution peak compared to other quark TMDs and does not show any symmetry around x. Even the $h_{1T}(x,\bfk)$ quark TMDs are coming out to be zero in the case of NJL model \cite{Ninomiya:2017ggn,Zhang:2024plq}, LFHM \cite{Kaur:2020emh} and ILM \cite{Liu:2025fuf}. The relation among $g_{1T}(x,\bfk)$ and $h_1(x,\bfk)$ quark TMDs as discussed in Refs. \cite{Zhang:2024plq,Ninomiya:2017ggn}
\begin{eqnarray}
    g_{1T}(x,\bfk)=\frac{m}{M_\rho}h_1(x,\bfk) 
\end{eqnarray}
is not followed by the S-1 type spin wave function, while satisfied in the S-2 type spin wave function. The quark TMDs in S-2 type spin wave function are also found to have a minimum distribution around $0.2 \le x \le 0.8$. While in the case of ILM results, the distributions are mostly found at $x \approx 0.5$ \cite{Liu:2025fuf}. For both the spin wave functions, the momentum distributions decrease with increasing transverse momenta. All the quark TMDs are found to be vanish after $\textbf{k}_\perp \ge 0.4$ GeV.
\par Now, looking into the tensor TMDs, which are absent for spin-$\frac{1}{2}$ nucleons, the quark distributions are coming to zero due to the symmetric spin wave function of S-1 type. These tensor TMDs are coming to be non-zero for the S-2 type spin wave function. In Fig. \ref{fig6} (f), \ref{fig6} (g) and \ref{fig6} (h), we have plotted the tensor quark TMDs for the case of S-2 type spin wave function. The tensor $f_{1LL}(x,\bfk)$ and $f_{1TT}(x,\bfk)$ quark TMDs are found to have negative and positive distributions all over the $x$ and $\textbf{k}_\perp$, respectively. While the $f_{1LT}(x,\bfk)$ quark TMD has both positive and negative momentum distributions in the region $0.5 \le x \le 1$ and $0 \le x \le 0.5$, respectively. These distributions of the tensor quark TMDs are indications of the non-zero quark OAM. Under the transformation $x \longleftrightarrow (1-x)$, the tensor $f_{1LL}(x,\bfk)$ and $f_{1TT}(x,\bfk)$ quark TMDs show symmetry, while $f_{1LT}(x,\bfk)$ shows anti-symmetry. In Refs. \cite{Ninomiya:2017ggn,Zhang:2024plq,Liu:2025fuf,Shi:2022erw}, the tensor $f_{1LL}(x,\bfk)$ quark TMDs have some positive distribution near $x \approx 0.5$, which is not seen in our case. In our case, the $f_{1LL}(x,\bfk)$ is found to have nearly zero distributions near $x=0.5$, which can also be seen in LFHM \cite{Kaur:2020emh}. We have also observed that $f_{1LT}(x,\bfk)$ is found to be zero at $x=0.5$ all over the $\textbf{k}_\perp$ values. 
\par The leading twist PDFs can also be extracted from the leading twist TMDs as
\begin{eqnarray}
    f(x)&=& \int d^2\textbf{k}_\perp f_1(x,\bfk), \\
     g(x)&=& \int d^2\textbf{k}_\perp g_{1L}(x,\bfk), \\
      h(x)&=& \int d^2\textbf{k}_\perp h_1(x,\bfk), \\
       f_{1LL}(x)&=& \int d^2\textbf{k}_\perp f_{1LL}(x,\bfk).
\end{eqnarray}
We have also predicted the average transverse momenta $\langle \textbf{k}_\perp \rangle$ carried by the valence quark for individual TMDs as 
\begin{eqnarray}
    \langle \textbf{k}_\perp \rangle= \frac{\int dx d^2 \textbf{k}_\perp \textbf{k}_\perp TMD(x,\bfk)}{\int dx d^2 \textbf{k}_\perp TMD(x,\bfk)}.
\end{eqnarray}
The $\langle \textbf{k}_\perp \rangle$ values of each TMD have been presented in Table \ref{transverse} for both the spin wave functions, along with comparison with different theoretical models \cite{Shi:2022erw,Kaur:2020emh,Ninomiya:2017ggn,Liu:2025fuf,Zhang:2024plq}. Our results are found to be in good agreement with others predictions. We observed that the $\langle \textbf{k}_\perp \rangle$ is found to have a higher value for S-2 type spin wave function compared to S-1 type spin wave function. As one can expect, the unpolarized $f_1(x,\bfk)$ should carry higher transverse momentum, which can also be seen in our case.
\subsection*{Positivity Constraints}
The spin-1 quark TMDs also obey some positivity constraints, which have been discussed in Refs. \cite{Ninomiya:2017ggn,Zhang:2024plq,Kaur:2020emh} as
\begin{widetext}
  \begin{align}
	f_1(x,\bfk) &\ge 0, && (PS\text{-}1) \\[3pt]
	f_1(x,\bfk) + \tfrac{2}{3} f_{1LL}(x,\bfk) &\ge 0, && (PS\text{-}2) \\[3pt]
	f_1(x,\bfk) - \tfrac{1}{3} f_{1LL}(x,\bfk) + g_{1L}(x,\bfk) &\ge 0, && (PS\text{-}3) \\[3pt]
	f_1(x,\bfk) - \tfrac{1}{3} f_{1LL}(x,\bfk) - g_{1L}(x,\bfk) &\ge 0, && (PS\text{-}4) \\[3pt]
	\Big(f_1 + \tfrac{2}{3}f_{1LL}\Big)
	\Big(f_1 - \tfrac{1}{3}f_{1LL} + g_{1L}\Big)
	&\ge 2\, h_1^2, && (PS\text{-}5) \\[3pt]
	\Big(f_1 + \tfrac{2}{3}f_{1LL}\Big)
	\Big(f_1 - \tfrac{1}{3}f_{1LL} + g_{1L}\Big)
	&\ge \frac{\bfk^2}{2M_\rho^2}\big(g_{1T}+f_{1LT}\big)^2, && (PS\text{-}6) \\[3pt]
	\Big(f_1 + \tfrac{2}{3}f_{1LL}\Big)
	\Big(f_1 - \tfrac{1}{3}f_{1LL} - g_{1L}\Big)
	&\ge \frac{\bfk^2}{2M_\rho^2}\big(g_{1T}-f_{1LT}\big)^2, && (PS\text{-}7) \\[3pt]
	\Big(f_1 + \tfrac{2}{3}f_{1LL}\Big)
	\Big(f_1 - \tfrac{1}{3}f_{1LL} - g_{1L}\Big)
	&\ge \frac{\bfk^4}{2M_\rho^4} h_{1T}^2, && (PS\text{-}8) \\[3pt]
	\Big(f_1 - \tfrac{1}{3}f_{1LL}\Big)^2 - g_{1L}^2
	&\ge \frac{\bfk^4}{M_\rho^4} f_{1TT}^2, && (PS\text{-}9) \\[3pt]
	\Big(f_1 - \tfrac{1}{3}f_{1LL}\Big)^2 - g_{1L}^2
	&\ge \frac{\bfk^2}{M_\rho^2} h_{1L}^2. && (PS\text{-}10)
\end{align}  
\end{widetext}
These positivity constraints have been plotted as a function of x and $\textbf{k}_\perp$ in Figs. \ref{tmdpcs1} and \ref{tmdpc} for S-1 and S-2 type spin wave function, respectively. We observed that all the positivity constraints are obeyed by both the spin wave functions. We also noticed that PS-9 and PS-7 positivity constraints are coming zero for S-1 and S-2 spin wave function, respectively.

\section*{Conclusion}
\label{Conclusion}
We have presented a comprehensive analysis of the quark structure of spin-$1$ vector mesons within the LFQM framework, in terms of both PDFs and TMDs. The longitudinal degree of freedom of the quarks are studied using the leading twist and subleading twist PDFs. While the transverse structure have been studied using the quark TMDs. The quark PDFs and TMDs have been calculated by considering the trace of spin-1 spin densities. We have presented the PDFs and TMDs through a matrix, which is equal to a LFWFs amplitude matrix. The LFWFs form of quark PDFs and TMDs have been presented along with the explicit forms. We have presented all the quark PDFs and TMDs through two dimensional and three dimensional plots. We have observed that the leading twist quark PDFs and TMDs have similar kind of behavior with available other theoretical models.
\par The leading twist PDFs have been evolved to higher scales through NLO DGLAP evolutions. These evolved PDFs are plotted through two dimensional pictures. To predict the average longitudinal momentum fraction or Mellin moment, we have used a simple NN frame. The calculated Mellin moment are found to have similar values with other model predictions. The PDF sum rule and positivity constraints are found to be obeyed by our model. The average transverse momenta carried by the quark TMDs also calculated in this model. The positivity constraints on the leading twist TMDs also obeyed by our model. Overall, our model predictions are found to have similar results with other model predictions. We will also looking for the spin asymmetries arises for the case of spin-1 hadrons in our next project.
The upcoming Jefferson Lab, Nuclotron-based Ion Collider fAcility (NICA) and electron ion collider (EIC) will provide more insight about the tensor structure of spin-1 systems. 
\section*{Acknowledgement}
We thank Prof. Wolfgang Bentz and Prof. Oleg V. Teryaev for fruitful discussions. S. P. gratefully acknowledges the Bogoliubov Laboratory of Theoretical Physics (BLTP), Joint Institute for Nuclear Research (JINR), Dubna, for providing the research facilities and support during his visit, where this work was carried out. H.D. would like to thank the Science and Engineering
 Research Board, Anusandhan-National Research Foundation,
 Government of India under the scheme SERB-POWERFellow
ship (Ref No. SPF/2023/000116) for financial support.
\section{Appendix}\label{appendixspin}
For spin-$1$ hadrons, there are a total of $36$ helicity matrix amplitudes depending upon the quark antiquark helicities ($i(j)$) and hadron polarizations ($\lambda$) , which can be arranged into a $6 \times 6$ matrix as
\begin{widetext}
\begin{eqnarray}
\Phi= \left(
\begin{array}{cccccc}
A_{\uparrow(+1),\uparrow(+1)} ~&~ A_{\uparrow(+1),\uparrow(0)} ~&~ A_{\uparrow(+1),\uparrow(-1)} ~&~ A_{\uparrow(+1),\downarrow(+1)} ~&~ A_{\uparrow(+1),\downarrow(0)} ~&~ A_{\uparrow(+1),\downarrow(-1)}\\
A_{\uparrow(0),\uparrow(+1)}& A_{\uparrow(0),\uparrow(0)}& A_{\uparrow(0),\uparrow(-1)} & A_{\uparrow(0),\downarrow(+1)}& A_{\uparrow(0),\downarrow(0)} & A_{\uparrow(0),\downarrow(-1)}\\
A_{\uparrow(-1),\uparrow(+1)}& A_{\uparrow(-1),\uparrow(0)} & A_{\uparrow(-1),\uparrow(-1)} & A_{\uparrow(-1),\downarrow(+1)} & A_{\uparrow(-1),\downarrow(0)} & A_{\uparrow(-1),\downarrow(-1)}\\
A_{\downarrow(+1),\uparrow(+1)} & A_{\downarrow(+1),\uparrow(0)} & A_{\downarrow(+1),\uparrow(-1)} & A_{\downarrow(+1),\downarrow(+1)} & A_{\downarrow(+1),\downarrow(0)} & A_{\downarrow(+1),\downarrow(-1)}\\
A_{\downarrow(0),\uparrow(+1)} & A_{\downarrow(0),\uparrow(0)} & A_{\downarrow(0),\uparrow(-1)} & A_{\downarrow(0),\downarrow(+1)} & A_{\downarrow(0),\downarrow(0)} & A_{\downarrow(0),\downarrow(-1)}\\
A_{\downarrow(-1),\uparrow(+1)} & A_{\downarrow(-1),\uparrow(0)} & A_{\downarrow(-1),\uparrow(-1)} & A_{\downarrow(-1),\downarrow(+1)} & A_{\downarrow(-1),\downarrow(0)} & A_{\downarrow(-1),\downarrow(-1)}
\end{array}
\right)\,.
\label{matrix-helicity}
\end{eqnarray}
\end{widetext}
Here, the general element of the helicity amplitude matrix $\Phi$ is expressed as 
\begin{eqnarray*}
    A_{i^{\prime}(\lambda^{\prime}),i(\lambda)}&=&\int \frac{d^2\textbf{k}_\perp}{(2\pi)^3}\sum_j \Psi^{\lambda^\prime *}_{i^\prime,j}(x,\bfk)\Psi^{\lambda}_{i,j}(x,\bfk)
\end{eqnarray*}
for quark PDF calculations, and as 
\begin{eqnarray*}
    A_{i^{\prime}(\lambda^{\prime}),i(\lambda)}&=& \frac{1}{(2\pi)^3}\sum_j \Psi^{\lambda^\prime *}_{i^\prime,j}(x,\bfk)\Psi^{\lambda}_{i,j}(x,\bfk)
\end{eqnarray*}
for TMD calculations. In this notation, $\lambda^{\prime}$ and $\lambda$ denote the final and initial hadron polarizations, respectively, while $i^{\prime}$ and $i$ correspond to the final and initial quark polarizations. The $6 \times 6$ light-front helicity matrix can be written in terms of the parametrization of spin-$1$ meson TMDs and quark PDFs as 
\begin{eqnarray*}
    \Phi=Tr[\rho_{\lambda^{\prime},\lambda}M_{i,j}],
\end{eqnarray*} 
where $\rho_{\lambda^{\prime},\lambda}$ is the spin density matrix for spin-1 mesons and $M_{i,j}$ is the scattering matrix encoding the hadron spin structure. For leading-twist quark PDFs, the corresponding $M_{i,j}$ is given by 
\begin{widetext}
\begin{eqnarray}
M_{i,j} &=& 
\left(
\begin{array}{cc}
M_{\uparrow,\uparrow} & M_{\uparrow,\downarrow} \\
M_{\downarrow,\uparrow} & M_{\downarrow,\downarrow} \\
\end{array}
\right)
\nonumber\\
&=&
\left(
\begin{array}{cc}
f(x) + \lambda S_{L}\, g(x) + S_{LL}\, f_{1LL}(x) &
(S_\perp^x + i S_\perp^y)\, h(x) \\[6pt]
(S_\perp^x - i S_\perp^y)\, h(x) &
f(x) - \lambda S_{L}\, g(x) + S_{LL}\, f_{1LL}(x) \\
\end{array}
\right).
\label{matrix-helicity}
\end{eqnarray}

Similarly, for the case of TMDs, the scattering matrix takes the form
\begin{eqnarray}
M_{i,j} &=& 
\left(
\begin{array}{cc}
A_1 & A_2 \\
A_3 & A_4 \\
\end{array}
\right),
\end{eqnarray}
where
\begin{align}
A_1 &= f_1(x,\bfk) + S_{LL} f_{1LL}(x,\bfk) 
+ \frac{S_{LT}\cdot \mathbf{k}_\perp}{M_\rho} f_{1LT}(x,\bfk) 
+ \frac{\mathbf{k}_\perp \cdot S_{TT} \cdot \mathbf{k}_\perp}{M_\rho^2} f_{1TT}(x,\bfk) 
\nonumber\\
&\quad+ \lambda \left( S_L g_{1L}(x,\bfk) + \frac{S_T \cdot \mathbf{k}_\perp}{M_\rho} g_{1T}(x,\bfk)\right), \nonumber\\[6pt]
A_2 &= \lambda (S_T^x+i S_T^y)\, h_1(x,\bfk) 
+ \frac{\lambda S_L (k_\perp^x+i k_\perp^x)}{M_\rho} h_{1L}(x,\bfk) \nonumber\\
&\quad+ \frac{\lambda}{2 M_\rho^2} 
\left[\,2 i k_\perp^x S_T^x k_\perp^y 
- \bfk(S_T^x+i S_T^y) 
+ 2 i (k_\perp^y)^2 S_T^y 
+ 2 k_\perp^x (\mathbf{k}_\perp \cdot S_T^{i})\,\right] h_{1T}(x,\bfk), \nonumber\\[6pt]
A_3 &= \lambda (S_T^x-i S_T^y)\, h_1(x,\bfk) 
+ \frac{\lambda S_L (k_\perp^x-i k_\perp^x)}{M_\rho} h_{1L}(x,\bfk) \nonumber\\
&\quad+ \frac{\lambda}{2 M_\rho^2} 
\left[\,-2 i k_\perp^x S_T^x k_\perp^y 
- \bfk (S_T^x-i S_T^y) 
- 2 i (k_\perp^y)^2 S_T^y 
+ 2 k_\perp^x (\mathbf{k}_\perp \cdot S_T^{i})\,\right] h_{1T}(x,\bfk), \nonumber\\[6pt]
A_4 &= f_1(x,\bfk) + S_{LL} f_{1LL}(x,\bfk) 
+ \frac{S_{LT}\cdot \mathbf{k}_\perp}{M_\rho} f_{1LT}(x,\bfk) 
+ \frac{\mathbf{k}_\perp \cdot S_{TT} \cdot \mathbf{k}_\perp}{M_\rho^2} f_{1TT}(x,\bfk) 
\nonumber\\
&\quad- \lambda \left( S_L g_{1L}(x,\bfk) + \frac{S_T \cdot \mathbf{k}_\perp}{M_\rho} g_{1T}(x,\bfk)\right).
\end{align}
\end{widetext}
Using the spin density matrix from Refs.~\cite{Bacchetta:2001rb,Cotogno:2017puy} and the above scattering amplitudes, the $6 \times 6$ leading-twist PDF matrix is obtained as
\begin{eqnarray}
&&\Phi(x)= \nonumber\\
&&\left(
\begin{array}{cccccc}
A(x) & 0 & 0 & 0 & \sqrt{2}h(x) & 0\\
0 & B(x) & 0 & 0 & 0 & \sqrt{2}h(x)\\
0 & 0 & C(x) & 0 & 0 & 0\\
0 & 0 & 0 & C(x) & 0 & 0\\
\sqrt{2}h(x) & 0 & 0 & 0 & B(x) & 0\\
0 & \sqrt{2}h(x) & 0 & 0 & 0 & A(x)
\end{array}
\right)\,,\nonumber\\
\label{maatrix}
\end{eqnarray}
where $A(x)=f(x)+g(x)-\tfrac{f_{1LL}(x)}{3}$, $B(x)=f(x)+\tfrac{2 f_{1LL}(x)}{3}$, and $C(x)=f(x)-g(x)-\tfrac{f_{1LL}(x)}{3}$. Similar results have also been reported in Refs.~\cite{Bacchetta:2001rb,Ninomiya:2017ggn,Kaur:2020emh}. In an analogous way, the $6\times6$ matrix for the leading-twist T-even TMDs is obtained as 
\begin{widetext}
    \begin{eqnarray}
\Phi(x,\bfk)=\left(
\begin{array}{cccccc}
f_1+f_3^- &  \frac{\textbf{k}_\perp^L}{\sqrt{2}\,M_\rho} g_{1T}^{(+)}
& \frac{{(\textbf{k}_\perp^L)}^2}{M_\rho^2}f_{1TT}    & \frac{\textbf{k}_\perp^R}{M_\rho} h_{1L}^{\perp}  & \sqrt{2} \, h_1  &  0\\  \\
\frac{\textbf{k}_\perp^R}{\sqrt{2}\,M_\rho}  g_{1T}^{(+)} & f_1+\frac{2}{3}f_{1LL}  &
\frac{\textbf{k}_\perp^L}{\sqrt{2}\,M_\rho}  g_{1T}^{(-)} & \frac{{(\textbf{k}_\perp^R)}^2}{\sqrt{2}\,M_\rho^2} h_{1T}^{\perp}  &  0  &  \sqrt{2} \, h_1  \\  \\
 \frac{{(\textbf{k}_\perp^R)}^2}{M_\rho^2}  f_{1TT} 
& \frac{\textbf{k}_\perp^R}{\sqrt{2}\,M_\rho} g_{1T}^{(-)} & f_1-f_4^+ & 0  &  \frac{{(\textbf{k}_\perp^R)}^2}{\sqrt{2}\,M_\rho^2}  h_{1T}^{\perp}  &  
- \frac{\textbf{k}_\perp^R}{M_\rho}   h_{1L}^{\perp}   \\   \\ 
\frac{\textbf{k}_\perp^L}{M_\rho}  h_{1L}^{\perp}  &  
\frac{{(\textbf{k}_\perp^L)}^2}{\sqrt{2}\,M_\rho^2}  h_{1T}^{\perp} & 0  & f_1-f_4^+ &  -\frac{\textbf{k}_\perp^L}{\sqrt{2}\, M_\rho} g_{1T}^{(-)}
&  \frac{{(\textbf{k}_\perp^L)}^2}{M_\rho^2}  f_{1TT} \\  \\
\sqrt{2} \, h_1  &   0 &  \frac{{(\textbf{k}_\perp^L)}^2}{\sqrt{2}\,M_\rho^2}  h_{1T}^{\perp}  & -\frac{\textbf{k}_\perp^R}{\sqrt{2}\, M_\rho} g_{1T}^{(-)} & f_1+\frac{2}{3}f_{1LL}  &
-\frac{\textbf{k}_\perp^L}{\sqrt{2}\, M_\rho} g_{1T}^{(+)} \\ \\
0  &  \sqrt{2} \, h_1  &  -\frac{\textbf{k}_\perp^L}{M_\rho}   h_{1L}^{\perp}  &  \frac{k^2_R}{M_\rho^2} f_{1TT} 
& -\frac{\textbf{k}_\perp^R}{\sqrt{2}\, M_\rho}  g_{1T}^{(+)} & f_1+f_3^-     
\label{matrix}
\end{array}
\right)\,,
\end{eqnarray}
\end{widetext}
with $f_3^{+}=g_{1L}-\frac{1}{3}f_{1LL}$, $f_4^+=g_{1L}+\frac{1}{3}f_{1LL}$, $g_{1T}^{\pm}=g_{1T}\pm f_{1LT}$. Things to note that, In Refs. \cite{Ninomiya:2017ggn,Kaur:2020emh}, the authors have the same matrix for the leading twist TMDs.

Furthermore, following the same procedure, we have derived the subleading-twist-quark PDFs for spin-1 mesons, for which the corresponding matrix is found to be
\begin{widetext}
    \begin{eqnarray}
&&\frac{\Phi(x)}{M_{\rho}}= \nonumber\\
&&\left(
\begin{array}{cccccc}
\frac{x e}{m}-\frac{x e_{LL}}{3 m}+\frac{ x h_L}{m} &2 x \frac{\textbf{k}_\perp^L}{\bfk}\frac{(f_{LT}+g_T)}{\sqrt{2}} & 0 & -x\frac{\textbf{k}_\perp^R}{\bfk}h_L & \sqrt{2}\frac{x}{m}g_T & 0\\
2 x \frac{\textbf{k}_\perp^R}{\bfk}\frac{(f_{LT}+g_T)}{\sqrt{2}} & \frac{x e}{m}+\frac{2 x e_{LL}}{3 m} & -2x \frac{\textbf{k}_\perp^L}{\bfk}\frac{(f_{LT}-g_T)}{\sqrt{2}} & 0 & 0 & \sqrt{2}\frac{x}{m}g_T\\
0 & -2x \frac{\textbf{k}_\perp^R}{\bfk}\frac{(f_{LT}-g_T)}{\sqrt{2}} & \frac{x e}{m}-\frac{x e_{LL}}{3 m}-\frac{ x h_L}{m} & 0 & 0 & x\frac{\textbf{k}_\perp^R}{\bfk}h_L\\
- x\frac{\textbf{k}_\perp^L}{\bfk}h_L & 0 & 0 & \frac{x e}{m}-\frac{x e_{LL}}{3 m}-\frac{ x h_L}{m} &2x \frac{\textbf{k}_\perp^L}{\bfk}\frac{(f_{LT}-g_T)}{\sqrt{2}} & 0\\
\sqrt{2}\frac{x}{m}g_T & 0 & 0 &2 x \frac{\textbf{k}_\perp^R}{\bfk}\frac{(f_{LT}-g_T)}{\sqrt{2}} & \frac{x e}{m}+\frac{2 x e_{LL}}{3 m} & -2x \frac{\textbf{k}_\perp^L}{\bfk}\frac{(f_{LT}+g_T)}{\sqrt{2}}\\
0 & \sqrt{2}\frac{x}{m}g_T & x\frac{\textbf{k}_\perp^L}{\bfk}h_L & 0 & -2x \frac{\textbf{k}_\perp^R}{\bfk}\frac{(f_{LT}+g_T)}{\sqrt{2}} & \frac{x e}{m}-\frac{x e_{LL}}{3 m}+
\frac{ x h_L}{m}
\end{array}
\right)\,.\nonumber\\
\label{twist3pdf}
\end{eqnarray}
\end{widetext}

\bibliographystyle{apsrev}  
\bibliography{ref} 
\end{document}